\documentclass[12pt]{article}
\usepackage{epsfig,amsmath,amsfonts,amssymb,makeidx,ifthen}
\newtheorem{theorem}{Theorem}[section]
\newtheorem{lemma}[theorem]{Lemma}

\newtheorem{pro}[theorem]{Proposition}
\newtheorem{definition}[theorem]{Definition}

\makeatletter
\@addtoreset{equation}{section}
\makeatother

\makeatletter
\long\def\@makecaption#1#2{{\small
\advance\leftskip1cm
\advance\rightskip1cm
\vskip\abovecaptionskip
\sbox\@tempboxa{#1: #2}%
\ifdim \wd\@tempboxa >\hsize
 #1: #2\par
\else
\global \@minipagefalse
\hb@xt@\hsize{\hfil\box\@tempboxa\hfil}%
\fi
\vskip\belowcaptionskip}}
\makeatother
\def\eq#1\en{\begin{equation}#1\end{equation}}  
\def\eqa#1\ena{\begin{align}#1\end{align}}
\def\eqg#1\eng{\begin{gather}#1\end{gather}}
\newcommand{\lb}[1]{\label{e:#1}}
\newcommand{\rlb}[1]{\eqref{e:#1}} 
\newcommand{\nl}{\notag\\}

\newcommand{\abs}[1]{\left|#1\right|}
\newcommand{\norm}[1]{\left\Vert#1\right\Vert}

\newcommand{\rbk}[1]{\left(#1\right)}

\newcommand{\bkt}[1]{\left\langle#1\right\rangle}
\newcommand{\sbkt}[1]{\langle#1\rangle}
\newcommand{\bbkt}[1]{\bigl\langle#1\bigr\rangle}
\newcommand{\Bbkt}[1]{\Bigl\langle#1\Bigr\rangle}
\newcommand{\set}[2]{\left\{#1\,\Bigl|\,#2\right\}}

\newcommand{\sumtwo}[2]%
{\mathop{\sum_{#1}}_{#2}}
\newcommand{\sumthree}[3]%
{\mathop{\mathop{\sum_{#1}}_{#2}}_{#3}}
\newcommand{\sumfour}[4]%
{\mathop{\mathop{\mathop{\sum_{#1}}_{#2}}_{#3}}_{#4}} 
\newcommand{\prodtwo}[2]%
{\mathop{\prod_{#1}}_{#2}}
\newcommand{\mintwo}[2]%
{\mathop{\min_{#1}}_{#2}}
\newcommand{\maxtwo}[2]%
{\mathop{\max_{#1}}_{#2}}
\newcommand{\maxthree}[3]%
{\mathop{\mathop{\max_{#1}}_{#2}}_{#3}}
\newcommand{\limtwo}[2]%
{\mathop{\lim_{#1}}_{#2}}
\newcommand{\suptwo}[2]%
{\mathop{\sup_{#1}}_{#2}}
\newcommand{\supthree}[3]%
{\mathop{\mathop{\sup_{#1}}_{#2}}_{#3}}
\newcommand{\supfour}[4]%
{\mathop{\mathop{\mathop{\sup_{#1}}_{#2}}_{#3}}_{#4}} 
\newcommand{\inftwo}[2]%
{\mathop{\inf_{#1}}_{#2}}
\newcommand{\infthree}[3]%
{\mathop{\mathop{\inf_{#1}}_{#2}}_{#3}}
\newcommand{\inffour}[4]%
{\mathop{\mathop{\mathop{\inf_{#1}}_{#2}}_{#3}}_{#4}} 

\newcommand\calG{{\cal G}}
\newcommand\calH{{\cal H}}

\newcommand\calK{{\cal K}}

\newcommand\calN{{\cal N}}

\newcommand\calS{{\cal S}}

\newcommand{\tim}{\tilde{m}}

\newcommand{\tiz}{\tilde{z}}
\newcommand{\tiD}{\tilde{D}}
\newcommand{\tiJ}{\tilde{J}}
\newcommand{\tiK}{\tilde{K}}
\newcommand{\tiM}{\tilde{M}}

\newcommand{\bsm}{\boldsymbol{m}}
\newcommand{\bsn}{\boldsymbol{n}}

\newcommand{\bsp}{\boldsymbol{p}}

\newcommand{\bsr}{\boldsymbol{r}}

\newcommand{\bssigma}{\boldsymbol{\sigma}}
\newcommand{\bstau}{\boldsymbol{\tau}}

\newcommand{\sfA}{\mathsf{A}}
\newcommand{\sfB}{\mathsf{B}}
\newcommand{\sfC}{\mathsf{C}}

\newcommand{\sfN}{\mathsf{N}}

\newcommand{\bbR}{\mathbb{R}}
\newcommand{\bbZ}{\mathbb{Z}}

\newcommand{\up}{\uparrow}

\newcommand{\Di}{\mathit{\Delta}}
\newcommand{\qedm}{\rule{1.5mm}{3mm}}

\newcommand{\Htot}{\calH_{V,\mathrm{tot}}}
\newcommand{\Ham}{\mathsf{H}_V}
\newcommand{\HVu}{\mathcal{H}_{V,u}}
\newcommand{\tHVu}{\tilde{\mathcal{H}}_{V,u}}
\newcommand{\HVtu}{\mathcal{H}_{\Vtot,u}}
\newcommand{\tHVtu}{\tilde{\mathcal{H}}_{\Vtot,u}}
\newcommand{\Du}{\Di u}
\newcommand{\Vtot}{{V_\mathrm{tot}}}
\newcommand{\hA}{\mathsf{A}}
\newcommand{\hM}{\mathsf{M}}
\newcommand{\hO}{\mathsf{O}}
\newcommand{\hP}{\mathsf{P}}
\newcommand{\hQ}{\mathsf{Q}}
\newcommand{\hS}{\mathsf{S}}
\newcommand{\Pneq}{\hP\!_\mathrm{neq}}
\newcommand{\bktmc}[1]{\sbkt{#1}^\mathrm{mc}_{V,u}}
\newcommand{\Bktmc}[1]{\Bbkt{#1}^\mathrm{mc}_{V,u}}
\newcommand{\JVu}{J_{V,u}}
\newcommand{\DVu}{D_{V,u}}
\newcommand{\Dneq}{D_\mathrm{neq}}
\newcommand{\Deff}{D_\mathrm{eff}}
\newcommand{\MiV}{\hM^{(i)}_V}
\newcommand{\tMV}{\tiM_V}
\newcommand{\tMiV}{\tilde{\hM}^{(i)}_V}

\newcommand{\MV}{\hM_V}
\newcommand{\hBV}{\hat{\mathsf{B}}_V}
\newcommand{\BV}{\mathsf{B}_V}

\newcommand{\miu}{m^{(i)}(u)}
\newcommand{\di}{\delta^{(i)}}

\newcommand{\bra}[1]{\langle#1|}
\newcommand{\ket}[1]{|#1\rangle}
\newcommand{\kpj}{\ket{\psi_j}}
\newcommand{\kph}{\ket{\varphi}}
\newcommand{\kpz}{\ket{\ph(0)}}
\newcommand{\kpt}{\ket{\ph(t)}}
\newcommand{\kxi}{\ket{\xi_i}}

\newcommand{\DU}{\Di U}

\newcommand{\ph}{\varphi}
\newcommand{\hH}{\mathsf{H}}
\newcommand{\hm}{\mathsf{m}}

\newcommand{\Ym}{Y_\mathrm{main}}
\newcommand{\Yr}{Y_\mathrm{rem}}

\newcommand{\ez}{\epsilon_0}
\newcommand{\eo}{\epsilon_1}
\newcommand{\kB}{k_\mathrm{B}}
\newcommand{\oL}{\{1,2,\ldots,L\}}

\newcommand{\hc}{\mathsf{c}}
\newcommand{\hcd}{\mathsf{c}^\dagger}
\newcommand{\ha}{\mathsf{a}}
\newcommand{\had}{\mathsf{a}^\dagger}

\newcommand{\vac}{\ket{\Phi_\mathrm{vac}}}
\newcommand{\sumx}{\sum_{x\in\Lambda}}
\newcommand{\sumk}{\sum_{k\in\calK}}
\newcommand{\zz}{z_0}
\newcommand{\zi}{z_{\rm coup}}
\newcommand{\ei}{\epsilon_{\rm coup}}

\newcommand{\hf}{\mathsf{f}}
\newcommand{\hg}{\mathsf{g}}

\newcommand{\limV}{\lim_{V\up\infty}}

\begin{document}
\noindent
{\bf
\Large Typicality of thermal equilibrium and thermalization

\vspace{1.5mm}
\noindent
in isolated macroscopic quantum systems
}
\par\bigskip

\noindent
Hal Tasaki\footnote{
Department of Physics, Gakushuin University, Mejiro, Toshima-ku, 
Tokyo 171-8588, Japan
}

%
%
\begin{abstract}
Based on the view that thermal equilibrium should be characterized through macroscopic observations, we develop a general theory about typicality of thermal equilibrium and the approach to thermal equilibrium in macroscopic quantum systems.
We first formulate the notion that a pure state in an isolated quantum system represents thermal equilibrium.  
Then by assuming, or proving in certain classes of nontrivial models (including that of two bodies in thermal contact), large-deviation type bounds (which we call thermodynamic bounds) for the microcanonical ensemble, we prove that to represent thermal equilibrium is a typical property for pure states in the microcanonical energy shell.  
We believe that the typicality, along with the empirical success of statistical mechanics, provides a sound justification of equilibrium statistical mechanics.
We also establish the approach to thermal equilibrium under two different assumptions; one is that the initial state has a moderate energy distribution, and the other is the energy eigenstate thermalization hypothesis.
\end{abstract}
%
%
\tableofcontents

\section{Introduction}

\subsection{Motivation and background}
The recent renewed interest in the foundation of quantum statistical mechanics and in the dynamics of isolated quantum systems has led to a revival of the old approach by von Neumann to address the foundation of equilibrium statistical mechanics in terms of quantum dynamics in an isolated system \cite{vonNeumann,GLTZ}.
It has been demonstrated in some general or concrete settings that a pure initial state evolving under quantum dynamics indeed approaches the equilibrium state \cite{Tasaki1998,Reimann,LindenPopescuShortWinter,
GLMTZ09b,Hal2010,ReimannKastner,Reimann2,SatoKanamotoKaminishiDeguchi,Reimann3}.
The important related idea that a single pure quantum state can fully describe thermal equilibrium has become much more concrete \cite{vonNeumann,PopescuShortWinter,GLTZ06,Sugita06,Sugita07,Reimann07,SugiuraShimizu12,SugiuraShimizu13,GHLT}, and the notion of ``energy eigenstate thermalization'', which indeed goes back to von Neumann \cite{vonNeumann}, has been investigated in various physical situations \cite{Deutsch1991,Srednicki1994,Horoi1995,Zelevinsky1996,Tasaki1998}.

In the present paper we discuss the foundation of equilibrium statistical mechanics by taking into account the above mentioned recent progress and also accumulated results in mathematical physics of many-body quantum systems.
Our theory closely follows those developed by von Neumann \cite{vonNeumann,GLTZ} and by Goldstein, Lebowitz, Mastrodonato, Tumulka, and Zangh\`\i\ \cite{GLMTZ09b}.

We take the view that thermal equilibrium should be characterized only through the observation of macroscopic quantities.
We focus on the microcanonical setting, and describe a macroscopic system as an isolated quantum system.
We also assume that a state of the system is described by a quantum mechanical pure state\footnote{
Throughout the present paper, ``pure state'' implies a quantum mechanical pure state, which is a vector in the Hilbert space (or the many-body wave function).
We treat mixed states in Appendix~\ref{s:mixed}.
}.

A pure state in an isolated quantum many-body system can be (approximately) realized in very limited situations including ultra cold atom systems\footnote{
To relate our theory to cold atom experiments may be an important future issue.
}.
We nevertheless believe it meaningful and fruitful to study such an idealized situation, and to learn which phenomena can be reproduced in this limit.
After doing that, we may study extra effects played by the interaction or entanglement with the external environment.

A crucial starting point of our theory is a formulation of the notion that a quantum mechanical pure state describes thermal equilibrium.
Briefly speaking, we say that a pure state represents thermal equilibrium if a single measurement (in the state) of macroscopic quantities yield, with probability extremely close to one, the corresponding equilibrium values with very high precision.
Assuming (or proving in some cases) the {\em thermodynamic bound}\/ which guarantees that the system behaves as a normal macroscopic system, we prove a theorem which provides a clear interpretation (and, hopefully, a partial justification) based on the typicality point of view of the microcanonical distribution.
A crucial point is that our theory does not rely only on abstract quantum mechanical argument, but also makes use of concrete properties of physically realistic systems.
We also discuss thermalization, i.e., the approach to thermal equilibrium, and prove two preliminary theorems.

We have tried to make the present paper self-contained; most part of the paper is accessible to graduate (or even undergraduate) students who have proper background in quantum mechanics and statistical mechanics.

\bigskip
The present paper is organized as follows.
In the following section~\ref{s:what}, we shall discuss  basic pictures about the typicality of thermal equilibrium and the approach to thermal equilibrium.
The purpose of the present paper is to provide concrete and mathematically rigorous foundation to this picture.

In the important section~\ref{s:setup}, we describe basic setup and important notions.
After fixing the  notations in section~\ref{s:MQS}, we define in section~\ref{s:pure} the notion of pure states representing thermal equilibrium.
Then in section~\ref{s:TDB} we introduce our essential assumption called thermodynamic bound.
In section~\ref{s:comparison}, we compare our formulation with those in the literature.

In section~\ref{s:examples}, we discuss three classes of examples where the thermodynamic bound is provable, and hence our theory is applicable.
All the theorems in this section are proved in section~\ref{s:proofTDB}.

Section~\ref{s:typicality} is the most important section of the paper, where we prove the typicality of thermal equilibrium assuming the thermodynamic bound.

Sections~\ref{s:thermalization}, \ref{s:moderate}, and \ref{s:ETH} are devoted to the problem of thermalization.
After stating a general condition for thermalization in section~\ref{s:thermalization}, we discuss two strategies for justifying the condition in the two sections that follow.
In section~\ref{s:moderate}, we focus on the assumption that the initial state has a moderate energy distribution.
Such an assumption has been used in various works, but we shall here give a careful analysis about the meaning of the assumption (at least in the present context).
In section~\ref{s:ETH}, we focus on the energy eigenstate thermalization hypothesis.

In section~\ref{s:discussion}, we summarize the paper, and discuss some open problems.

In Appendix~\ref{s:toy}, we discuss three simple solvable models in which one can study the typicality of thermal equilibrium and thermalization explicitly.

There are three appendices which discuss extensions.
In Appendix~\ref{s:small}, we explain how one can extend the present theory to cover quantities which are not macroscopic.
We show that one can treat correlation functions and the probability distribution in a small system.
In Appendix~\ref{s:mixed}, we discuss thermalization when the initial state is a mixed state.
In Appendix~\ref{s:canonical}, we briefly discuss the justification of the canonical distribution in the similar spirit as in the main body of the paper.

\subsection{What is thermal equilibrium and how do we get there?}
\label{s:what}
Before developing the theory for macroscopic quantum systems, we shall briefly discuss basic pictures about thermal equilibrium and thermalization.

\paragraph{Macroscopic picture:}
Thermodynamics is founded on several premises which are justified empirically through macroscopic observations and operations on macroscopic systems.
Among the most important premises is that a macroscopic system, when isolated from the outside world, approaches a unique thermal equilibrium after a sufficiently long time.
The thermal equilibrium is characterized by a small number of macroscopic quantities; in the case of a system consisting of a single substance, only the volume $V$, the amount of substance $N$, and the total energy $U$ are sufficient for the full characterization.

\paragraph{Microscopic picture:}
Consider, for example, a system of $N$ molecules confined in a region with volume $V$.
Then, from the macroscopic point of view, one only needs to specify the total energy $U$ to fully characterize thermal equilibrium.
From the microscopic point of view, however, the information about the total energy is far from enough to characterize the state of the system.

Take, for simplicity, the classical description, and let $\Gamma=(\bsr_1,\ldots,\bsr_N,\bsp_1,\ldots,\bsp_N)$ be the microscopic state of the system, where $\bsr_i$ and $\bsp_i$ denote the coordinate and the momentum, respectively, of the $i$-the particle.
The total energy of the system is given by
\eq
E(\Gamma):=\sum_{i=1}^N\frac{|\bsp_i|^2}{2m}+V(\bsr_1,\ldots,\bsr_N),
\lb{claE}
\en
where $m$ is the mass of the molecules, and $V(\bsr_1,\ldots,\bsr_N)$ is the potential energy.

Given the information that the total energy is (almost) $U$, we see that the state $\Gamma$ belongs to the energy shell 
\eq
\calS_U:=\{\Gamma\,|\,U-\DU\le E(\Gamma)\le U+\DU\},
\lb{SU}
\en
where $\DU$ is an energy interval which is negligible from the macroscopic point of view\footnote{
The corresponding energy shell $\HVu$ for for quantum case is defined in section~\ref{s:MQS}.
See also section~\ref{s:typclassical} for a comparison between classical and quantum cases.
}.
Since there are enormous variety of states in the space $\calS_U$, it is not at all obvious how the unique thermal equilibrium characterized by $U$ is related to $\calS_U$.

Of course the desired relation is given by the principle of equal weights, i.e.,  one should uniformly average over all the states in $\calS_U$ to represent the thermal equilibrium.
It has been confirmed repeatedly in the history that equilibrium statistical mechanics, which is based on this principle, reproduces a wide range of phenomena in nature with great accuracy.
Since there is no doubt about the applicability of the principle, the main question is to understand the physical picture behind the principle, and hopefully the reason why it works.

\begin{figure}
\begin{center}
\includegraphics[width=8cm]{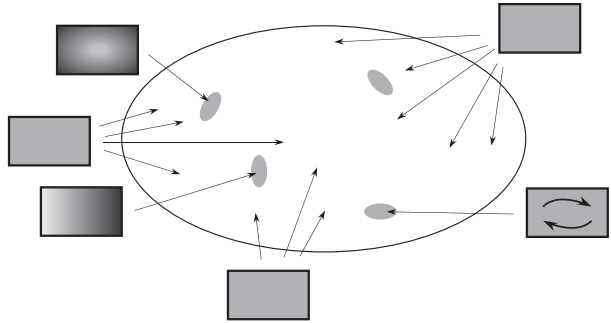}
\end{center}
\caption[dummy]{
The big elliptic region represents the energy shell, the space of states that have a macroscopic energy $U$.
Based on the {\em fact}\/ that an overwhelming majority of the states in the energy shell look almost the same from the macroscopic point of view, we {\em postulate}\/ that those states in the majority represent the unique thermal equilibrium.
Exceptional states which belong to very small shaded regions have macroscopic properties different from the thermal equilibrium.

Note that this is a very crude picture of the energy shell, which indeed have extremely high dimension.
Also note that the exceptional regions must be much smaller than depicted.
The figure is taken from \protect{\cite{TasakiSM}}.
}
\label{f:AS}
\end{figure}

\paragraph{Typicality of thermal equilibrium:}
We believe that a physically natural interpretation of the principle of equal weights is provided by an argument based on {\em typicality}\/\footnote{
We believe that the ergodic property of (classical) dynamical systems, although being quite interesting and deep, has little to do with the justification of equilibrium statistical mechanics.
} \cite{Lebowitz93B,Lebowitz07}.
See Figure~\ref{f:AS}.

The argument starts with the (mathematical) {\em fact}\/ that, in a normal macroscopic system, an overwhelming majority of the states in the energy shell $\calS_U$ are almost indistinguishable if one only makes macroscopic measurement.
Of course this fact should be established for concrete microscopic models of macroscopic systems.
Then we shall make a (physical) {\em postulate}\/ that thermal equilibrium is nothing but the collection of properties shared by all these overwhelming majority of states.
We shall say that a state which belong to the majority {\em represents thermal equilibrium}\/.

If one picks up a state from $\calS_U$, the chance is big\footnote{
We must note, however, that one does not choose a state randomly in reality.
See Section~\ref{s:typphys} for further discussion about the preparation of states.
} that it belongs to the overwhelming majority, unless there are special reasons to expect otherwise.
In other words, to represent thermal equilibrium is a {\em typical property}\/ for the states in the energy shell $\calS_U$.

This picture solves the above mentioned puzzle about the uniqueness and robustness of the thermal equilibrium characterized by $U$.
Although the microscopic states in the energy shell $\calS_U$ are far from unique, they look essentially unique from the macroscopic point of view.
{\em Thermal equilibrium is observed in a robust manner simply because it is represented by the majority of states}\/.\footnote{
We note in passing that numerical simulations (based, e.g., on the molecular dynamics or the Markov process Monte Carlo methods) for equilibrium states of a macroscopic system work effectively because they are designed to generate states which belong to the majority (and hence represent thermal equilibrium).
In this sense the Monte Carlo simulation in statistical mechanics is essentially different from Monte Carlo calculations of low-dimensional integrals.
}

Although the ultimate justification of equilibrium statistical mechanics must come from the collection of empirical facts, we believe that this theoretical picture about the typicality of thermal equilibrium provides a clear and convincing interpretation.
One of the purposes of the present paper is to give a precise formulation to the picture for isolated macroscopic quantum systems, and {\em prove}\/ the typicality of thermal equilibrium in certain important concrete settings.

\paragraph{Thermalization:}
Once accepting this picture of typicality, we see why it is natural that the state of the system, in the long run, should approach thermal equilibrium  \cite{Lebowitz93B,Lebowitz07}.

\begin{figure}
\begin{center}
\includegraphics[width=12cm]{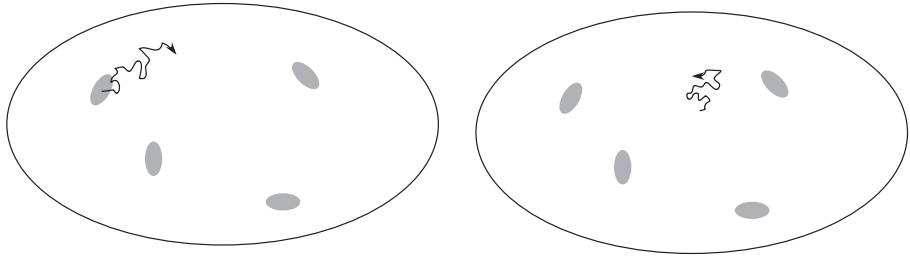}
\end{center}
\caption[dummy]{
The basic picture of thermalization based on the typicality point of view.
Left:~If the state was initially in one of the exceptional regions, it eventually moves out of the  region and evolves into a state which belongs to the overwhelming majority.
Right:~If the state initially belongs to the majority, it rarely wanders into the exceptional regions, and keeps representing thermal equilibrium.
The figure is taken from \protect{\cite{TasakiSM}}.
}
\label{f:IR}
\end{figure}

Suppose that the initial state describes a physical situation which is very far from equilibrium, e.g., two bodies at different temperatures in contact.
The initial state certainly is an ``exceptional'' state which does not belong to the overwhelming majority, but belongs to a very small region of exceptional states.
The region is by definition surrounded by states which belong to the majority.

Then, unless the dynamics of the system has a special property (in relation with the exceptional region), it is expected that the state won't stay in this small exceptional region for a long time.
When the state gets out of the region, it joins the majority of states which represent thermal equilibrium.
From the macroscopic point of view, this process can be interpreted as  approach to thermal equilibrium, or thermalization.
See Figure~\ref{f:IR}.
Note that no special properties of the time evolution, such as chaoticity,  is assumed in this rough argument.
The argument is of course consistent with the reversibility of the microscopic dynamics since we are only discussing the escape from the very special initial region.

Since we are dealing with macroscopic isolated quantum system in this paper, let us discuss the corresponding dynamics.
By using the notation defined in section~\ref{s:MQS}, the state of the system at time $t$ is written as
\eq
\kpt=\sum_{j}c_j\,e^{-iE_jt}\,\kpj,
\lb{pt}
\en
where $\kpj$ is the energy eigenstate.
In order for the initial state $\ket{\ph(0)}$ to be out of thermal equilibrium, the coefficients $c_j$ should be chosen with an extreme care and precision so that the linear combination does not belong to the great majority.
As $t$ grows, each coefficient gets individual phase factor as in \rlb{pt}, and the very delicate balance of the coefficients realized at $t=0$ will soon be lost.
It is likely that, after a sufficiently long time, the state $\kpt$ is no longer exceptional, and belongs to the overwhelming majority which represent thermal equilibrium.

Again we shall make this idea precise and prove some results which (although partially) justify this picture.

\section{Setup and main assumptions}
\label{s:setup}
We describe our setup in detail, and introduce essential assumptions.
Note that we take the microcanonical point of view throughout the present paper (except for  Appendix~\ref{s:canonical}, where we discuss the canonical setting).

\subsection{Macroscopic quantum systems}
\label{s:MQS}
We consider an isolated macroscopic quantum system characterized by its volume $V$.
A typical example is a system of $N$ particles confined in a box, where the density $\rho=N/V$ is kept constant as $V$ varies.
We can also treat quantum systems on a lattice, such as quantum spin systems or interacting particles on a lattice.
We denote by $\Htot$ the total Hilbert space whose dimension may be finite or countably infinite.

Let $\Ham$ be the Hamiltonian, and denote by $E_j$ and and $\kpj\in\Htot$ the eigenvalue and  the corresponding normalized eigenstate, respectively, i.e., $\Ham\kpj=E_j\kpj$, where  $j=1,2,\ldots$.
Let the number of states $\Omega_V(U)$ be the number of $j$'s such that $E_j\le U$.
We assume the normal behavior, which can be proved in a large class of systems \cite{Ruelle},
\eq
\log\Omega_V(U)=V\,\sigma(U/V)+o(V),
\lb{logO}
\en
where the entropy density $\sigma(u)$ is a concave function independent of $V$.
We further assume that $\sigma(u)$ is strictly increasing\footnote{
\label{fn:negativeT}
In a system (such as a quantum spin system) where the energy is bounded from above, there is a range of energy (corresponding to ``negative temperatures'') where the density of states $d\Omega_V(U)/dU$ is strictly decreasing in $U$.
In such a region the entropy density $\sigma(u)$ defined from the number of states (rather than the density of states) as in \rlb{logO} becomes constant, and the assumption does not hold.
But this is not an essential problem.
The region with  ``negative temperatures'' becomes a normal one by simply switching the sign of the Hamiltonian; we can apply our theory to this reversed Hamiltonian.
}, differentiable, and grows sublinearly\footnote{
In concrete models like quantum spin systems, we shall prove these properties.
}.
For the final assumption, see \rlb{Osb}.
We then denote by\footnote{
By $A:=B$ (or, equivalently, $B=:A$) we mean that $A$ is defined in terms of $B$.
}
\eq
\beta(u):=\sigma'(u)>0,
\lb{betau}
\en
the inverse temperature corresponding to the energy density $u$.

Let $u$ be an energy density and $\Du>0$ be a small energy width\footnote{Small in the sense that $\sigma(u)-\sigma(u-\Du)\ll\sigma(u)$}.
We then denote by $\JVu$ the set of $j$'s such that\footnote{
We have set the upper bound in \rlb{uEVu} to $u$ in order to make some formula simple.
One can change it to, say, $u+\Du$.
}
\eq
u-\Du<\frac{E_j}{V}\le u,
\lb{uEVu}
\en
and by $\DVu:=\Omega_V(Vu)-\Omega_V(V(u-\Du))$ the number of elements in $\JVu$.
We define the energy shell $\HVu$ as the $\DVu$ dimensional subspace of $\Htot$ spanned by $\kpj$ with $j\in\JVu$.
The microcanonical average of any operator $\hO$ on $\Htot$ is defined as usual by
\eq
\bktmc{\hO}:=\frac{1}{\DVu}\sum_{j\in\JVu}\sbkt{\psi_j|\hO|\psi_j}
=\operatorname{Tr}[{\rho}^\mathrm{mc}_{V,u}\,\hO],
\lb{mc}
\en
where
\eq
{\rho}^\mathrm{mc}_{V,u}:=\frac{1}{\DVu}\sum_{j\in\JVu}\kpj\bra{\psi_j}
\lb{rhomc}
\en
is the microcanonical density matrix. 

Finally let $\hA$ be an arbitrary self-adjoint operator on $\Htot$, and let $a_k$ and $\ket{\xi_k}$ (with $k=1,2,\ldots$) be the corresponding eigenvalues and normalized eigenstates, respectively.
For any $a\in\bbR$, we denote by
\eq
\hP[\hA\ge a]:=\sumtwo{k}{(a_k\ge a)}\ket{\xi_k}\bra{\xi_k}
\en
the projection onto the subspace spanned by the eigenstates of $\hA$ corresponding to the eigenvalues not less than $a$.

\subsection{Pure states which represent thermal equilibrium}
\label{s:pure}
We shall precisely formulate the notion that a pure state of a macroscopic quantum system represents thermal equilibrium.

The basic philosophy behind our formulation is that thermal equilibrium is an intrinsically macroscopic notion, and should be characterized operationally in terms of the observation of macroscopic physical quantities\footnote{
But we can also treat quantities which are not macroscopic.
See Appendix~\ref{s:small}.
}.
In this sense our formulation of thermal equilibrium differs in an essential manner from that in many existing works, especially those with quantum information theoretic background.
The formulation by Goldstein, Lebowitz, Mastrodonato, Tumulka, and Zangh\`\i\ \cite{GLMTZ09b}, which is based on the early work by von Neumann \cite{vonNeumann,GLTZ}, is very close to ours.
See section~\ref{s:comparison} for comparison of our formulation with others.

\paragraph{Single quantity:}
For simplicity, we first treat the case where one is interested only in a single extensive quantity which is represented by a self-adjoint operator $\MV$ on $\Htot$ for each $V>0$.
This simple (or oversimplified) case indeed contains the essence of our theory.

Let
\eq
m(u):=\lim_{V\up\infty}\frac{1}{V}\bktmc{\MV}
\lb{mu}
\en
be the equilibrium value of the density $\MV/V$.
We also assume that one examines $\MV/V$ with a certain fixed precision $\delta>0$ which is independent\footnote{
Although it is physically natural that one uses a fixed precision for the density, the critical reader might argue that the precision can be made higher for larger $V$.
In fact our convention to use a  fixed precision is also motivated by the (theoretical) fact that it leads us to an upper bound for the fluctuation, the thermodynamic bound \rlb{tdb}, which is exponentially small in $V$.
} of $V$.
We then define the ``nonequilibrium projection operator'' by\footnote{
This projection, which acts on the whole Hilbert space $\Htot$, should not be confused with the similar projection operator (also denoted as $1-P_\mathrm{eq}$ or $\hat{P}_\mathrm{neq}$) which appears in \cite{GLMTZ09b,GHT13,GHT14short,GHT14long}.
The latter is the projection onto a subspace of the energy shell $\HVu$.
See section~\ref{s:comparison}.
}
\eq
\Pneq:=\hP\Bigl[\bigl|(\MV/V)-m(u)\bigr|\ge\delta\Bigr].
\lb{Pneq1}
\en

The following definition is essential.
\begin{definition}
\label{d:eq}
Let us choose (and fix) a constant $\alpha>0$.
Suppose that, for some $V>0$ and for a normalized pure state $\kph\in\HVu$, one has
\eq
\bra{\varphi}\Pneq\kph
\le e^{-\alpha V}.
\lb{eq}
\en
Then we say that the pure state $\kph$ represents thermal equilibrium.
\end{definition}

The bound \rlb{eq} says that, when one performs projective measurement of $\MV/V$ in the state $\kph$, the measurement result must lie in the range $m(u)\pm\delta$ with probability not less than $1-e^{-\alpha V}$.
If the volume $V$ is large this means that one (almost) certainly observes the equilibrium value (within the precision).
We also stress that, as in the case for experiments in macroscopic systems, only a single measurement is enough to get the equilibrium value of $\MV/V$.
Though the pure state $\kph$ and the standard equilibrium state described, e.g., by the microcanonical density matrix ${\rho}^\mathrm{mc}_{V,u}$ (see \rlb{rhomc}) are different, they are essentially indistinguishable when one only measures the single thermodynamic quantity $\MV$.

Although our characterization of thermal equilibrium depends on the choice of the constants $\delta$ and $\alpha$, we believe that this ambiguity causes no problems.
We also note that one is likely to find ``natural'' values of $\delta$ and $\alpha$ for a given concrete problem\footnote{
The value of $\alpha$ must be set to be smaller than the constant $\gamma$, which appears in the thermodynamic bound \rlb{tdb}.
The optimal value of $\gamma$ is model specific, and there is no room for our choice.
}.

\paragraph{Multiple quantities:}
Let us move onto a more general case with multiple quantities\footnote{
The quick reader can skip the treatment of multiple quantities.
}.
Suppose that we are interested in $n$ extensive quantities $\MiV$ with $i=1,\ldots,n$.
We assume that $n$ is not too large, and is independent of $V$.
Each $\MiV$ is a self-adjoint operator on $\Htot$ for each $V>0$.
We denote by 
\eq
\miu:=\lim_{V\up\infty}\frac{1}{V}\bktmc{\MiV}
\lb{mav}
\en
the equilibrium value of  their densities in the thermodynamic limit.
We also fix, for each $i=1,\ldots,n$, the ($V$-independent) precision $\di>0$ for the density $\MiV/V$.

We shall again make an operational characterization of thermal equilibrium based on the measurement of the $n$ quantities.
This is not trivial since  the operators $\MV^{(1)},\ldots,\MV^{(n)}$ do not commute in general, and hence are not simultaneously measurable.
On the other hand, we have good reasons to expect that this won't be a serious problem since it is believed (from empirical facts) that one can determine the values of multiple macroscopic quantities simultaneously with sufficient accuracy.

The following definition\footnote{
This characterization of thermal equilibrium is referred to as TMATE in \cite{GHLT}.
} relies on this belief.
\begin{definition}
\label{d:eq2}
Let us choose (and fix) a constant $\alpha>0$.
Suppose that, for some $V>0$ and for a normalized pure state $\kph\in\HVu$, one has
\eq
\Bigl\langle\varphi\Bigr|\,\hP\Bigl[\bigl|(\MiV/V)-\miu\bigr|\ge\di\Bigr]\,\Bigl|\varphi\Bigr\rangle
\le e^{-\alpha V},
\lb{PeaVsimple}
\en
for each $i=1,\ldots,n$.
Then we say that the pure state $\kph$ represents thermal equilibrium.
\end{definition}

The definition implies that, if we measure any of $\MV^{(1)},\ldots,\MV^{(n)}$, the measurement result must be close to the equilibrium value with probability close to one.

When using this definition, we shall redefine the operator $\Pneq$ as
\eq
\Pneq:=\sum_{i=1}^n\hP\Bigl[\bigl|(\MiV/V)-\miu\bigr|\ge\di\Bigr],
\lb{Pneqnew}
\en
which is not a projection, but a non-negative operator.
Then we see that the condition $\bra{\varphi}\Pneq\kph
\le e^{-\alpha V}$, which is exactly the same as \rlb{eq} but with a new definition of $\Pneq$, is a sufficient condition for \rlb{PeaVsimple}.
Therefore, in what follows, one can still use \rlb{eq} as the definition of a pure state representing thermal equilibrium, and use the definition \rlb{Pneq1} or \rlb{Pneqnew} of $\Pneq$ depending on the situation.

\paragraph{Multiple quantities (with commuting approximants):}
Let us explain a more sophisticated treatment of multiple quantities, which goes back to von Neumann's idea in \cite{vonNeumann}, and is similar to that used by Goldstein, Lebowitz, Mastrodonato, Tumulka, and Zangh\`\i\ \cite{GLMTZ09b}.

Since physically natural extensive quantities are the sums (or the integrals) of local quantities, we expect to have in general that $[\MiV,\hM^{(j)}_V]=O(V)$ for $i\ne j$.
The densities $\MiV/V$ then ``almost commute'' in the sense that $[\MiV/V,\hM^{(j)}_V/V]\to0$ as $V\uparrow\infty$ for all $i,j=1,\ldots,n$.
This fact suggests that they can be well approximated by mutually commuting self-adjoint operators.
More precisely one expects that there exist self-adjoint operators $\tMV^{(1)},\ldots,\tMV^{(n)}$ such that
\eq
\bigl[\tMiV,\tilde{\hM}^{(j)}_V\bigr]=0\quad\text{for any $i,j=1,\ldots,n$, and any $V>0$},
\lb{VMM}
\en
and
\eq
\lim_{V\up\infty}\frac{1}{V}\bigl\Vert\MiV-\tMiV\bigr\Vert=0\quad\text{for any $i=1,\ldots,n$}.
\lb{MtM}
\en
As for general quantum spin systems on the $d$-dimensional hypercubic lattice with $\MiV$ chosen as a translationally invariant sum of local operators, Ogata \cite{Ogata2011} proved the existence of $\tMiV$ with the desired properties\footnote{
The problem whether $n$ Hermitian matrices which ``almost commute''  can be approximated by $n$ mutually commuting Hermitian matrices has a long history.
The case $n=2$ was solved affirmatively by Lin \cite{Lin}, while it is known \cite{Davidson} that the statement does not hold in general for $n\ge3$.
In this sense Ogata's result \cite{Ogata2011} for quantum spin systems is nontrivial and important.
}.
Here we shall assume that such operators exist for general cases.

Since $\tMV^{(1)},\ldots,\tMV^{(n)}$ are simultaneously measurable, we can define the nonequilibrium projection operator as
\eqa
\Pneq&:=
\hP\Bigl[\bigl|(\tMiV/V)-\miu\bigr|\ge\di\ 
\text{for at least one $i\in\{1,2,\ldots,n\}$}\Bigr]\nl
&=\mathsf{1}-\prod_{i=1}^n\hP\Bigl[\bigl|(\tMiV/V)-\miu\bigr|<\di\Bigr].
\lb{Pneqn}
\ena
Then we shall define the notion of pure states representing thermal equilibrium precisely by Definition~\ref{d:eq} with $\Pneq$ defined in \rlb{Pneqn}.
The physical interpretation is clear; if one performs simultaneous projective measurement of $\tMV^{(1)},\ldots,\tMV^{(n)}$, then with with probability not less than $1-e^{-\alpha V}$, the measurement result for each $i=1,\ldots,n$ lies in the range $\miu\pm\di$.
From the operational point of view, the quantum mechanical pure state $\kph$ is nothing but the thermal equilibrium\footnote{
We are not arguing that one should literally measure the operators $\tMV^{(1)},\ldots,\tMV^{(n)}$.
We have discussed this mathematical construction because it ensures that simultaneous approximate measurement of the $n$ quantities is possible.
}.

\subsection{Thermodynamic bound}
\label{s:TDB}
In sections~\ref{s:typicality}, \ref{s:thermalization}, \ref{s:moderate}, and \ref{s:ETH}, we shall show that pure states which appear in certain situations represent equilibrium in the sense of Definition~\ref{d:eq}.
To do that we need to ensure that our quantum system, along with the set of macroscopic observables, behaves as a normal thermodynamic system in the energy shell $\HVu$.
We shall characterize the normal behavior in terms of the following {\em thermodynamic bound}\/, which is a property of equilibrium statistical mechanics.

\begin{definition}
\label{d:tdb}
The system satisfies the thermodynamic bound (for the energy density $u$) if there are constants $\gamma>0$ and $V_0$ such that
\eq
\bktmc{\Pneq}\le e^{-\gamma V},
\lb{tdb}
\en
holds for any $V\ge V_0$.
\end{definition}
Here the nonequilibrium projection operator $\Pneq$ is defined by \rlb{Pneq1}, \rlb{Pneqnew} or \rlb{Pneqn}  depending on the situation and the treatment.

The bound \rlb{tdb} simply says that a large fluctuation (proportional to $V$) of the quantity of interest from its equilibrium value is exponentially rare in the thermal equilibrium.
This property, which is closely related to the large deviation property in probability theory\footnote{
The bound \rlb{tdb} may be called the ``global large deviation upper bound for the microcanonical ensemble''.
}, is expected to be valid for any equilibrium ensemble which corresponds to a single thermodynamic phase.

Our general results in sections~\ref{s:typicality}, \ref{s:thermalization}, \ref{s:moderate}, and \ref{s:ETH} are based on the assumption that the thermodynamic bound \rlb{tdb} is valid.
To apply these conclusions to concrete quantum systems, one therefore has to justify the validity of the bound \rlb{tdb}.
This indeed turns out to be a nontrivial task, but should be possible for a large class of models.
In section~\ref{s:examples}, we shall discuss some important examples where the bound can be proved.

\subsection{Comparison with other formulations}
\label{s:comparison}
It may be useful to compare our formulation with other definitions of pure states representing (thermal) equilibrium.
We stress that these differences in the basic notion are directly reflected in the interpretations of other notions such as the typicality of equilibrium and the approach to equilibrium.

See also the illuminating discussion by Goldstein, Huse, Lebowitz and Tumulka \cite{GHLT}.

\paragraph{Expectation values:}
A very common definition, which can be found, e.g., in \cite{Tasaki1998,Reimann,ReimannKastner,Reimann2,Reimann3,SugiuraShimizu12,SugiuraShimizu13}, deals with the expectation values of certain selected observables $\hA_1,\ldots,\hA_n$.
A normalized pure state $\kph$ is said to represent ``equilibrium'' if
\eq
\bra{\varphi}\hA_i\kph\simeq\sbkt{\hA_i}_\infty:=\operatorname{Tr}[\hA_i\,{\rho}_\infty],
\lb{AsimA}
\en
for $i=1,\ldots,n$, 
where ${\rho}_\infty$ is the density matrix of the ``equilibrium'' state that one wishes to reproduce.
In general ${\rho}_\infty$ need not describe thermal equilibrium, and one sometimes allows ${\rho}_\infty$ to depend on the initial state.

In this formulation, the observables  $\hA_i$ may exhibit large fluctuation (compared with\footnote{
Throughout the present paper $\norm{\cdot}$ denotes the operator norm, i.e., $\norm{\hO}:=\sup_{\kph\ne0}\norm{\hO\kph}/\norm{\kph}$.
} $\norm{\hA_i}$) in the states $\rho_\infty$ or $\kph$.
This happens when $\hA_i$ are not macroscopic observables, or when ${\rho}_\infty$ does not represent a pure thermodynamic phase\footnote{
Consider, for example, the canonical distribution for the classical three dimensional ferromagnetic Ising model at very low temperature without external magnetic field.
If the model is defined on a finite but large lattice, the canonical distribution represents the mixture of the two phases where the spins align upward or downward.
}.
Note that, in such a case with large fluctuation, one generally needs to make repeated measurement (in a single fixed state $\kph$) of a quantity $\hA_i$ in order to determine the expectation value $\bra{\varphi}\hA_i\kph$.

We note that our requirement \rlb{eq} is stronger than \rlb{AsimA}. 
But it does not only imply near identities like \rlb{AsimA} for $\MV^{(1)},\ldots,\MV^{(n)}$, but also implies that (quantum) fluctuation around the equilibrium value is negligible.
This guarantees that one (almost certainly) gets the equilibrium value (with sufficient accuracy) after a single quantum mechanical measurement.
We believe that this formulation is suited for the purpose of reproducing thermodynamic behavior.

\paragraph{Canonical setting:}
In another common formulation \cite{LindenPopescuShortWinter,PopescuShortWinter,GLTZ06}, one assumes that the whole system is divided into the large reservoir (heat bath or particle bath) and the relatively small system of interest, and the Hilbert space is correspondingly decomposed as $\calH_{\rm tot}=\calH_{\rm res}\otimes\calH_{\rm sys}$.
Then one says that a pure state $\kph\in\calH_{\rm tot}$ represents equilibrium if the reduced density matrix of the system ${\rho}_{\rm sys}:=\operatorname{Tr}_{\calH_{\rm res}}\bigl[\kph\bra{\varphi}\bigr]$ satisfies
\eq
{\rho}_{\rm sys}\simeq{\rho}_\mathrm{can},
\lb{rho=rho}
\en
where ${\rho}_\mathrm{can}$ is the density matrix for the canonical distribution\footnote{
This is similar to MITE formulated in \cite{GHLT}.  But we have here fixed the decomposition $\calH_{\rm res}\otimes\calH_{\rm sys}$, while various decompositions are considered in \cite{GHLT}.
}.
It satisfies ${\rho}_\mathrm{can}\simeq\operatorname{Tr}_{\calH_{\rm res}}[\rho_{\rm mc}]$, where $\rho_{\rm mc}$ is the density matrix for the microcanonical distribution of the whole systems.

There is an obvious difference that this formalism is based on the canonical view point while ours on the microcanonical view.
We also note that \rlb{rho=rho} requires $\kph$ to reproduce the expectation values of any observables on $\calH_{\rm sys}$, which is in contrast to our formalism which deals only with a limited number of macroscopic observables\footnote{
It is indeed a deep problem to determine to what extent the prediction of equilibrium statistical mechanics should be regarded as realistic.
See section~\ref{s:issues} and \cite{GHLT} for related issues.
}.

See Appendix~\ref{s:canonical} for our approach to the canonical distribution.

\paragraph{The nonequilibrium subspace:}
Finally let us discuss the formulation of Goldstein, Lebowitz, Mastrodonato, Tumulka, and Zangh\`\i\ \cite{GLMTZ09b}, which is a modern version of the earlier proposal by von Neumann \cite{vonNeumann,GLTZ}.
We have also used this formalism when we discussed the time scale required for thermalization in \cite{GHT13,GHT14short,GHT14long}.
This characterization of thermal equilibrium is called MATE (macroscopic thermal equilibrium) in \cite{GHLT}.

Here one deals with the same energy shell $\HVu$ as we have defined (but see below), and postulates that it is decomposed into the equilibrium and the nonequilibrium subspaces as $\HVu=\calH_{\rm eq}\oplus\calH_{\rm neq}$, where the dimensions of the subspaces satisfy
\eq
\operatorname{dim}[\calH_{\rm eq}]\gg\operatorname{dim}[\calH_{\rm neq}].
\lb{d>>d}
\en
Then a pure state $\kph\in\HVu$ is said to represent equilibrium if
\eq
\bra{\varphi}\hP(\calH_{\rm neq})\kph\ll1,
\lb{Pneq<<1}
\en
where $\hP(\calH_{\rm neq})$, which should not be confused with our $\Pneq$ defined by \rlb{Pneq1}, \rlb{Pneqnew} or \rlb{Pneqn} , is the projection onto $\calH_{\rm neq}$.

The idea behind this formalism is essentially the same as ours.
Take the Hamiltonian $\Ham$, and the macroscopic observables $\hM_V^{(1)},\ldots,\hM_V^{(n)}$ as before.
Assume that there are operators $\hat{\hH}_V$ and ${\hat{\hM}}_V^{(1)},\ldots,{\hat{\hM}}_V^{(n)}$ which commute with each other, and satisfy $\hat{\hH}_V\simeq\Ham$, and  ${\hat{\hM}}_V^{(i)}\simeq\hM^{(i)}$ for $i=1,\ldots,n$.
One then redefines the energy shell $\HVu$ to be that determined from the new Hamiltonian $\hat{\hH}_V$.
One can define $\calH_{\rm eq}$ as the subspace of (the redefined) $\HVu$ spanned by the simultaneous eigenstates of ${\hat{\hM}}_V^{(1)},\ldots,{\hat{\hM}}_V^{(n)}$ where the eigenvalues lie in the ranges $(m^{(1)}(u)\pm\delta^{(1)})V,\ldots,(m^{(n)}(u)\pm\delta^{(n)})V$, respectively.
Then it is clear that our requirement \rlb{eq} is almost the same as \rlb{Pneq<<1}, and our thermodynamic bound \rlb{tdb} corresponds\footnote{
In fact we can prove \rlb{d>>d} for some models by modifying our proof of the thermodynamic bound.
See \rlb{dd<<1}.
} to the inequality \rlb{d>>d}.

In fact we have followed this formulation of Goldstein, Lebowitz, Mastrodonato, Tumulka, and Zangh\`\i\ \cite{GHLT} rather faithfully in the present paper, and the difference is mostly technical\footnote{
An essential difference appears when we consider the time evolution.
(Note that time evolution is not considered in \cite{GHLT}.)
In the prescription of \cite{GHLT}, the energy shell $\HVu$ is redefined according to the modified Hamiltonian $\hat{\hH}_V$.
Since the time evolution must be determined by the original Hamiltonian $\Ham$, the (redefined) energy shell $\HVu$ is not invariant under the time evolution.
Note that in our formalism neither the Hamiltonian nor the energy shell is redefined.
}.
Our formalism does not introduce the equilibrium and the nonequilibrium subspaces, since our projection operator $\Pneq$ does not commute with the projection onto $\HVu$ in general.
We also note that while von Neumann or Goldstein, Lebowitz, Mastrodonato, Tumulka, and Zangh\`\i\ were not clear about the role of the volume $V$ in the relations like \rlb{Pneq<<1}, we here give formulas (which are reminiscent of the large deviation theory) about how small certain expectation values should be when $V$ becomes large.

\section{Examples where the thermodynamic bound is provable}
\label{s:examples}
Let us discuss examples where the thermodynamic bound can be established rigorously.
The reader may skip some of the examples depending on his/her interest.
The proof is presented in section~\ref{s:proofTDB}.

\begin{figure}[btp]
\begin{center}
\includegraphics[width=8cm]{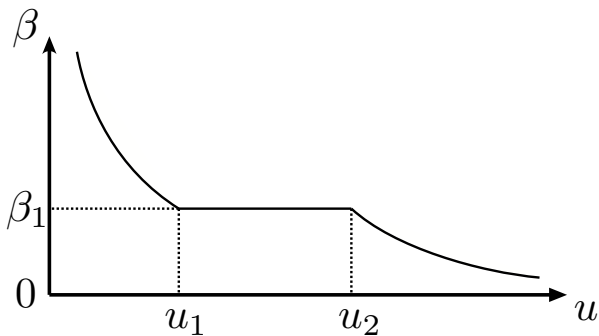}
\end{center}
\caption[dummy]{
A typical relation between $u$ and $\beta$ when a first order phase transition takes place.
The condition \rlb{ucond} does not hold for $u$ in the range $u_1\le u\le u_2$, where two phase coexist.
The inverse function $u(\beta)$ is discontinuous at $\beta_1$.
}
\label{fig:ubeta}
\end{figure}

In all the examples below it is necessary that the energy density $u$ satisfies
\eq
\text{$\beta(u)\ne\beta(u')$ for any $u'\ne u$}.
\lb{ucond}
\en
This condition is not obvious, and is not satisfied when the first order phase transition (or, more precisely, a phase coexistence) takes place at $u$.
See Figure~\ref{fig:ubeta}.
In the following examples, the condition \rlb{ucond} is known to be valid, or simply assumed\footnote{
It can be shown quite generally that there are plenty of $u$ which satisfies the condition \rlb{ucond}.
Note that, in terms of the inverse function $u(\beta)$, the condition \rlb{ucond} is equivalent to the continuity of $u(\beta)$.
But since $u(\beta)$ is monotonically nonincreasing, there are at most countably many values of $\beta$ at which $u(\beta)$ is discontinuous.
}.

\subsection{Heat conduction between two identical bodies in contact}
\label{s:EHC}
We start from the standard (and realistic) model for heat conduction between two identical bodies in contact.
Fortunately we can establish the thermodynamic bound in a general setting by only using  elementary techniques in the large deviation theory for classical statistical mechanics.

We assume that the system with volume $V$ consists of two identical subsystems with volume $V/2$.
For example one can take a system which consists of two identical boxes with volume $V/2$ each containing $N/2$ particles (with $\rho=N/V$ fixed).

The Hilbert spaces for the subsystems are denoted as $\calH^{(1)}_{V/2}$ and $\calH^{(2)}_{V/2}$, respectively, and are assumed to be identical.
The Hamiltonians of the subsystems are $\mathsf{H}^{(1)}_{V/2}$ and $\mathsf{H}^{(2)}_{V/2}$, respectively, and again assumed to be identical for simplicity.
We assume that the density of states $\Omega^{(1)}_{V/2}(U')$ and $\Omega^{(2)}_{V/2}(U')$ of the subsystems satisfy
\eq
\log\Omega^{(1)}_{V/2}(U')=\log\Omega^{(2)}_{V/2}(U')=V\,\sigma(2U'/V)+o(V),
\lb{logO12}
\en
as in \rlb{logO},
with the entropy density $\sigma(u)$ satisfying the same properties.

The whole Hilbert space is then $\Htot=\calH^{(1)}_{V/2}\otimes\calH^{(2)}_{V/2}$, and we write the total Hamiltonian as 
\eq
\Ham=\mathsf{H}^{(1)}_{V/2}\otimes\mathsf{1}+\mathsf{1}\otimes\mathsf{H}^{(2)}_{V/2}+\mathsf{H}_\mathrm{int}.
\lb{HamHC}
\en
The interaction Hamiltonian $\mathsf{H}_\mathrm{int}$ acts on the whole Hilbert space $\Htot$.
Again for simplicity we assume that $\mathsf{H}_\mathrm{int}$ acts symmetrically on the two subspaces  $\calH^{(1)}_{V/2}$ and $\calH^{(2)}_{V/2}$.
Since the interaction usually takes place at the boundary of the two subsystems, we assume that $\Vert\mathsf{H}_\mathrm{int}\Vert\le h_0\,V^{(d-1)/d}$, where $h_0$ is a constant and $d=1,2,3\ldots$ is the dimension.
It then follows from the standard argument \cite{Ruelle} that the density of states $\Omega_V(U)$ for the whole system determined from the total Hamiltonian $\Ham$ satisfies \rlb{logO} with the same entropy density $\sigma(u)$ as in \rlb{logO12}.

Let us focus on the energy difference
\eq
\MV:=\mathsf{H}^{(1)}_{V/2}\otimes\mathsf{1}-\mathsf{1}\otimes\mathsf{H}^{(2)}_{V/2},
\lb{MVHC}
\en
which is an important thermodynamic quantity when one is interested in heat conduction.
Since the two subsystems are assumed to be identical, the equilibrium value of $\MV$ is $\bktmc{\MV}=0$ for any $u$ and $V$.

As for the fluctuation around the equilibrium value, we can prove the following.

\begin{pro}
\label{p:TDBHC}
Assume\footnote{
Physically speaking the condition should hold in general except at the triple point.
The triple point must be excluded since energy is not distributed equally between the two subsystems at the point.
} the condition \rlb{ucond} for the energy density $u$.
Take an arbitrary $\delta>0$.
Then, for sufficiently large $V_0$, there exists $\gamma(u,\delta,V_0)>0$ such that
\eq
\Bktmc{\hP\bigl[|\MV|\ge V\delta\bigr]}\le e^{-\gamma(u,\delta,V_0)\,V}
\lb{TDBHC}
\en
holds for any $V\ge V_0$.

Suppose further that $\beta'(u)$ exists.
Then, for small $\delta$ and large $V_0$, one has
\eq
\gamma(u,\delta,V_0)\simeq-\frac{\beta'(u)}{2}\delta^2=\frac{\delta^2}{2\kB T^2\,c(T)},
\lb{gammaHC}
\en
where $T=\kB/\beta(u)$ is the temperature corresponding to $u$, and $c(T)=du/dT$ is the specific heat.
\end{pro}

As we shall see below, the value of $\gamma$ in \rlb{gammaHC} seems to be optimal.
We also note that the thermodynamic bound \rlb{gammaHC} itself is valid for the decoupled system with $\mathsf{H}_\mathrm{int}=0$, where heat conduction cannot take place.

\paragraph{Heuristic derivation:}
It is useful to see a heuristic justification of Proposition~\ref{p:TDBHC}.
This discussion will be used later in section~\ref{s:moderate}.
A complete proof of the proposition will be given in section~\ref{s:PHC}.

Fix $V$, and take the energy eigenstates for the two subsystems, i.e.,
\eq
\mathsf{H}^{(1)}_{V/2}\ket{\psi^{(1)}_{j_1}}=E^{(1)}_{j_1}\ket{\psi^{(1)}_{j_1}},\quad
\mathsf{H}^{(2)}_{V/2}\ket{\psi^{(2)}_{j_2}}=E^{(2)}_{j_2}\ket{\psi^{(2)}_{j_2}}.
\en
Obviously the tensor products
\eq
\ket{\Psi_{j_1,j_2}}:=\ket{\psi^{(1)}_{j_1}}\otimes\ket{\psi^{(2)}_{j_2}},
\lb{pp12}
\en
with all possible $(j_1,j_2)$, span the whole Hilbert space $\Htot$.
Let $\calH_{V,u}^{(0)}$ be the subspace of $\Htot$ spanned by the states $\ket{\Psi_{j_1,j_2}}$ with $(j_1,j_2)$ such that
\eq
u-\Du\le\frac{E^{(1)}_{j_1}+E^{(2)}_{j_2}}{V}\le u.
\lb{uE1E2}
\en
Note that $\HVu^{(0)}$ gives a good approximation to the subspace (energy shell) $\HVu$ since the interaction $\mathsf{H}_\mathrm{int}$ is small compared to the main part of the Hamiltonian.

Let $U=uV$.
The dimension of $\HVu^{(0)}$, which is the number of $(j_1,j_2)$ satisfying \rlb{uE1E2}, is roughly estimated as\footnote{
In general we write $a\simeq b$ when $a$ and $b$ are almost equal, and $a\sim b$ when they are roughly equal or of the same order.
For $V$ dependent quantities, we write $f(V)\simeq g(V)$ when $\lim_{V\up\infty}f(V)/g(V)=1$, and $f(V)\sim g(V)$ when $\lim_{V\up\infty}V^{-1}\log[f(V)/g(V)]=0$.
}
\eqa
\DVu&:=\mathrm{dim}[\HVu]\sim\mathrm{dim}[\HVu^{(0)}]
\nl
&\simeq\int dU_1\int dU_2\,\chi\bigl[U-V\Du< U_1+U_2\le U\bigr]
\,\rho^{(1)}_{V/2}(U_1)\,\rho^{(2)}_{V/2}(U_2)
\nl
&\sim
\mathop{\max_{U_1,U_2}}_{(U_1+U_2=U)}
\rho^{(1)}_{V/2}(U_1)\,\rho^{(2)}_{V/2}(U_2)
\nl&\sim\rho^{(1)}_{V/2}(U/2)\,\rho^{(2)}_{V/2}(U/2)
\nl&\sim e^{V\,\sigma(u)},
\ena
where $\rho^{(i)}_{V/2}(\tilde{U})=d\Omega^{(i)}_{V/2}(\tilde{U})/d\tilde{U}$ is the density of states, and the indicator function $\chi[\cdot]$ is defined by $\chi[\text{true}]=1$ and $\chi[\text{false}]=0$.
We noted that $\rho^{(i)}_{V/2}(\tilde{U})\sim\Omega^{(i)}_{V/2}(\tilde{U})$, and used  the standard saddle point approximation (or, more precisely, Laplace's method) to get the third line.

We next consider the subspace $\HVu^{(\delta)}$ spanned by the basis states $\ket{\Psi_{j_1,j_2}}$ of \rlb{pp12} with $(j_1,j_2)$ which satisfy \rlb{uE1E2} and
\eq
\bigl|E^{(1)}_{j_1}-E^{(2)}_{j_2}\bigr|\ge V\delta.
\lb{E1E2d}
\en
Note that $\HVu^{(\delta)}$ is the subspace of $\HVu^{(0)}$ in which the energy densities in the two subsystems are considerably different.
The dimension of the subspace is again roughly estimated as
\eqa
\mathrm{dim}[\HVu^{(\delta)}]&\sim
\mathop{\max_{U_1,U_2}}_
{\rbk{\substack{U_1+U_2=U\\|U_1-U_2|\ge V\delta}}}\rho^{(1)}_{V/2}(U_1)\,\rho^{(2)}_{V/2}(U_2)
\nl&\sim
\rho^{(1)}_{V/2}\rbk{\frac{U\pm V\delta}{2}}\,
\rho^{(2)}_{V/2}\rbk{\frac{U\mp V\delta}{2}}
\nl&\sim
\exp\Bigl[\frac{V}{2}\bigl\{\sigma(u+\delta)+\sigma(u-\delta)\bigr\}\Bigr]
\nl&\sim
\exp\Bigl[V\,\sigma(u)+\frac{V}{2}\sigma''(u)\,\delta^2\Bigr].
\ena
We also note that, within the subspace $\HVu^{(0)}$, the projection $\hP\bigl[|\MV|\ge V\delta\bigr]$ is identified with the projection onto the subspace $\HVu^{(\delta)}$.
Thus we get the desired thermodynamic bound (within the present approximation)  as 
\eq
\Bktmc{\hP\bigl[|\MV|\ge V\delta\bigr]}\sim
\frac{\mathrm{dim}[\HVu^{(\delta)}]}{\mathrm{dim}[\HVu^{(0)}]}
\sim e^{-\gamma V},
\lb{PMDD}
\en
with $\gamma=-\sigma''(u)\,\delta^2/2=-\beta'(u)\,\delta^2/2$.

We stress that the thermodynamic bound in this case is obtained by relying on the similarity of the system to the decoupled system with $\mathsf{H}_\mathrm{int}=0$, and using the standard estimates on the number of states.

\subsection{Quantum spin systems}
\label{s:EQL}
As for quantum systems on a lattice, such as quantum spin systems, we can make use of existing results on the large deviation properties \cite{NetocnyRedig,LenciRey-Bellet,HiaiMosonyiOgawa,Ogata2010,OgataRey-Bellet} to prove the thermodynamic bound.
For simplicity we shall here concentrate on quantum spin systems, but some results apply to lattice fermion systems \cite{LenciRey-Bellet}.

We consider general quantum spin systems.
Let $\Lambda$ be the $d$-dimensional $L\times\cdots\times L$ hypercubic lattice with periodic boundary conditions.
We identify the volume $V$ with the number of sites $L^d$.
With each site $x\in\Lambda$ we associate a finite dimensional Hilbert space $\calH_x\cong\mathbb{C}^\nu$, where $\nu$ is a constant independent of $L$.
The total Hilbert space is $\Htot=\bigotimes_{x\in\Lambda}\calH_x\cong\mathbb{C}^{V\nu}$.

We take a quite general translationally invariant Hamiltonian.
Let $\mathsf{h}_o$ (where $o$ is the origin) be an arbitrary self-adjoint operator which acts only on a finite number of sites and is independent of $L$.
Then our Hamiltonian is
\eq
\Ham:=\sum_{x\in\Lambda}\mathsf{h}_x,
\en
where $\mathsf{h}_x$ is the translation of $\mathsf{h}_o$ by $x$.
Then the asymptotic behavior \rlb{logO} of the number of states and the concavity of $\sigma(u)$ can be proved by the standard method \cite{Ruelle}.
The differentiability of the entropy density $\sigma(u)$, i.e., the existence of the inverse temperature $\beta(u)$ can be proved in the models treated in the following Proposition~\ref{p:TDBQS}.

The thermodynamic quantity of interest is also defined as
\eq
\MV:=\sum_{x\in\Lambda}\mathsf{m}_x,
\lb{Mbx}
\en
where $\mathsf{m}_o$ is an arbitrary ($L$-independent) self-adjoint operator which acts only on a finite number of sites, and $\mathsf{m}_x$ is its translation.

\begin{pro}
\label{p:TDBQS}
For $d=1$ let $u$ be such that $\beta(u)>0$, and for $d\ge2$ let $u$ be such that $0<\beta(u)\le\beta_0$, where $\beta_0>0$ is a constant which depends on the model.
Take an arbitrary $\delta>0$.
Then, for sufficiently large $V_0$, there exists $\gamma(u,\delta,V_0)>0$ such that
\eq
\Bktmc{\hP\Bigl[\bigl|(\MV/V)-m(u)\bigr|\ge\delta\Bigr]}\le e^{-\gamma(u,\delta,V_0)\, V},
\lb{TDBQS}
\en
holds for any $V\ge V_0$.
\end{pro}

For $d=1$, thanks to Ogata's complete large deviation theory \cite{Ogata2010}, the proposition covers the whole range of the energy density $u$ which corresponds to nonnegative $\beta(u)$.
As we have explained in the footnote~\ref{fn:negativeT}, this essentially means that the whole range of $u$ can be covered.

For $d\ge2$, on the other hand, the proposition is valid only in the limited range of $u$ which corresponds to very  high temperatures.
This is because expansions in $\beta(u)$ are used to prove the bound.
This is unfortunate since we believe that the thermodynamic bound is valid for {\em any}\/ range of energy.
It is quite important to look for a better argument which naturally covers a wider range of energy.

We can also prove the thermodynamic bound when we treat multiple thermodynamic quantities $\hM^{(1)}_V,\ldots,\hM^{(n)}_V$.
See the beginning of section~\ref{s:Pgeneral}.

\subsection{Ising model under transverse magnetic field}
\label{s:EIsing}
This is a specific model in the previous category, but we can prove a stronger bound in a wider range of energy by using correlation inequalities.

We associate an $S=1/2$ quantum spin with each site $x\in\Lambda$, and denote by ${\boldsymbol{\mathsf{S}}}_x=(\mathsf{S}^{(1)}_x,\mathsf{S}^{(2)}_x,\mathsf{S}^{(3)}_x)$ the corresponding spin operator.
We take the standard orthonormal basis $\{\ket{\varphi^+_x},\ket{\varphi^-_x}\}$ of the local Hilbert space $\calH_x\cong\mathbb{C}^2$.
The basis states are characterized by
\eq
\hS^{(3)}_x\ket{\varphi^\pm_x}=\pm\frac{1}{2}\ket{\varphi^\pm_x}.
\lb{Sphipm}
\en
Let $\bssigma=(\sigma_x)_{x\in\Lambda}$ with $\sigma_x=\pm$ be a (classical) spin configuration on $\Lambda$.
We define the corresponding basis state by
\eq
\ket{\Phi_{\bssigma}}:=\bigotimes_{x\in\Lambda}\ket{\varphi^{\sigma_x}_x}.
\lb{Phisigma}
\en
The whole Hilbert space $\Htot$ is spanned by $\ket{\Phi_{\bssigma}}$ with all possible $\bssigma$.

We take the general Hamiltonian of the ferromagnetic Ising model under transverse magnetic field
\eq
\Ham=
-\sumtwo{x,y\in\Lambda}{(x>y)}J_{x,y}\,\mathsf{S}^{(3)}_x\mathsf{S}^{(3)}_y
+h\sum_{x\in\Lambda}\mathsf{S}^{(1)}_x,
\lb{HamIsingTF}
\en
where we have introduced an arbitrary (but fixed) ordering in $\Lambda$ to avoid double counting in the first sum.
The interaction is translationally invariant, i.e., $J_{x,y}=J_{x+u,y+u}$ for any $x$, $y$, and $u$.
It also satisfies $J_{x,y}=0$ for any $x$, $y$ such that $|x-y|\ge R$ (where the range of interaction $R$ is an arbitrary constant independent of $L$), and $J_{x,y}\ge0$ for any $x$, $y$.
We also assume $h>0$.

We again focus on the range of energy where $\sigma(u)$ is strictly increasing, and assume that $\sigma(u)$ is differentiable\footnote{
We believe that $\sigma(u)$ is differentiable for all $u$, but cannot prove it in general.
}.
For $\beta>0$, let us define
\eq
\tilde{\chi}(\beta):=\lim_{V\up\infty}\sum_{x\in\Lambda}\sbkt{\mathsf{S}^{(3)}_o\mathsf{S}^{(3)}_x}^\mathrm{can}_{V,\beta},
\en
where $o$ is the origin, and $\sbkt{\cdots}^\mathrm{can}_{V,\beta}$ is the canonical expectation of the model \rlb{HamIsingTF}.
Although $\tilde{\chi}(\beta)$ is reminiscent of the formula for the susceptibility in classical spin systems, this quantity is not the susceptibility of this quantum spin system.

We focus on the total magnetization
\eq
\MV=\sum_{x\in\Lambda}\mathsf{S}^{(3)}_x.
\en
From the symmetry we readily see that $\bktmc{\MV}=0$ for any $V$ and $u$.
As for the fluctuation we can prove the following.

\begin{pro}
\label{p:TDBIsing}
Let $u$ be such that $\sigma(u)$ is differentiable, $\beta(u)=\sigma'(u)>0$, the condition \rlb{ucond} holds, and $\tilde{\chi}(\beta(u))<\infty$.
Take an arbitrary $\delta>0$.
Then, for sufficiently large $V_0$, there exists $\gamma(u,\delta,V_0)>0$ such that
\eq
\Bktmc{\hP\bigl[|\MV|\ge V\delta\bigr]}\le e^{-\gamma(u,\delta,V_0)\,V},
\lb{TDBIsing}
\en
holds for any $V\ge V_0$.
When $V_0$ is sufficiently large, we have
\eq
\gamma(u,V,\delta)\simeq\frac{\delta^2}{4\,\tilde{\chi}(\beta(u))}.
\lb{gammaIsing}
\en
\end{pro}

\section{Typicality of Thermal Equilibrium}
\label{s:typicality}
We shall return to the general setting of section~\ref{s:setup}, and discuss the typicality of thermal equilibrium.
We believe that the result in this section provides a rather satisfactory foundation of the description of thermal equilibrium in terms of the microcanonical ensemble.

We note that the argument here is essentially an application of standard results \cite{vonNeumann,PopescuShortWinter,GLTZ06,Sugita06,Sugita07,Reimann07,SugiuraShimizu12,SugiuraShimizu13,GHLT} to our setting and our definition of pure states representing equilibrium.
For completeness we shall describe a full derivation.

\subsection{Main statement}
\label{s:typMain}
Let us define the notion of typicality in general.
Let $\calS$ be a set equipped with a measure.
We say that a property for elements of $\calS$ is {\em typical}\/ if it is satisfied by an overwhelming majority of elements in $\calS$.
For example if $\calS$ is the interval $[0,1]\subset\mathbb{R}$ equipped with the standard Lebesgue measure, irrationality is a typical property.

Note that we cannot (and should not) say whether a given element of $\calS$ is typical or not.
In the above example of real numbers, transcendentality is also a typical property; we can never decide whether $\sqrt{2}/2$, which is irrational but algebraic, is typical or not.

Let $\tHVu$ be the space of all normalized states in $\HVu$.
To discuss typicality,
we need to introduce a measure on the space $\tHVu$.
Since $\kph\in\tHVu$ is expanded as $\kph=\sum_{j\in\JVu}c_j\,\kpj$ with $c_j\in\mathbb{C}$ and $\sum_{j\in\JVu}|c_j|^2=1$, the space $\tHVu$ is identified with the unit sphere in the $\DVu$ dimensional complex space $\mathbb{C}^{\DVu}$.
A mathematically natural measure is then the uniform measure on the unit sphere.
Note that this is the unique measure on the space $\tHVu$ which is independent of the choice of orthonormal basis of $\HVu$.

We thus define the average (with respect to the uniform measure) over $\tHVu$ as
\eq
\overline{F[\,\kph\,]}
:=\frac{\displaystyle\int\displaystyle\Bigl(\prod_{j\in\JVu}dc_j\Bigr)\,\delta\bigl(\displaystyle\sum_{j\in\JVu}|c_j|^2-1\bigr)\,F}
{\displaystyle\int\displaystyle\Bigl(\prod_{j\in\JVu}dc_j\Bigr)\,\delta\bigl(\displaystyle\sum_{j\in\JVu}|c_j|^2-1\bigr)},
\lb{avphi}
\en
where $F$ is an arbitrary function of $\kph$ (and hence of $(c_j)_{j\in\JVu}$), and $dc_j=d(\operatorname{Re}c_j)\,d(\operatorname{Im}c_j)$.

Noting that 
\eq
\overline{c_j^*\,c_k}=\frac{\delta_{j,k}}{\DVu},
\lb{cjck}
\en
which immediately follows from the symmetry, we readily find that 
\eq
\overline{\sbkt{\varphi|\hA|\varphi}}
=\sum_{j,k\in\JVu}\overline{c_j^*\,c_k}\,\bra{\psi_j}\hA\ket{\psi_k}
=\bktmc{\hA},
\lb{MC2}
\en
for any operator $\hA$, where $\bktmc{\cdots}$ is the microcanonical average defined in \rlb{mc}.
The relation \rlb{MC2} is mathematically trivial, but may be illuminating.
According to the standard definition \rlb{mc} in statistical mechanics, the microcanonical average is an average over a finite number of energy eigenstates in the energy shell.  
But \rlb{MC2} shows that exactly the same average is also written as an average over continuously infinite states in $\tHVu$.

Let $\Pneq$ be defined by \rlb{Pneq1}, \rlb{Pneqnew} or \rlb{Pneqn}  depending on the situation and the treatment.
The following theorem says that to represent thermal equilibrium is a typical property for  states in the energy shell $\HVu$.

\begin{theorem}[Typicality of thermal equilibrium]
\label{t:thtyp}
Assume that the thermodynamic bound \rlb{tdb} is valid with $\gamma$ such that $\gamma>\alpha$.
Fix an arbitrary $V\ge V_0$, and choose a normalized state $\kph\in\tHVu$ randomly according to the uniform measure on the unit sphere.
Then with probability\footnote{
The probability is with respect to the random choice of $\kph$.
} larger than $1-e^{-(\gamma-\alpha) V}$, we have
\eq
\sbkt{\varphi|\Pneq|\varphi}\le e^{-\alpha V},
\lb{typbound}
\en
which means that $\kph$ represents thermal equilibrium in the sense of Definition~\ref{d:eq}.
\end{theorem}

\noindent
{\em Proof\/\footnote{
One can indeed get a stronger bound by using the Chebyshev estimate (see, e.g., \cite{Sugita06,Sugita07})
\newline
${\rm Prob}[|\Pneq-\bktmc{\Pneq}|\ge s]\le n^2/(s^2\,\DVu)$.
}:}\/
Note that for any nonnegative (random) variable $A$ and a positive constant $a$, we have $\chi[A> a]\le A/a$, where $\chi[\text{true}]=1$ and $\chi[\text{false}]=0$.
Then
\eqa
{\rm Prob}\Bigl[\sbkt{\varphi|\Pneq|\varphi}>e^{-\alpha V}\Bigr]
&=
\overline{\chi\Bigl[\sbkt{\varphi|\Pneq|\varphi}>e^{-\alpha V}\Bigr]}
\nl&
\le e^{\alpha V}
\,\overline{\sbkt{\varphi|\Pneq|\varphi}}
\nl&
=e^{\alpha V}\,\bktmc{\Pneq}\le e^{-(\gamma-\alpha)V},
\ena
where we used \rlb{MC2} and the thermodynamic bound \rlb{tdb}.~\qedm

\bigskip
Recall that we have established the validity of the thermodynamic bound  \rlb{tdb} for some physical systems in section~\ref{s:examples}.
Since the only assumption in Theorem~\ref{t:thtyp} is the thermodynamic bound, we now have {\em concrete examples in which the typicality of thermal equilibrium has been proved}\/ rigorously.

\subsection{Discussion}
\label{s:typphys}

\paragraph{Typicality and justification of the microcanonical ensemble:}
Let us discuss the implication of Theorem~\ref{t:thtyp} on foundation of equilibrium statistical mechanics, in particular, justification of microcanonical ensemble.
Although we have already defined (in Definitions~\ref{d:eq} and \ref{d:eq2}) the notion of pure states representing thermal equilibrium by using the microcanonical average $\bktmc{\cdots}$, we shall (temporarily) forget about the physical interpretation that $\bktmc{\cdots}$ gives the expectation value in thermal equilibrium.
We shall regard $\bktmc{\cdots}$ as purely mathematical objects.

Since the discussion is rather delicate, we shall carefully distinguish between mathematical facts and physical postulates.

We start from a {\em mathematical fact}\/ implied by Theorem~\ref{t:thtyp}.
The theorem, along with the thermodynamic bound \rlb{tdb}, states that it is typical for states in the energy shell $\HVu$ to satisfy the bound \rlb{typbound}.
In other words, we see that an overwhelming majority of states in $\HVu$ are almost indistinguishable if one is interested only in the values of the macroscopic quantities $\MV^{(1)},\ldots,\MV^{(n)}$.
We stress that this mathematical fact is far from trivial.
It does not follow only from abstract quantum mechanics\footnote{
When the notion of pure states representing (thermal) equilibrium is defined through the expectation values as in \rlb{AsimA}, the typicality of (thermal) equilibrium follows only from the fact that the dimension of the relevant Hilbert space is large.
See, e.g., \cite{vonNeumann,Sugita06,Sugita07,Reimann07}.
}, but also requires nontrivial information (summarized as the thermodynamic bound) that the system has a normal thermodynamic behavior.

Given this mathematical fact, we shall make a {\em physical postulate}\/ that those macroscopic properties shared by the overwhelming majority of states correspond to thermal equilibrium observed in reality.
Of course typicality does not necessarily imply reality, but it may be  natural to assume what we normally observe are typical.

If we accept this postulate the use of the microcanonical ensemble is readily justified.
Since we know (for sure) that most states in the energy shell $\HVu$ are essentially identical from a macroscopic point of view, it suffices to take the average over the states in $\HVu$ in order to extract typical behaviors.
As is shown in \rlb{MC2}, this precisely leads to the microcanonical average.
We believe that this justification of equilibrium statistical mechanics  is directly relevant to the characterization of thermal equilibrium from a macroscopic and operational point of view.

It should be stressed that, although the thermodynamic bound \rlb{tdb} is stated in terms of the microcanonical expectation value, our argument does not involve circular logic (to make use of statistical mechanics for its own foundation).
When using the bound \rlb{tdb} in the proof of Theorem~\ref{t:thtyp}, we never make use of the interpretation that the left-hand side represents a thermal expectation value.
We simply take the quantity as it is, and derive the conclusion by only using basic rules of quantum mechanics.

The validity of any physical postulate should finally be verified empirically.
Needless to say the validity of equilibrium statistical mechanics (which is based on the microcanonical ensemble) has been tested experimentally through the long history.

To summarize, we believe that the argument based on the typicality of thermal equilibrium (which is grounded on the mathematical fact summarized in Theorem~\ref{t:thtyp}), along with the empirical success of  statistical mechanics, provides us with a sound justification of equilibrium statistical mechanics.

\paragraph{Physical meaning of typicality:}
The uniform measure on the unit sphere of $\mathbb{C}^{\DVu}$ is essentially the unique mathematically natural measure that we can associate with the space $\tHVu$ of normalized states in the energy shell.
This does not mean, however, that a state of the system is chosen according to this measure in physically realistic situations.
In reality, a state of the system is first prepared through a highly nontrivial process (which involves interactions with external systems) and then evolves according to the Hamiltonian time evolution.
The implication of the typicality is not immediately clear.

Nevertheless the typicality ensures that there are plenty of states in $\HVu$ which represent thermal equilibrium.
The typicality also suggests that one likely finds the system in thermal equilibrium unless there are special reasons to keep the system away from thermal equilibrium.

It is true that we are not able to make any definite physical conclusions from typicality alone.
But typicality, combined with empirical facts, can be a useful guide for finding the correct physical postulate.

\paragraph{Criticisms to the typicality argument:}
There are some criticisms to the typicality argument.
Let us discuss two points.

Although nonequilibrium states are atypical, it is (at least logically) possible that some of them are associated with extraordinarily long relaxation time.
Then if the system is once trapped in such nonequilibrium state, there is no chance of getting out within a reasonable amount of time.
In this sense justification of statistical mechanics should also deal with the problem of the time scale of thermalization.
See section~\ref{s:issues}.

Another criticism deals with the entanglement property \cite{EisertCramerPlenio}.
It is known that a state $\kph$ in $\HVu$ typically has strong entanglement within it\footnote{
Take a spacial region A in the system, and consider the density matrix $\rho_{\rm A}:={\rm Tr}_{\calH_{\bar{\rm A}}}[\kph\bra{\varphi}]$, where the trace is taken over the subspace corresponding to the region out side A.
Then the canonical typicality \cite{PopescuShortWinter,GLTZ06,Sugita06} implies that the entropy $S_{\rm A}=-{\rm Tr}[\rho_{\rm A}\log\rho_{\rm A}]$ is typically close to that of the canonical distribution, and is hence proportional to the volume of A.
}.
Since it has been argued that easily preparable states usually have much smaller entanglement\footnote{
A method for preparing a nonequilibrium state (in a numerical or a cold-atom experiment) is to start from the ground state of a certain Hamiltonian, and then quickly change the Hamiltonian (see the next part).
Since a ground state generally has small entanglement, the state cannot have too strong entanglement after a finite time.
} \cite{EisertCramerPlenio}, one might question if a realistic equilibrium state can be typical.
We still do not understand whether this point is essential, especially when our main goal is to reproduce macroscopic properties of thermal equilibrium starting from quantum mechanics.

\paragraph{Preparation of nonequilibrium states:}
Given the fact that an overwhelming majority of states in the energy shell represent thermal equilibrium, one might wonder why it is possible to prepare a state which is out of equilibrium.
Here we shall argue that one can generate a nonequilibrium state by a sudden change of Hamiltonian\footnote{
Physically speaking, the change of Hamiltonian is caused by an external agent, who must be in a nonequilibrium state to perform operations.
On Earth, such nonequilibrium states are prepared by using energy from the sun.
In the larger time scale, the origin of nonequilibrium goes back to the Big Bang.
}, as is often done in numerical works.
Although the argument should be standard, let us present it here in our setting\footnote{
A quench from the ground state may be discussed in a similar manner.
}.

We consider two translationally invariant Hamiltonians $\Ham$ and $\Ham'$, which are both the sums (or the integrals) of local operators.

First consider the energy shell $\HVu$ defined (as in section~\ref{s:MQS}) with respect to $\Ham$.
Then $\Ham'$ may be regarded as a macroscopic quantity.
We expect (and can prove in some situations) that the bound
\eq
\Bktmc{\hP\bigl[|(\Ham'/V)-u'|\ge\Du\bigr]}\le e^{-\gamma'V},
\lb{H'tdb}
\en
holds with a constant $\gamma'>0$ where $u':=\lim_{V\up\infty}\bktmc{(\Ham'/V)}$.
Note that \rlb{H'tdb} is nothing but the thermodynamic bound \rlb{tdb} for $\Ham'$.

The bound \rlb{H'tdb} implies that almost all state $\kph$ from $\HVu$ satisfies
\eq
\bra{\varphi}\hP\bigl[|(\Ham'/V)-u'|\ge\Du\bigr]\kph\le e^{-\alpha'V},
\lb{H'ph}
\en
with $0<\alpha'<\gamma'$.
This means that the energy shell $\HVu$ is essentially contained in another shell $\calH'_{V,u'}$ defined with respect to the Hamiltonian $\Ham'$ and the energy density $u'$.
If we denote the dimensions of the energy shells $\HVu$ and $\calH'_{V,u'}$ as $\DVu$ and $D'_{V,u'}$, respectively, this implies the inequality $\DVu\lesssim D'_{V,u'}$.

It is expected that $D'_{V,u'}$ should become much larger than $\DVu$
unless the two Hamiltonians $\Ham$ and $\Ham'$ are related in a special manner.
We expect (and can prove for some simple models) that $D'_{V,u'}$ is usually exponentially larger than $\DVu$, i.e., there is a constant $\Di\sigma>0$, and
\eq
\DVu e^{\Di\sigma\,V}\le D'_{V,u'}.
\lb{DD}
\en
In other words, the first energy shell $\HVu$ occupies an exponentially small proportion of the new energy shell $\calH'_{V,u'}$.
Let us assume in the following that the inequality \rlb{DD} is valid\footnote{
In terms of the Boltzmann entropy $S=\kB\log D$, the inequality \rlb{DD} means $S'(u')\ge S(u)+\kB\Di\sigma\,V$.
It is quite normal that the entropy increases after a sudden change of Hamiltonian.
}.

We imagine that the Hamiltonian is initially $\Ham$, and take a state $\kph\in\HVu$, which is very likely to represent thermal equilibrium with respect to $\Ham$ (but we do not assume this).
We then imagine that the Hamiltonian is suddenly changed to $\Ham'$, but the state remains to be $\kph$.
We assume that $\kph$ satisfies the bound \rlb{H'ph} (which is quite likely), and hence is essentially contained in the new energy shell $\calH_{V,u'}'$.
Since $\HVu$ occupies an exponentially small fraction of $\calH_{V,u'}'$, there is a chance that the state $\kph$ is regarded to be nonequilibrium with respect to the new Hamiltonian $\Ham'$.
We argue that this is indeed the case.

Note that $\Ham$ is no longer the Hamiltonian, but a macroscopic quantity.
We claim that the equilibrium value $u''$ of $\Ham/V$ in $\calH_{V,u'}'$ satisfies $u''>u$.
Then, since the state $\kph$ satisfies $\bra{\varphi}(\Ham/V)\kph\simeq u$ by definition, it is definitely not in thermal equilibrium.

To show the above claim, let $u'':=\lim_{V\up\infty}\sbkt{(\Ham/V)}_{V,u'}^{\rm mc'}$, where $\sbkt{\cdots}_{V,u'}^{\rm mc'}$ is the microcanonical expectation corresponding to $\calH_{V,u'}'$.
Then by repeating the same argument as before, one can say that $\calH_{V,u'}'$ is essentially contained in $\calH_{V,u''}$, the energy shell defined in terms of $\Ham$ and $u''$.
Since the dimension must satisfy $D'_{V,u'}\lesssim D_{V,u''}$, the inequality \rlb{DD} implies that $u''$ must be strictly larger than $u$.

\section{Thermalization: General statement}
\label{s:thermalization}
We shall move onto the issue of thermalization, or, equivalently, the approach to thermal equilibrium.
Although the results in this direction are not as satisfactory as that of the typicality, we can show that, in some situations, the unitary time evolution of the isolated quantum system can describe thermalization.
Here we take the setting of section~\ref{s:setup}, and discuss a general statement.

Let $\kpz\in\HVu$ be a normalized initial state\footnote{
We can extend our results about thermalization to the case where the system is initially in a mixed state.
See Appendix~\ref{s:mixed}.
}, and
\eq
\kpt=e^{-i\Ham t}\kpz,
\lb{kpt}
\en
be the corresponding state at $t\ge0$.
Note that we are considering purely Hamiltonian time evolution in the isolated quantum system.
We wish to show that, when we start from the initial state $\kpz$ which does not necessarily represent thermal equilibrium, the state $\kpt$ approaches thermal equilibrium, i.e., represents thermal equilibrium for sufficiently large and most $t$.

The following lemma shows what we mean by $\kpz$ approaches thermal equilibrium in the time scale $\tau$.
Again $\Pneq$ is defined by \rlb{Pneq1}, \rlb{Pneqnew} or \rlb{Pneqn} .
\begin{lemma}
\label{l:th}
Suppose that there is (sufficiently large) $\tau>0$ and it holds that
\eq
\frac{1}{\tau}\int_0^\tau dt\,\sbkt{\varphi(t)|\Pneq|\varphi(t)}
\le e^{-(\alpha+\nu) V},
\lb{taV}
\en
with a constant $\nu>0$.
Then there exists a collection of intervals $\calG\subset[0,\tau]$ such that\/\footnote{
$|\calG|$ stands for the total length (i.e., the Lebesgue measure) of the intervals in $\calG$.
} $|\calG|/\tau\ge1-e^{-\nu V}$, and we have for any $t\in\calG$ that
\eq
\sbkt{\varphi(t)|\Pneq|\varphi(t)}\le e^{-\alpha V},
\lb{thermalizationbound}
\en
which means that $\kpt$ represents thermal equilibrium in the sense of Definition~\ref{d:eq}.
\end{lemma}

Here $\calG$ stands for the ``good" subset of $[0,\tau]$, in which the time-evolved state $\kpt$ represents equilibrium.
Since $\calG$ occupies an overwhelming majority of the whole time interval $[0,\tau]$ when $V$ is large, one who makes measurement at an arbitrary moment almost certainly falls into the set $\calG$, and hence almost certainly observes thermal equilibrium.
We can therefore say that the system is in thermal equilibrium for sufficiently long and most $t\in[0,\tau]$, provided that the condition of the lemma is satisfied.

Note that one can never expect a complete relaxation to thermal equilibrium since the time evolution \rlb{kpt} is quasi periodic.
See also \rlb{pt}.
To find the system in thermal equilibrium for most $t$ is the most we can expect.

\bigskip\noindent
{\em Proof of Lemma~\ref{l:th}\/:}\/
The proof is trivial.
Define the ``good" set by
\eq
\calG:=\set{t\in[0,\tau]}{\sbkt{\varphi(t)|\Pneq|\varphi(t)}
\le e^{-\alpha V}}.
\en
Then, noting that $e^{-\alpha V}\chi[t\not\in\calG]\le \sbkt{\varphi(t)|\Pneq|\varphi(t)}$, we see from \rlb{taV} that
\eq
1-\frac{|\calG|}{\tau}=\frac{1}{\tau}\int_0^\tau dt\,\chi[t\not\in\calG]
\le\frac{1}{\tau}\int_0^\tau dt\,e^{\alpha V}\,\sbkt{\varphi(t)|\Pneq|\varphi(t)}\le e^{-\nu V}.\quad\qedm
\en

\bigskip

The essential task then is to justify the bound \rlb{taV} for the time average, which is the only assumption in the lemma.
For the moment two complementary strategies for justification are known; the first applies to any system but relies on the assumption that the initial state has a moderate energy distribution, and the second works for any initial state but assumes the ``energy eigenstate thermalization hypothesis''.
We shall discuss them separately in the following two sections.

\paragraph{The law of entropy increase:}
According to (nonequilibrium) thermodynamics, the entropy should increase when a nonequilibrium initial state approaches thermal equilibrium.
This may look puzzling since, in our case where the state is always pure, the von Neumann entropy $S_{\rm vN}(t)=-\kB{\rm Tr}[\rho(t)\log\rho(t)]$ with $\rho(t)=\ket{\varphi(t)}\bra{\varphi(t)}$ is always vanishing, and hence is time-independent.

Recall, however, that we can define multiple essentially different entropies for a macroscopic quantum system\footnote{
Note that we do not have such freedom (or ambiguity) for geometric quantities such as the volume or mechanical quantity such as the energy.
Entropy is a delicate quantity.
}.
The von Neumann entropy is the ``most microscopic" entropy, which may or may not be relevant to macroscopic physics\footnote{
For the Gibbs state $\rho_{\beta}:=e^{-\beta\Ham}/{\rm Tr}[e^{-\beta\Ham}]$, the von Neumann entropy $-{\rm Tr}[\rho_\beta\log\rho_\beta]$ coincides with the (most macroscopic) thermodynamic entropy.
}.
There are other entropies which reflect certain coarse grained points of view.
The most macroscopic is the thermodynamic entropy, which is a function of some macroscopic quantities.
Consider a system of particles, and take the energy $U$, the volume $V$, the particle number $N$, and the (value of the) macroscopic quantity $M$ as the parameters.
Then the thermodynamic entropy\footnote{
Note that the standard thermodynamic entropy $S(U,V,N)$ must be a function of controllable parameters.
Since the value of $\MV$ cannot be controlled, $S(U,V,N,M)$ should be regarded as a nonequilibrium entropy.
Such an entropy may be defined microscopically, for example, as $S(U,V,N,M):=\kB\log{\rm Tr}_{\HVu}[\hP[|\MV-M|\le V\delta]]$ in the spirit of Boltzmann.
} $S(U,V,N,M)$ should take its maximum at the equilibrium value of $M$ when $U$, $V$, and $N$ are fixed.
It is expected (but not easy to prove) that the entropy $S_{\rm macro}(t)=S(U,V,N,M(t))$ with $M(t)=\bra{\varphi(t)}\MV\ket{\varphi(t)}$ increases in time when the system approaches thermal equilibrium.
See \cite{GL2004,GGL2004} for a related research for classical systems.

\paragraph{Systems which do not thermalize:}
Thermalization is not always expected to take place.
It has been clarified that certain isolated quantum systems fail to thermalize, or relax to states which are different from the thermal equilibrium.

An important class consists of those models which exhibit many body localization \cite{PH,Imbrie}, which prohibits the system from relaxing to equilibrium states.
See also \cite{GHLT} for a discussion about the implication of many body localization.

Another important class is that of exactly solvable models, which relax to macroscopic states described by the Generalized Gibbs Ensemble (GGE) \cite{Rigol2007,Ilievski}.
It is not clear whether our notion of thermal equilibrium distinguishes between the standard thermal equilibrium states and states described by the GGE.
It is also known that thermalization is lost near integrable points \cite{Rigol2009,Rigol2009B}.

\section{Thermalization: Initial states with moderate energy distributions}
\label{s:moderate}
Let us discuss the first strategy for the proof of the bound \rlb{taV}.
It is based on the assumption that the initial state has a moderate energy distribution.
To our knowledge, such a strategy was first discussed in \cite{Tasaki1998}, and used in a variety of works including \cite{Reimann,LindenPopescuShortWinter,ReimannKastner,Reimann2}.

\subsection{Main results}

\paragraph{Assumptions:}
We fix the volume $V$.
We assume that the thermodynamic bound \rlb{tdb} is valid, and that there is no degeneracy in the energy eigenvalues, i.e., $E_j\ne E_{k}$ if $j\ne k$.
In fact our results hold when there is some degeneracy.  See the discussion below Theorem~\ref{t:th1}.

Take a normalized initial state $\kpz\in\tHVu$, and expand it as 
\eq
\kpz=\sum_{j\in\JVu}c_j\,\kpj.
\lb{exp}
\en
Then we assume that, for some constant $\eta$ such that $0<\eta<\gamma$ and some $V$, the coefficients satisfy
\eq
\Deff:=\Bigl(\sum_{j\in\JVu}|c_j|^4\Bigr)^{-1}\ge e^{-\eta V}\DVu.
\lb{caV}
\en
Note that $\Deff$, which is called the effective dimension, can be interpreted as the effective number of basis states which contribute to the expansion \rlb{exp}.

The bound \rlb{caV} essentially says that the initial state $\kpz$ is not too sharply concentrated on a small number of energy eigenstates.
If $\kpz$ is a linear combination of $n$ energy eigenstates, then one has $\Deff\le n$, and hence \rlb{caV} is never satisfied.
If, on the other hand, $|c_j|=1/\sqrt{\DVu}$ for all $j\in\JVu$ in the expansion \rlb{exp}, the effective dimension takes the maximum possible value $\Deff=\DVu$, and
the condition \rlb{caV} is satisfied.

Note that $\Deff=\DVu$ means $|c_j|=1/\sqrt{\DVu}$ for all $j\in\JVu$, only leaving the freedom to choose phase factor of each $c_j$.
It is crucial that we have an extra small factor $e^{-\eta V}$ in the assumed bound \rlb{caV}.
The factor $e^{-\eta V}$ allows much more freedom in the choice of $\kpz$.

We expect that many (or, hopefully, most) nonequilibrium initial states in a generic quantum many body system satisfy the bound \rlb{caV}.
It is indeed easy to see\footnote{
An explicit calculation shows $\overline{\sum_{j\in\JVu}|c_j|^4}=2/(\DVu+1)$, where the bar indicates the average over all normalized states as in section~\ref{s:typicality}.
If we again choose $\kpz\in\tHVu$ randomly, we see that
$\mathrm{Prob}[\sum_j|c_j|^4\ge e^{\eta V}/\DVu]
=\overline{\chi[\sum_j|c_j|^4\ge e^{\eta V}/\DVu]}
\le\overline{\sum_j|c_j|^4\,e^{-\eta V}\,\DVu}
\le2\,e^{-\eta V}$.
} that an overwhelming majority of states in $\tHVu$ satisfy the bound \rlb{caV}; but this fact is useless since most of the states in $\tHVu$ are already known to represent thermal equilibrium.
Whether a nonequilibrium initial state $\kpz$ generally satisfies the bound \rlb{caV} is a nontrivial issue, which crucially depends on the nature of the system.
For the moment, the validity of  \rlb{caV} for general nonequilibrium states is known for rather artificial examples.
See sections~\ref{s:freefermions} and \ref{s:toytwobody}.
We shall discuss this important issue about the assumption  \rlb{caV} after stating and proving the general theorem.

\paragraph{Theorem and the proof:}
The following theorem is a variation of a statement due to Goldstein, Hara and Tasaki\footnote{
It was stated first in the footnote of the unpublished work \cite{GHT2ndLaw} and then as Theorem~A.2 of \cite{GHT14long}.
}.

\begin{theorem}
\label{t:th1}
If the thermodynamic bound \rlb{tdb} and the bound \rlb{caV} for the initial state are valid with $\gamma$ and $\eta$ such that\footnote{
Given $\gamma$ and $\eta$ with $\gamma>\eta$, one may choose $\alpha$ and $\nu$ such that $\alpha+\nu<(\gamma-\eta)/2$.
} $\gamma-\eta>2(\alpha+\nu)$, then we have the desired bound \rlb{taV}.  Thus the state $\kpz$ approaches thermal equilibrium (in the sense of Lemma~\ref{l:th}). 
\end{theorem}

Recall again that the thermodynamic bound \rlb{tdb} has been proved in some concrete systems.
In these cases we have established the approach to thermal equilibrium from initial states $\kpz$ satisfying \rlb{exp} and \rlb{caV}.
We believe that this is a rather strong result in the foundation of equilibrium statistical mechanics, although we still need to develop a better understanding of the assumption \rlb{caV}.

\bigskip
\noindent{\em Proof:}\/
Since we have $\kpt=\sum_{j\in\JVu}e^{-iE_j t}c_j\kpj$ from the expansion \rlb{exp}, we see that the long-time average is given by
\eqa
\lim_{\tau\uparrow\infty}\frac{1}{\tau}\int_0^\tau dt\,
\sbkt{\varphi(t)|\Pneq|\varphi(t)}
&=\lim_{\tau\uparrow\infty}\frac{1}{\tau}\int_0^\tau dt\,\sum_{j,k\in\JVu}(c_j)^*\,c_{k}\,e^{i(E_j-E_k)t}\sbkt{\psi_j|\Pneq|\psi_k}
\nl
&=\sum_{j\in\JVu}|c_j|^2\sbkt{\psi_j|\Pneq|\psi_j},
\lb{thupper0}
\ena
where we used the nondegeneracy.
By using the Schwarz inequality and noting that $\sbkt{\psi_j|\Pneq|\psi_j}\le1$, we bound the right-hand side as
\eqa
\sum_{j\in\JVu}|c_j|^2\sbkt{\psi_j|\Pneq|\psi_j}
&\le
\sqrt{\Bigl(\sum_{j\in\JVu}|c_j|^4\Bigr)
\Bigl(\sum_{j\in\JVu}\sbkt{\psi_j|\Pneq|\psi_j}^2\Bigr)}
\nl&\le
\sqrt{\Bigl(\sum_{j\in\JVu}|c_j|^4\Bigr)
\Bigl(\sum_{j\in\JVu}\sbkt{\psi_j|\Pneq|\psi_j}\Bigr)}
=\sqrt{\frac{\DVu\,\bktmc{\Pneq}}{\Deff}},
\ena
where we used the definitions \rlb{mc} and \rlb{caV}.
Recalling the bounds \rlb{tdb} and \rlb{caV}, we find 
\eq
\lim_{\tau\uparrow\infty}\frac{1}{\tau}\int_0^\tau dt\,
\sbkt{\varphi(t)|\Pneq|\varphi(t)}\le e^{-\{(\gamma-\eta)/2\}V},
\en
where the right-hand side is strictly smaller than $e^{-(\alpha+\nu) V}$ provided that $\gamma-\eta>2(\alpha+\nu)$.  This means that the desired \rlb{taV} is valid for sufficiently large $\tau$.~\qedm

\paragraph{Treatment of degeneracy:}
It might be obvious that the assumption about the nondegeneracy of the energy eigenvalues can be replaced by a milder condition.
Although we expect that a a generic Hamiltonian has no degeneracy, let us describe how one can take into account some degeneracy.

To treat a degenerate Hamiltonian, fix $V$ and $u$, and decompose the index set $\JVu$ as $\JVu=\bigcup_{k=1}^K\tiJ_k$ in such a way that $E_j=E_{j'}$ if $j,j'\in\tiJ_k$, and $E_j\neq E_{j'}$ if $j\in\tiJ_k$ and $j'\in\tiJ_{k'}$ with $k\ne k'$.

Then the long-time average in \rlb{thupper0} becomes
\eq
\lim_{\tau\uparrow\infty}\frac{1}{\tau}\int_0^\tau dt\,
\sbkt{\varphi(t)|\Pneq|\varphi(t)}
=\sum_{k=1}^K\sum_{j,j'\in\tiJ_k}(c_j)^*c_{j'}\sbkt{\psi_j|\Pneq|\psi_{j'}}.
\lb{thuDeg}
\en
By using the Schwarz inequality the right-hand side is bounded as
\eqa
&\le\sum_{k=1}^K\sum_{j,j'\in\tiJ_k}
\sqrt{|c_{j}|^2\sbkt{\psi_{j}|\Pneq|\psi_{j}}}
\sqrt{|c_{j'}|^2\sbkt{\psi_{j'}|\Pneq|\psi_{j'}}}
\nl
&\le\frac{1}{2}\sum_{k=1}^K\sum_{j,j'\in\tiJ_k}
\Bigl\{
|c_{j}|^2\sbkt{\psi_{j}|\Pneq|\psi_{j}}+
|c_{j'}|^2\sbkt{\psi_{j'}|\Pneq|\psi_{j'}}\Bigr\}
\nl
&\le\sum_{k=1}^K|\tiJ_k|\sum_{j\in\tiJ_k}|c_{j}|^2\sbkt{\psi_{j}|\Pneq|\psi_{j}},
\lb{thuDeg2}
\ena
where we used the trivial inequality $ab\le(a^2+b^2)/2$ for $a,b\in\bbR$ to get the second line.

Now assume that the maximum degree of degeneracy of the energy eigenvalues is $\bar{d}_{V,u}$, i.e., we have $|\tiJ_k|\le\bar{d}_{V,u}$ for any $k=1,\ldots,K$.
Then, from \rlb{thuDeg} and \rlb{thuDeg2}, we get
\eq
\lim_{\tau\uparrow\infty}\frac{1}{\tau}\int_0^\tau dt\,
\sbkt{\varphi(t)|\Pneq|\varphi(t)}
\le\bar{d}_{V,u}\sum_{j\in\JVu}|c_j|^2\sbkt{\psi_j|\Pneq|\psi_j},
\lb{thuDeg3}
\en
which is the same as the bound \rlb{thupper0} except for the extra degeneracy factor $\bar{d}_{V,u}$.
This means that Theorems~\ref{t:th1} and \ref{t:th2}, which are both based on the bound \rlb{thupper0}, are valid as they are if there are positive constants $a$, $b$ (which may depend on $u$) and the degeneracy is bounded as $\bar{d}_{V,u}\le a V^b$.

\paragraph{On the assumption \protect\rlb{caV}:}
Note that Theorem~\ref{t:th1} is proved by only assuming the non-degeneracy of the energy eigenvalues, the thermodynamic bound, and the bound \rlb{caV} for the initial state.
The theorem therefore applies to a wide class of systems.
Indeed it applies also to such trivial systems like two bodies {\em not}\/ in contact (obtained by setting $\mathsf{H}_\mathrm{int}=0$ in the model of section~\ref{s:EHC}) or the Ising model {\em without}\/ external magnetic field (obtained by setting $h=0$ in the model of section~\ref{s:EIsing}), where thermalization clearly does note take place.
This may sound puzzling since the theorem states the approach to thermal equilibrium.

The key lies in the condition \rlb{caV} for the initial state to have a moderate energy distribution.
To avoid contradiction, we must conclude that, in a trivial system without thermalization, any state that satisfies the condition \rlb{caV} represents thermal equilibrium to begin with\footnote{
Note that we are here talking about the thermal equilibrium characterized by the particular $\Pneq$.
Even in the system of two bodies not in contact, there can be nontrivial thermalization within each body.
}.

This observation suggests that the validity of the condition \rlb{caV} for a {\em non}\/equilibrium initial state $\kpz$ is a much more delicate issue than it seems.
Theorem~\ref{t:th1} implies that the system must be nontrivial (so as to exhibit thermalization) in order for a nonequilibrium state $\kpz$ satisfying \rlb{caV} to be possible.

In the following two subsections, we shall examine this picture in two concrete examples.
As for the example of two bodies in contact (section~\ref{s:condHC}), we argue that the existence of a nonequilibrium state satisfying \rlb{caV} may be regarded as a criterion for the two subsystems to be truly coupled.
Then in section~\ref{s:suff}, we state a general sufficient condition for the bound \rlb{caV}.

\subsection{Ising model under transverse magnetic field}
We briefly discuss the Ising model under (or without) transverse magnetic field introduced in section~\ref{s:EIsing}.
We first fix a classical spin configuration $\bssigma^\mathrm{neq}$ which can be regarded as nonequilibrium, and set the initial state as 
\eq
\kpz=\ket{\Phi_{\bssigma^\mathrm{neq}}}, 
\lb{kpzPhi0}
\en
where $\ket{\Phi_{\bssigma}}$ is the basis state defined in \rlb{Phisigma}.

Let us consider the trivial model with $h=0$.
The model is nothing but the classical Ising model, where the total magnetization $\MV=\sum_{x\in\Lambda}\mathsf{S}^{(3)}_x$ is a constant of motion; we never have thermalization (in which $\MV$ decays to zero).

In this model each $\ket{\Phi_{\bssigma}}$ is an energy eigenstate.
Therefore when we expand our initial state \rlb{kpzPhi0} in terms of the energy eigenstate as in \rlb{exp}, there is only one nonvanishing term in the sum.
We find that $D_\mathrm{eff}=1$, and the assumption \rlb{caV} can never be satisfied.

Let us then consider nontrivial models with sufficiently large $h>0$.
We expect that the energy eigenstates are linear combinations of various $\ket{\Phi_{\bssigma}}$'s, and hence many energy eigenstates contribute in the expansion \rlb{exp}.
Then it is likely that the assumption \rlb{caV} is satisfied.
This expectation may be justified by the following two simple observations.

Let $E_0=-\sum_{x,y}J_{x,y}\sigma^\mathrm{neq}_x\sigma^\mathrm{neq}_y$.
By recalling that $\sbkt{\varphi_x^{\pm}|\mathsf{S}^{(1)}_x|\varphi_x^{\pm}}=0$ and $(\mathsf{S}^{(1)}_x)^2=1/4$, we readily find that
\eqg
\sbkt{\Phi_{\bssigma^\mathrm{neq}}|\{\Ham-E_0\}|\Phi_{\bssigma^\mathrm{neq}}}=0,\\
\sbkt{\Phi_{\bssigma^\mathrm{neq}}|\{\Ham-E_0\}^2|\Phi_{\bssigma^\mathrm{neq}}}=\frac{h^2}{4}V.
\eng
This implies that, in the state $\Phi_{\bssigma^\mathrm{neq}}$, the energy is distributed roughly in the range $E_0\pm h\sqrt{V}$.
Since there are a large number of energy eigenstates in this range, we expect (but cannot yet prove) that $\kpz=\ket{\Phi_{\bssigma^\mathrm{neq}}}$ is a linear combination of many energy eigenstates.

The second observation is based on the translation invariance of the model.
Let $T_y$ be the translation by $y$, and assume that $T_y[\bssigma^\mathrm{neq}]$ with $y\in\Lambda$ are all distinct.
There are many spin configurations with this property.
Now assume that the energy eigenvalues are nondegenerate, and let $\ket{\psi_j}$ be an arbitrary energy eigenstate.
The nondegeneracy implies that $\ket{\psi_j}$ is translationally invariant, and hence
\eq
\Bigl|\bbkt{\Phi_{\bssigma^\mathrm{neq}}|\psi_j}\Bigr|=
\Bigl|\bbkt{\Phi_{T_y[\bssigma^\mathrm{neq}]}|\psi_j}\Bigr|,
\en
for any $y\in\Lambda$.
Since $\ket{\Phi_{T_y[\bssigma^\mathrm{neq}]}}$ with $y\in\Lambda$ are all distinct, we find from the normalization condition that
\eq
\Bigl|\bbkt{\Phi_{\bssigma^\mathrm{neq}}|\psi_j}\Bigr|\le\frac{1}{\sqrt{V}}.
\en
This is an extremely crude bound (since we expect the left-hand side to be exponentially small in $V$), but at least proves that $\kpz=\Phi_{\bssigma^\mathrm{neq}}$ is a linear combination of a large number of energy eigenstates, and that the corresponding effective dimension satisfies $D_\mathrm{eff}\ge V$.

\subsection{Two bodies in contact}
\label{s:condHC}
Next we focus on the problem of two bodies in contact formulated in section~\ref{s:EHC}.

\paragraph{Trivial model without thermalization:}
Let us examine in detail the trivial case where two bodies are {\em not}\/ in contact.
This consideration sheds light on the assumption \rlb{caV}.

Consider the model of section~\ref{s:EHC}, but set\footnote{
The energy eigenvalues are then degenerate since the two subsystems are identical.
But this is not essential since the degeneracy may be lifted by making very small difference between the two Hamiltonians $\mathsf{H}^{(1)}_{V/2}$ and $\mathsf{H}^{(2)}_{V/2}$.
} $\mathsf{H}_\mathrm{int}=0$.
The two subsystems are completely decoupled, and there can be no thermalization (where the energy difference $\MV$ of \rlb{MVHC} relaxes).
We recall that the thermodynamic bound \rlb{TDBHC} is still valid for this model.

Note that, in this case, the tensor product state $\ket{\Psi_{j_1,j_2}}$ defined in \rlb{pp12} is an exact energy eigenstate.
Let us expand the initial state $\kpz\in\HVu$ as
\eq
\kpz=\sum_{j_1,j_2}c_{j_1,j_2}\,\ket{\Psi_{j_1,j_2}}.
\lb{pcP12}
\en
We also note that the projection operator which characterizes the nonequilibrium behavior (i.e., the difference in the energy densities in the two subsystems) is written exactly as
\eq
\hP\bigl[|\MV|\ge V\delta\bigr]=\sum_{(j_1,j_2)\in\JVu^{(\delta)}}
\ket{\Psi_{j_1,j_2}}\bra{\Psi_{j_1,j_2}},
\lb{PMPP}
\en
where $\JVu^{(\delta)}$ is the set of $(j_1,j_2)$ which satisfies both \rlb{uE1E2} and \rlb{E1E2d}.

We first look at a genuine nonequilibrium state $\kpz$ characterized by
\eq
\bkt{\varphi(0)\Bigl|\,\hP\bigl[|\MV|\ge V\delta\bigr]\,\Bigr|\varphi(0)}=1.
\lb{genuine}
\en
Clearly such a state is a linear combination of $\ket{\Psi_{j_1,j_2}}$ as in \rlb{pcP12} where $c_{j_1,j_2}\ne0$ only for $(j_1,j_2)\in\JVu^{(\delta)}$.
Since the effective dimension of such a state cannot exceed the number of elements in $\JVu^{(\delta)}$, we see
\eq
D_\mathrm{eff}\le\bigl|\JVu^{(\delta)}\bigr|=\mathrm{dim}[\HVu^{(\delta)}]=e^{-\gamma V}\,\DVu,
\en
where we used \rlb{PMDD}.
This means that the desired bound $D_\mathrm{eff}\ge e^{-\eta V}\,\DVu$ can be valid only when $\eta\ge\gamma$.
Since our requirement is $\gamma>\eta$, we find that any genuine nonequilibrium state with \rlb{genuine} fails to satisfy the condition \rlb{caV}  (with $0<\eta<\gamma$).

Let us turn to the case of a general nonequilibrium state.
Suppose that a state $\kpz$ does {\em not}\/ represent thermal equilibrium in the sense that
\eq
\bkt{\varphi(0)\Bigl|\,\hP\bigl[|\MV|\ge V\delta\bigr]\,\Bigr|\varphi(0)}\ge\epsilon(V),
\lb{pPpep}
\en
where $\epsilon(V)$ is any quantity (like $V^{-1}$) which exceeds $e^{-\alpha V}$ with any $\alpha>0$ when $V$ grows.
Recalling \rlb{pcP12} and \rlb{PMPP}, we see from the bound \rlb{pPpep} that
\eq
\sum_{(j_1,j_2)\in\JVu^{(\delta)}}|c_{j_1,j_2}|^2\ge\epsilon(V).
\en
Since the Schwarz inequality implies
\eq
\sum_{(j_1,j_2)\in\JVu^{(\delta)}}|c_{j_1,j_2}|^2
\le\sqrt{\bigl|\JVu^{(\delta)}\bigr|\sum_{(j_1,j_2)\in\JVu^{(\delta)}}|c_{j_1,j_2}|^4},
\en
we have
\eq
\sum_{j_1,j_2}|c_{j_1,j_2}|^4
\ge
\sum_{(j_1,j_2)\in\JVu^{(\delta)}}|c_{j_1,j_2}|^4
\ge\frac{\{\epsilon(V)\}^2}{\bigl|\JVu^{(\delta)}\bigr|}
=\frac{\{\epsilon(V)\}^2}{e^{-\gamma V}\,\DVu},
\en
and hence
\eq
D_\mathrm{eff}\le\{\epsilon(V)\}^{-2}\,e^{-\gamma V}\,\DVu,
\en
Since $\{\epsilon(V)\}^{-2}$ decays faster (as $V$ grows) than $e^{\alpha V}$ (with any $\alpha$), we again see that the condition \rlb{caV} (with $0<\eta<\gamma$) can never be satisfied.

\paragraph{Nontrivial model which (probably) shows thermalization:}
The reason for the failure of the condition \rlb{caV} in the above example is that the nonequilibrium state $\ket{\Psi_{j_1,j_2}}$ itself happens to be an energy eigenstate.
This is of course a very special situation which is found only when the two subsystems are  decoupled.

If the two subsystems are fully coupled, one expects that a general energy eigenstate of the total Hamiltonian $\Ham$ is written as
\eq
\ket{\Psi_j}=\sum_{j_1,j_2}\gamma^{(j)}_{j_1,j_2}\ket{\Psi_{j_1,j_2}},
\lb{demoexp1}
\en
where the amplitudes $|\gamma^{(j)}_{j_1,j_2}|^2$ are (naively) expected to be nonvanishing and comparable for most $(j_1,j_2)$ such that 
\eq
\bigl|E_j-(E^{(1)}_{j_1}+E^{(2)}_{j_2})\bigr|\lesssim \delta E,
\lb{EE1E2}
\en
where $\delta E$ is a small energy width determined by the interaction $\hH_\mathrm{int}$.
It is then expected that for a general pair $(j_1,j_2)$, one has an expansion
\eq
\ket{\Psi_{j_1,j_2}}=\sum_j c^{(j_1,j_2)}_j\ket{\Psi_j},
\lb{demoexp2}
\en
where the amplitudes $|c^{(j_1,j_2)}_j|^2$ are nonvanishing and comparable for most $j$ which satisfy \rlb{EE1E2}.
If this is the case, we can choose  $(j_1,j_2)$ such that $E^{(1)}_{j_1}-E^{(2)}_{j_2}\ge V\delta$ to define an nonequilibrium initial state as $\kpz=\ket{\Psi_{j_1,j_2}}$, which have the effective dimension $D_\mathrm{eff}\sim\DVu$.
The condition \rlb{caV} is satisfied.

The above picture, which relies on the assumption of ``democracy'' in the expansions \rlb{demoexp1} and \rlb{demoexp2}, may be too crude and naive to be valid in arbitrary macroscopic systems\footnote{
It holds for the highly artificial toy model for heat conduction discussed in section~\ref{s:toytwobody}.
See \cite{Tasaki1998} for another artificial example.
}.
But let us stress that the absence of a nonequilibrium state satisfying the condition \rlb{caV} implies that essentially a finite number of terms contribute in the expansion \rlb{demoexp1}, which means that the coupling between the two subsystems is extremely small.
We believe that the existence of a nonequilibrium state satisfying \rlb{caV} can be regarded as a criterion for the two subsystems to be truly coupled.
It would be extremely useful to have concrete and nontrivial examples where this picture can be justified.

\subsection{A sufficient condition for the bound \protect{\rlb{caV}}}
\label{s:suff}
Given the fact that the condition \rlb{caV} cannot be satisfied by any nonequilibrium state in a trivial system (which does not exhibit thermalization), it is desirable to have complementary results for nontrivial systems where thermalization is expected.

Here we state a sufficient condition for the bound \rlb{caV} for a moderate energy distribution to be valid for a large number of nonequilibrium initial states.
Let $\kxi\in\HVu$ with $i=1,\ldots,\Dneq$ be mutually orthogonal normalized states (i.e., $\sbkt{\xi_i|\xi_{i'}}=\delta_{i,i'}$) with the property that their linear combination
\eq
\kpz=\sum_{i=1}^{\Dneq}\alpha_i\kxi
\lb{p0neq}
\en
with any $\alpha_i\in\mathbb{C}$ such that $\sum_{i=1}^{\Dneq}|\alpha_i|^2=1$ does {\em not}\/ represent thermal equilibrium.
We expect that the maximum possible $\Dneq$ is given by $\Dneq\sim e^{-\gamma V}\DVu\ll\DVu$, where the factor $e^{-\gamma V}$ is that appears in the thermodynamic bound \rlb{tdb}.

Let us expand the nonequilibrium basis states by the energy eigenstate basis as
\eq
\kxi=\sum_{j\in\JVu}g_{i,j}\kpj,
\lb{xig}
\en
where $\sum_{j\in\JVu}|g_{i,j}|^2=1$ for any $i=1,\ldots,\Dneq$.

We shall then assume that, for some $V>0$, the coefficients $g_{i,j}$ satisfy
\eq
D_{\mathrm{eff},i}:=\Bigl(\sum_{j\in\JVu}|g_{i,j}|^4\Bigr)^{-1}\ge e^{-(\eta-\epsilon) V}\DVu,
\lb{c4B}
\en
for any $i=1,\ldots,\Dneq$ with some $\epsilon>0$.
The bound, which should be compared with the desired bound \rlb{caV}, says that any nonequilibrium basis state $\kxi$ is a linear combination of a large number of energy eigenstates with moderately distributed coefficients.
The validity of the bound \rlb{c4B} is of course highly nontrivial, and depends on the nature of the system.
It is a challenging important problem to justify the bound in a concrete quantum many-body system.
For the moment, we are able to justify the bound only for rather simple toy models.  See sections~\ref{s:freefermions} and \ref{s:toytwobody}.

\begin{theorem}
\label{t:moderatecondition}
Suppose that the bound \rlb{c4B} is satisfied.
Choose $\alpha_i\in\mathbb{C}$ with $i=1,\ldots,\Dneq$ with $\sum_{i=1}^{\Dneq}|\alpha_i|^2=1$ in a random manner according to the uniform measure on the unit sphere in $\mathbb{C}^{\Dneq}$.
Then with probability larger than $1-2e^{-\epsilon V}$, the desired bound \rlb{caV} is valid for the initial state $\kpz$ of \rlb{p0neq}.
\end{theorem}

The theorem says that there are plenty of nonequilibrium initial states that satisfy the desired condition \rlb{caV}.
When the nonequilibrium basis states $\kxi$ with $i=1,\ldots,\Dneq$ are chosen maximally so that any nonequilibrium state is written as a linear combination of $\kxi$'s and a small correction, the theorem says that an overwhelming majority of nonequilibrium initial states satisfy \rlb{caV}.
We must stress again that the condition of the theorem is rather strong and remains to be justified for concrete systems.

\bigskip\noindent
{\em Proof\/:}\/
By substituting \rlb{xig} into \rlb{p0neq}, we have
\eq
\kpz=\sum_{j\in\JVu}\Bigl(\sum_{i=1}^{\Dneq}\alpha_i\,g_{i,j}\Bigr)\kpj,
\en
which, compared with \rlb{exp}, means $c_j=\sum_{i=1}^{\Dneq}\alpha_i\,g_{i,j}$.
Then we shall compute the random average of $\sum_{j\in\JVu}|c_j|^4$ by using the formula (see, e.g., \cite{Sugita06,Sugita07})
\eq
\overline{(\alpha_{i_1}\alpha_{i_2})^*\,\alpha_{i_3}\alpha_{i_4}}
=
\begin{cases}
\dfrac{2}{\Dneq(\Dneq+1)}&\text{if $i_1=i_2=i_3=i_4$}\\
\dfrac{1}{\Dneq(\Dneq+1)}&\text{if $i_1=i_3\ne i_2=i_4$ or $i_1=i_4\ne i_2=i_3$}\\
0&\text{otherwise},
\end{cases}
\en
to get
\eqa
\overline{\sum_{j\in\JVu}|c_j|^4}&=
\sum_{j\in\JVu}\overline{\biggl|\sum_{i=1}^{\Dneq}\alpha_i\,g_{i,j}\biggr|^4}
\nl
&=\sum_{j\in\JVu}\sum_{i=1}^{\Dneq}\overline{|\alpha_i|^4}\,|g_{i,j}|^4
+2\sum_{j\in\JVu}\mathop{\sum_{i,i'=1}^{\Dneq}}_{(i\ne i')}
\overline{|\alpha_i|^2\,|\alpha_{i'}|^2}\,\,|g_{i,j}|^2\,|g_{i',j}|^2
\nl
&\le\frac{1}{\Dneq+1}\max_{i\in\{1,\ldots,\Dneq\}}\sum_{j\in\JVu}|g_{i,j}|^4
+2\,\frac{\Dneq-1}{\Dneq+1}\,\max_{i,i'\in\{1,\ldots,\Dneq\}}
\sum_{j\in\JVu}|g_{i,j}|^2\,|g_{i',j}|^2.
\ena
Noting that $\sum_j|g_{i,j}|^2\,|g_{i',j}|^2\le\sqrt{(\sum_j|g_{i,j}|^4)(\sum_j|g_{i',j}|^4)}$, we finally get
\eq
\overline{\sum_{j\in\JVu}|c_j|^4}\le
2\,\max_{i\in\{1,\ldots,\Dneq\}}\sum_{j\in\JVu}|g_{i,j}|^4
\le2\frac{e^{(\eta-\epsilon)V}}{\DVu}=2\,e^{-\epsilon V}\,\frac{e^{\eta V}}{\DVu}.
\en
The claimed bound for the probability follows readily from the Markov inequality as in the proof of Theorem~\ref{t:thtyp}.\quad\qedm

\section{Thermalization: The energy eigenstate thermalization hypothesis}
\label{s:ETH}
Let us discuss the second strategy for the proof of the bound \rlb{taV}.
It is based on a plausible but nontrivial assumption that every energy eigenstate in the energy shell $\HVu$ represents thermal equilibrium.

Such an assumption, which is usually called the energy eigenstate thermalization hypothesis or the eigenstate thermalization hypothesis (ETH), was first introduced by von Neumann in 1929 \cite{vonNeumann,GLTZ} (see \cite{RigolSrednicki}), and discussed later in many works including \cite{Deutsch1991,Srednicki1994,Horoi1995,Zelevinsky1996,Tasaki1998}.
Since there are various notions of (thermal) equilibrium as we have discussed at the end of section~\ref{s:setup}, the precise meaning of the assumption depends on the context.

We fix the volume $V$.
We assume that the thermodynamic bound \rlb{tdb} is valid, and that there is no degeneracy in the energy eigenvalues, i.e., $E_j\ne E_{k}$ if $j\ne k$.
Again we can take into account some degeneracy as is explained below Theorem~\ref{t:th1}.

In the present context, the energy eigenstate thermalization hypothesis is an assumption that, for some $V\ge V_0$, there exists a constant $\kappa>0$, and we have
\eq
\sbkt{\psi_j|\Pneq|\psi_j}\le e^{-\kappa V}\ 
\text{for  any $j\in\JVu$.}
\lb{EET}
\en
This assumption is strongly motivated by the fact that an overwhelming majority of states $\kph$ in $\HVu$ satisfies the bound\footnote{
This statement is a trivial variation of Theorem~\ref{t:thtyp}.
When the thermodynamic bound \rlb{tdb} is valid with $\gamma>\kappa$, the bound $\sbkt{\varphi|\Pneq|\varphi}\le e^{-\kappa V}$ is valid with probability larger than $1-e^{-(\gamma-\kappa)V}$.
} $\sbkt{\varphi|\Pneq|\varphi}\le e^{-\kappa V}$.
It is then expected that the chance that $\DVu$ energy eigenstates fail to belong to this majority is very small.

The following theorem is essentially contained in von Neumann's seminal paper \cite{vonNeumann,GLTZ}.  We should emphasize that this is nothing more than a trivial lemma in this highly nontrivial work of von Neumann's.

\begin{theorem}
\label{t:th2}
Suppose that the bound \rlb{EET} is valid for some $\kappa$ such that $\kappa>\alpha+\nu$.
Then for any normalized initial state $\kpz\in\HVu$, we have the desired bound \rlb{taV}.  
Thus any state from $\HVu$ approaches thermal equilibrium (in the sense of Lemma~\ref{l:th}). 
\end{theorem}

\noindent{\em Proof:}\/
One gets \rlb{taV} simply by substituting \rlb{EET} into the right-hand of \rlb{thupper0}.~\qedm

\bigskip

Note that the theorem states the approach to thermal equilibrium from {\em any}\/ initial state.
This is in contrast with Theorem~\ref{t:th1} and many other works on thermalization or equilibration where the initial state has to satisfy nontrivial condition about the effective dimension.
When Theorem~\ref{t:th2} is applicable, we can be sure that any nonequilibrium initial state (from $\HVu$) does thermalize\footnote{
If this is the case, there is little (or no) question about our conclusion in Section~\ref{s:typphys} that typical properties correspond to thermal equilibrium.
}.

We should note, however, that the energy eigenstate thermalization hypothesis (ETH) is a nontrivial assumption about macroscopic quantum systems.
Although we can prove the hypothesis in some easily solvable examples discussed in Appendix~\ref{s:toy}, we do not know of any truly nontrivial examples in which the ETH has been established.

In most of the literature, the notion of ETH is associated with the characterization of thermal equilibrium in terms of expectation values as in \rlb{AsimA}.
Thus discussions about validity of ETH do not directly apply to our notion.

At the end of section~\ref{s:thermalization}, we have discussed some cases where thermalization does not take place.
In such a situation ETH is likely to be invalid.
For further discussions about the validity of ETH, see, e.g., \cite{Rigol2008,Santos2010,Polkovnikov2011,Beugeling2014}.

\section{Proof of the thermodynamic bound}
\label{s:proofTDB}

Here we shall prove the thermodynamic bound, which plays an essential role in the present work.
We start by describing the general strategy in section~\ref{s:Pgeneral}, and then treat specific models in the following three sections.

\subsection{General consideration}
\label{s:Pgeneral}
Let us first explain how we treat multiple quantities $\hM^{(1)}_V,\ldots,\hM^{(n)}_V$.
In the first treatment where $\Pneq$ is defined by \rlb{Pneqnew}, one simply proves the bound
\eq
\Bigl\langle\,\hP\Bigl[\bigl|(\MiV/V)-\miu\bigr|\ge\di\Bigr]\,\Bigr\rangle^\mathrm{mc}_{V,u}
\le \frac{e^{-\gamma V}}{n},
\lb{tdbmulti}
\en
for each $i=1,\ldots,n$.
By summing these up, one immediately gets the desired thermodynamic bound \rlb{tdb}.

In the second treatment where $\Pneq$ is defined by \rlb{Pneqn}, we can prove the thermodynamic bound by using the bound\footnote{
This simple strategy may not be the optimal way for the proof, especially when $n$ is large.
}
\eq
\Pneq\le\sum_{i=1}^n\hP\Bigl[\bigl|(\tMiV/V)-\miu\bigr|\ge\di\Bigr].
\lb{P<nP}
\en
Recalling \rlb{MtM}, we also see that
\eq
\hP\Bigl[\bigl|(\tMiV/V)-\miu\bigr|\ge\di\Bigr]
\le
\hP\Bigl[\bigl|(\MiV/V)-\miu\bigr|\ge\di_V\Bigr],
\lb{PtMPM}
\en
for some $\di_V$ such that $\di_V\uparrow\di$ as $V\uparrow\infty$.
Then if we prove the bound \rlb{tdbmulti}, with $\delta^{(i)}$ replaced by $\di_V$, for each $i=1,\ldots,n$, we have the desired thermodynamic bound \rlb{tdb}.

We shall thus set $n=1$ in the following.

To prove the thermodynamic bound, we make use of two standard strategies.
First, instead of dealing with the microcanonical average $\bktmc{\hP\bigl[|\BV|\ge V\delta\bigr]}$, which is in general not easy to control, we shall treat the corresponding canonical average $\sbkt{\hP\bigl[|\BV|\ge V\delta\bigr]}^\mathrm{can}_{V,\beta(u)}$ for a suitable inverse temperature $\beta(u)$.
This is justified by the standard theory of equivalence of ensembles \cite{Ruelle}.
Secondly, instead of treating the expectation value $\sbkt{\hP\bigl[|\BV|\ge V\delta\bigr]}^\mathrm{can}_{V,\beta}$ of the projection operator, we first bound the expectation value $\sbkt{e^{\lambda\BV}}^\mathrm{can}_{V,\beta}$ where $\lambda$ is a real variable, and use the Markov inequality.
This is a standard procedure in the large deviation theory \cite{Elis,DemboZeitouni}.

Define the operator
\eq
\BV:=\MV-m(u)\,V,
\lb{MV}
\en
which satisfies $\limV\bktmc{\BV}/V=0$.
We wish to bound the quantity 
\eq
\Bktmc{\hP\bigl[|\BV|\ge V\delta\bigr]}=\frac{1}{\DVu}\sum_{j\in\JVu}
\Bbkt{\psi_j\Bigl|\,\hP\bigl[|\BV|\ge V\delta\bigr]\,\Bigr|\psi_j}.
\lb{PBMC}
\en
Observe that, for any $\beta>0$, the canonical expectation of $\hP[|\BV|\ge V\delta]$ satisfies
\eqa
\Bbkt{\hP\bigl[|\BV|\ge V\delta\bigr]}^\mathrm{can}_{V,\beta}
&:=\sum_{j}\frac{e^{-\beta E_j}}{Z_V(\beta)}
\Bbkt{\psi_j\Bigl|\,\hP\bigl[|\BV|\ge V\delta\bigr]\,\Bigr|\psi_j}
\nl&\ge
\frac{e^{-\beta uV}}{Z_V(\beta)}\sum_{j\in\JVu}
\Bbkt{\psi_j\Bigl|\,\hP\bigl[|\BV|\ge V\delta\bigr]\,\Bigr|\psi_j}
\nl&=\frac{\DVu\,e^{-\beta uV}}{Z_V(\beta)}\,\Bktmc{\hP\bigl[|\BV|\ge V\delta\bigr]}.
\ena
Take the energy density $u$ which satisfies the condition \rlb{ucond}, i.e., $\beta(u)\ne\beta(u')$ for any $u'\ne u$.
Then it is standard in the theory of the equivalence of ensembles \cite{Ruelle}  that one has
\eq
\frac{Z_V(\beta(u))}{\DVu\,e^{-\beta(u)\, uV}}\le\eta(u)\,V,
\lb{ZDeV}
\en
where $\eta(u)$ is a positive constant.
See the end of the present section for a proof.
We thus get
\eq
\Bktmc{\hP\bigl[|\BV|\ge V\delta\bigr]}\le\eta(u)\,V\,\Bbkt{\hP\bigl[|\BV|\ge V\delta\bigr]}^\mathrm{can}_{V,\beta(u)}.
\lb{A}
\en
The next step is to bound the right-hand side in the spirit of the large deviation theory.

In some models (with suitable $\beta$), it can be shown, or has already be shown that 
\eq
\sbkt{e^{\lambda\BV}}^\mathrm{can}_{V,\beta}\le e^{V\,\phi_{V,\beta}(\lambda)},
\lb{elaB}
\en
for any $V$ and any $\lambda\in\bigl(\lambda_-(\beta),\lambda_+(\beta)\bigr)$ with $\lambda_-(\beta)<0<\lambda_+(\beta)$, where the limit
\eq
\phi_\beta(\lambda)=\lim_{V\up\infty}\phi_{V,\beta}(\lambda)
\lb{glim}
\en
exists and defines a convex differentiable function of $\lambda\in\bigl(\lambda_-(\beta),\lambda_+(\beta)\bigr)$ with $\phi_\beta(\lambda)\ge0$ and $\phi_\beta(0)=0$.
We allow the cases where $\lambda_-(\beta)=-\infty$ or $\lambda_+(\beta)=\infty$.
When \rlb{elaB} holds as an equality, $\phi_\beta(\lambda)$ is the moment generating function, which is a standard tool in the large deviation theory.

\begin{figure}[btp]
\begin{center}
\includegraphics[width=5cm]{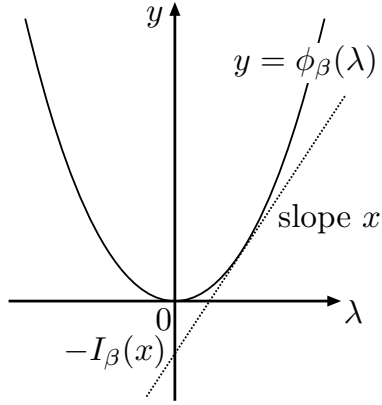}
\end{center}
\caption[dummy]{
The standard graphical interpretation of the definition \rlb{Ibx}.
It can be easily found that $-I_\beta(x)$ is the $y$-intercept of the tangent line with slope $x$ to the graph of $y=\phi_\beta(\lambda)$.
From the figure it should be obvious that $I(x)>0$ for $x\ne0$ when $\phi_\beta(\lambda)$ is differentiable in $\lambda$ in an open interval containing $\lambda=0$.
}
\label{fig:Ix}
\end{figure}

Let us also define, for $x\in\bbR$, the corresponding rate function\footnote{
This is the proper rate function of the large deviation theory when \rlb{elaB} holds as an equality.
} $I_\beta(x)$ by the Legendre transformation
\eq
I_\beta(x):=\sup_{\lambda\in(\lambda_-(\beta),\lambda_+(\beta))}\{\lambda x-\phi_\beta(\lambda)\}
=
\begin{cases}
\displaystyle\sup_{\lambda\in[0,\lambda_+(\beta))}\{\lambda x-\phi_\beta(\lambda)\}&\text{if $x\ge0$}\\
\displaystyle\sup_{\lambda\in(\lambda_-(\beta),0]}\{\lambda x-\phi_\beta(\lambda)\}&\text{if $x\le0$},\\
\end{cases}
\lb{Ibx}
\en
where the final expression follows from the assumption that the convex function $\phi_\beta(\lambda)$ attains its minimum at $\lambda=0$.
We easily see $I_\beta(x)\ge0$ since $\phi_\beta(0)=0$.
By using the fact that $\phi_\beta(\lambda)$ is differentiable in $\lambda$ (in an open interval containing $\lambda=0$), it follows from the standard results in Legendre transformation that  $I_\beta(x)>0$ for any $x\ne0$.
See Figure~\ref{fig:Ix}.

Let $\delta>0$.
For any $\lambda\in\bigl[0,\lambda_+(\beta)\bigr)$, we find 
\eq
\bbkt{\hP[\BV\ge V\delta]\,}^\mathrm{can}_{V,\beta}\le
\bbkt{e^{\lambda(\BV-V\delta)}}^\mathrm{can}_{V,\beta}\le
e^{V\{\phi_{V,\beta}(\lambda)-\lambda\delta\}},
\lb{B00}
\en
where we noted that $\hP[\BV\ge V\delta]\le e^{\lambda(\BV-V\delta)}$, and used the bound \rlb{elaB}.
By noting \rlb{Ibx}, we see that the bound can be optimized to give
\eq
\bbkt{\hP[\BV\ge V\delta]\,}^\mathrm{can}_{V,\beta}\le e^{-I_\beta(\delta)\,V+o(V)}.
\lb{B}
\en
Similarly we have, again for $\delta>0$, that
\eq
\bbkt{\hP[\BV\le -V\delta]\,}^\mathrm{can}_{V,\beta}\le
e^{-I_\beta(-\delta)\,V+o(V)}.
\lb{C}
\en

We now come back to our purpose, which is to bound the quantity \rlb{PBMC}.
Take $u$ satisfying \rlb{ucond}, and consider the corresponding rate function $I_{\beta(u)}(x)$.
Combining \rlb{A}, \rlb{B}, and \rlb{C}, we find
\eq
\Bktmc{\hP\bigl[|\BV|\ge V\delta\bigr]\,}\le e^{-\min\{I_{\beta(u)}(\delta),I_{\beta(u)}(-\delta)\}\,V+o(V)}.
\en
By taking sufficiently large  $V_0>0$ and $\gamma$ such that $0<\gamma<\min\{I_{\beta(u)}(\delta),I_{\beta(u)}(-\delta)\}$, we finally get
\eq
\Bktmc{\hP\bigl[|\BV|\ge V\delta\bigr]\,}\le e^{-\gamma V},
\en
for any $V\ge V_0$.
This is nothing but the desired thermodynamic bound \rlb{tdb}.
Note that $\gamma$ can be made as close to $\min\{I_{\beta(u)}(\delta),I_{\beta(u)}(-\delta)\}>0$ as one wishes by letting $V_0$ large.

The remaining task is to prove the bound \rlb{elaB} with $\phi_\beta(\lambda)$ that has the desired properties.
We shall discuss this for each model in the following sections.

\bigskip\noindent
{\em Proof of \rlb{ZDeV}:}\/
Although the relation is standard (see \cite{Ruelle}) we give a proof for completeness.
Let us assume that for any $\tilde{\beta}>0$ there is $\tilde{\sigma}>0$ such that the number of states satisfies
\eq
\Omega_V(U)\le e^{\tilde{\sigma}V+\tilde{\beta}U}
\lb{Osb}
\en
for any $V$ and $U$.
For simplicity we shall redefine $\Ham$ so that the ground state energy is zero.

Fix an arbitrary $\beta>0$, and let $\delta\ge 2/\beta$ be a constant independent of $V$.
We rewrite and bound the partition function as
\eq
Z_V(\beta)=\sum_j e^{-\beta E_j}
=\sum_{n=0}^\infty\sumtwo{j}{((n-1)\delta<E_l\le n\delta)}e^{-\beta E_j}
\le\sum_{n=0}^\infty\tiD_V(n\delta)\,e^{-\beta(n-1)\delta},
\lb{ZVYY1}
\en
where $\tiD_V(n\delta):=\Omega_V(n\delta)-\Omega_V((n-1)\delta)$.
We write the right-hand side of \rlb{ZVYY1} as $\Ym+\Yr$ with
\eq
\Ym:=\sum_{n\le aV}\tiD_V(n\delta)\,e^{-\beta(n-1)\delta},\quad
\Yr:=\sum_{n> aV}\tiD_V(n\delta)\,e^{-\beta(n-1)\delta},
\en
where $a>0$ is a constant which will be determined later.

We bound $\Ym$ from above simply by the product of the number of the summands and the maximum value as
\eq
\Ym=e^{\beta\delta}\sum_{n\le aV}\tiD_V(n\delta)\,e^{-\beta n \delta}\le 
e^{\beta\delta} aV\max_{n}\tiD_V(n\delta)\,e^{-\beta n \delta}
\le e^{\beta\delta} aV \max_{\tilde{u}} D_{V,\tilde{u}}\,e^{-\beta\tilde{u}V},
\en
where we noted that $\tiD_V(n\delta)\le D_{V,\,n\delta/V}=\Omega_V(n\delta)-\Omega_V(n\delta-V\Du)$.

To bound $\Yr$, we use \rlb{Osb} with $\tilde{\beta}=\beta/2$ as
\eq
\Yr\le\sum_{n>aV}\Omega_V(n\delta)\,e^{-\beta(n-1)\delta}
\le e^{\tilde{\sigma}V+\beta\delta}\sum_{n>aV}e^{-\beta n\delta/2}
\le 2\,e^{\tilde{\sigma}V+\beta\delta-\beta aV\delta/2},
\en
where we used $\beta\delta\ge2$ to bound the sum.
By choosing $a$ properly we see that the right-hand side does not exceed 1 for sufficiently large $V$.
Since $\Ym\ge1$ (because the ground state energy is zero), we see that $\Yr\le\Ym$, and hence
\eq
Z_V(\beta)\le2\Ym\le2e^{\beta\delta} aV \max_{\tilde{u}} D_{V,\tilde{u}}\,e^{-\beta\tilde{u}V}.
\lb{ZVDe}
\en

Let $u$ be such that \rlb{ucond} holds, and let $\beta=\beta(u)$.
Noting that $D_{V,\tilde{u}}\sim e^{V\sigma(\tilde{u})}$ when $\sigma(\tilde{u})$ is strictly increasing, one finds that the maximum in \rlb{ZVDe} is attained at $\tilde{u}=u$.~\qedm

\paragraph{On the formulation of Goldstein, Lebowitz, Mastrodonato, Tumulka, and Zangh\`\i:}
By using the same method as in the present section, we can prove bounds which are crucial for the formulation of Goldstein, Lebowitz, Mastrodonato, Tumulka, and Zangh\`\i\ \cite{GLMTZ09b} discussed in section~\ref{s:comparison}.
For the same class of models as we treat here, we can show for sufficiently large $V$ that
\eq
\frac{\operatorname{dim}[\calH_{\rm neq}]}{\operatorname{dim}[\HVu]}\le e^{-\gamma V},
\lb{dd<<1}
\en
which corresponds to our thermodynamic bound \rlb{tdb}, and implies the desired inequality \rlb{d>>d}.

To be specific, let $\Ham$ be the Hamiltonian, and suppose that we are interested in a single quantity $\MV$.
Multiple quantities can be treated in the same manner as we explained in the beginning of the present section.
From $\Ham$ and $\MV$, we construct commuting operators $\hat{\hH}_V$ and $\hat{\hM}_V$ which approximate them.
Let $\HVu$ be the energy shell defined in terms of $\hat{\hH}_V$.
Then the ratio of the dimensions can be written as
\eq
\frac{\operatorname{dim}[\calH_{\rm neq}]}{\operatorname{dim}[\HVu]}=
\bbkt{\hP(\calH_{\rm neq})}_{V,u}^{\mathrm{mc},\hat{\hH}_V},
\en
where the right-hand side is the expectation value in the microcanonical distribution defined with respect to $\hat{\hH}_V$.
The projection operator is
$\hP(\calH_{\rm neq})=\hP\bigl[|\hBV|\ge V\delta\bigr]$ with $\hBV=\hat{\hM}_V-m(u)V$.

Suppose that the original and the approximate Hamiltonians satisfy 
\eq
\Vert{\Ham-\hat{\hH}_V}\Vert\le h_V,
\lb{HH}
\en
where $h_V=o(V)$.
From the minimax principle, we see that the number of states $\hat{\Omega}_V(U)$ corresponding to $\hat{\hH}_V$ satisfies
\eq
\Omega_V(U-h_V)\le\hat{\Omega}_V(U)\le\Omega_V(U+h_V).
\en
This means that one has $\log\hat{\Omega}_V(U)=V\,\sigma(U/V)+o(V)$ with the same $\sigma(u)$ as in \rlb{logO}.
This in particular means that the bound \rlb{ZDeV}, which is based on the equivalence of ensembles, is valid for the corresponding quantities for $\hat{\hH}_V$.
Thus, as in \rlb{A}, we have
\eqa
\bbkt{\hP\bigl[|\hBV|\ge V\delta\bigr]}_{V,u}^{\mathrm{mc},\hat{\hH}_V}
&\le
\eta(u)V\bbkt{\hP\bigl[|\hBV|\ge V\delta\bigr]}_{V,\beta}^{\mathrm{can},\hat{\hH}_V},
\intertext{where the right-hand side is the expectation value in the canonical distribution for $\hat{\hH}_V$.
This can be bonded by the expectation value in the canonical distribution for the original Hamiltonian as
}
&\le
\eta(u)V\,e^{2\beta h_V}\bbkt{\hP\bigl[|\hBV|\ge V\delta\bigr]}_{V,\beta}^{\mathrm{can}}\ ,
\intertext{where we used \rlb{HH}.
Finally, by noting that $\Vert\MV-\hat{\hM}_V\Vert=o(V)$, we have}
&\le
\eta(u)V\,e^{2\beta h_V}\bbkt{\hP\bigl[|\BV|\ge V\delta_V\bigr]}_{V,\beta}^{\mathrm{can}}\ ,
\ena
where $\delta_V$ approaches $\delta$ as $V\uparrow\infty$.
This bound corresponds to our \rlb{A}.
The rest of the proof is exactly the same.

In the model of two bodies in contact (see sections~\ref{s:EHC} and \ref{s:PHC}), one can define the modified Hamiltonian as $\hat{\hH}_V=\mathsf{H}^{(1)}_{V/2}\otimes\mathsf{1}+\mathsf{1}\otimes\mathsf{H}^{(2)}_{V/2}$, which clearly commutes with $\MV$ of \rlb{MVHC}.
Then the problem becomes that of two decoupled systems, and the proof of \rlb{dd<<1} becomes quite elementary (the argument in the ``heuristic derivation'' in section~\ref{s:EHC} is essentially a proof).

\subsection{Heat conduction between two identical bodies}
\label{s:PHC}
Let us prove Proposition~\ref{p:TDBHC}.
The problem becomes almost trivial in the canonical formulation\footnote{
The problem of two subsystems exchanging particles as well as energy can be treated in almost the same manner.
}.
Note that $\MV=\BV$ since the average of $\MV$ is vanishing.

Since the interaction Hamiltonian $\mathsf{H}_\mathrm{int}$ satisfies $\Vert\mathsf{H}_\mathrm{int}\Vert\le h_0V^\zeta$ (where $\zeta:=(d-1)/d<1$), we can bound the total Hamiltonian $\Ham$ of \rlb{HamHC} as
\eq
\mathsf{H}^{(1)}_{V/2}\otimes\mathsf{1}+\mathsf{1}\otimes\mathsf{H}^{(2)}_{V/2}-h_0V^\zeta\le
\Ham\le
\mathsf{H}^{(1)}_{V/2}\otimes\mathsf{1}+\mathsf{1}\otimes\mathsf{H}^{(2)}_{V/2}+h_0V^\zeta.
\lb{HHH}
\en

Let $\sbkt{\cdots}^{\text{can}}_{V,\beta}$ be the canonical expectation of the whole system with the Hamiltonian $\Ham$.
Then we find
\eqa
\sbkt{e^{\lambda\MV}}^{\text{can}}_{V,\beta}
=\frac{\mathrm{Tr}[e^{\lambda\mathsf{H}^{(1)}_{V/2}}\,e^{-\lambda\mathsf{H}^{(2)}_{V/2}}\,e^{-\beta\Ham}]}{\mathrm{Tr}[e^{-\beta\Ham}]}
\le
\frac{e^{\beta h_0V^\zeta}\,\mathrm{Tr}[e^{\lambda\mathsf{H}^{(1)}_{V/2}}\,e^{-\lambda\mathsf{H}^{(2)}_{V/2}}\,
e^{-\beta\{\mathsf{H}^{(1)}_{V/2}+\mathsf{H}^{(2)}_{V/2}\}}]}
{e^{-\beta h_0V^\zeta}\,\mathrm{Tr}[e^{-\beta\{\mathsf{H}^{(1)}_{V/2}+\mathsf{H}^{(2)}_{V/2}\}}]},
\lb{elaMHC00}
\ena
where we used the first inequality in \rlb{HHH} to the numerator and the second to the denominator.
Denoting the trace in the subspace $\calH^{(j)}_{V/2}$ as $\mathrm{Tr}_j[\cdots]$ (where $j=1,2$), we have
\eq
\sbkt{e^{\lambda\MV}}^{\text{can}}_{V,\beta}
\le e^{2\beta h_0V^\zeta}\frac{\mathrm{Tr}_1[e^{-(\beta-\lambda)\mathsf{H}^{(1)}_{V/2}}]\,\mathrm{Tr}_2[e^{-(\beta+\lambda)\mathsf{H}^{(2)}_{V/2}}]}{\mathrm{Tr}_1[e^{-\beta\mathsf{H}^{(1)}_{V/2}}]\,\mathrm{Tr}_2[e^{-\beta\mathsf{H}^{(2)}_{V/2}}]}.
\lb{elaMHC1}
\en
This is indeed the desired bound \rlb{elaB} in the present case.

For any $\tilde{\beta}>0$, we define the free energy in the infinite volume limit by
\eq
f(\tilde{\beta}):=-\lim_{V\up\infty}\frac{2}{V\tilde{\beta}}\log
\mathrm{Tr}_j[e^{-\beta\mathsf{H}^{(j)}_{V/2}}],
\en
where the result is independent of $j=1,2$.
Then, for any $\lambda\in(-\beta,\beta)$, we can rewrite the bound \rlb{elaMHC1} as
\eq
\sbkt{e^{\lambda\MV}}^{\text{can}}_{V,\beta}
\le 
e^{V\phi_\beta(\lambda)+o(V)},
\lb{elaMHC2}
\en
where 
\eq
\phi_\beta(\lambda)=\beta\,f(\beta)-\frac{(\beta-\lambda)\,f(\beta-\lambda)}{2}
-\frac{(\beta+\lambda)\,f(\beta+\lambda)}{2},
\en
which is convex in $\lambda$ since $\tilde{\beta} f(\tilde{\beta})$ is concave in $\tilde{\beta}$.
It clearly satisfies $\phi_\beta(0)=0$.
From the convexity and the symmetry $\phi_\beta(\lambda)=\phi_\beta(-\lambda)$, we also find  $\phi_\beta(\lambda)\ge0$ 
As for the differentiability, we note that the relation
\eq
u=\frac{d}{d\tilde{\beta}}\{\tilde{\beta}f(\tilde{\beta})\}\Bigr|_{\tilde{\beta}=\beta(u)}
\en
implies that the condition assumed for $u$ is equivalent to the differentiability of $\tilde{\beta}f(\tilde{\beta})$ at $\tilde{\beta}=\beta(u)$.
This means that $\phi_\beta(\lambda)$ is differentiable in $\lambda$ in an interval containing $\lambda=0$.

To show \rlb{gammaHC}, we observe that
\eq
\phi_\beta(\lambda)=-\frac{(\beta\,f(\beta))''}{2}\lambda^2+O(\lambda^3)
=-\frac{u'(\beta)}{2}\lambda^2+O(\lambda^3),
\en
which, with \rlb{Ibx}, implies
\eq
I_\beta(\delta)\simeq I_\beta(-\delta)\simeq\sup_\lambda\{\lambda\delta+\frac{u'(\beta)}{2}\lambda^2\}
=-\frac{\delta^2}{2u'(\beta)}=-\frac{\beta'(u)}{2}\delta^2.
\en

We note in passing that the same argument proves
\eq
\phi_\beta(\lambda)=\limV\frac{1}{V}\log\sbkt{e^{\lambda\MV}}^{\text{can}}_{V,\beta},
\en
which means that $\phi_\beta(\lambda)$ is the proper moment generating function (for the canonical distribution).

\subsection{Quantum spin systems}
\label{s:PQL}
Proposition~\ref{p:TDBQS}, i.e., the thermodynamic bound for quantum spin systems, is an easy corollary of the large deviation principle established in \cite{NetocnyRedig,LenciRey-Bellet,HiaiMosonyiOgawa,Ogata2010,OgataRey-Bellet}.

Assume that the conditions for  Proposition~\ref{p:TDBQS} are satisfied.
Then it was shown for general quantum spin chains\footnote{
The quantity $\phi_\beta(\lambda)=\lim_{V\up\infty}V^{-1}\log \omega_\beta(e^{\lambda\BV})$, where $\omega_\beta(\cdot)$ denotes the equilibrium state (or, more precisely, the KMS state) for the infinite lattice, is treated in \cite{NetocnyRedig,Ogata2010}.
But by using the property called asymptotically decoupledness \cite{OgataRey-Bellet}, which is satisfied in the present models, it can be shown that this defines the same quantity as \rlb{MGF}.
See \cite{LenciRey-Bellet,HiaiMosonyiOgawa,OgataRey-Bellet}.
} by Ogata \cite{Ogata2010} and for higher dimensional systems\footnote{
In \cite{NetocnyRedig,LenciRey-Bellet}, the large deviation principle (for small enough $\beta(u)$) was proved when the operator $\hm_x$ (see \rlb{Mbx}) acts only on a single site $x$.
But by combining the derivations in \cite{LenciRey-Bellet} with the cluster expansion technique developed in section 3.1 of \cite{FroelichUeltschi}, one can prove the desired results for a general local operator $\hm_x$
(Rey-Bellet, private communication).
} with small enough $\beta(u)$ by Netocny and Redig \cite{NetocnyRedig} and by Lenci and Rey-Bellet \cite{LenciRey-Bellet} that the limit
\eq
\phi_\beta(\lambda)=\lim_{V\up\infty}\frac{1}{V}\log\bbkt{e^{\lambda\BV}}^\mathrm{can}_{V,\beta}
\lb{MGF}
\en
exists and is convex and analytic in $\lambda$ in an open interval containing $\lambda=0$.
Clearly this can be identified with our $\phi_\beta(\lambda)$ defined in \rlb{glim}.

Since \rlb{MGF} implies that $\phi_\beta(0)=0$, we only need to show that $\phi_\beta(\lambda)\ge0$, which is easy.
Note that the existence of the limit \rlb{MGF} implies the large deviation upper bound
\eq
\bktmc{\hP[\BV/V\simeq x]}\le e^{o(V)}\,\sbkt{\hP[\BV/V\simeq x]}^\mathrm{can}_{V,\beta(u)}
\le e^{-I_{\beta(u)}(x)V+o(V)},
\lb{LDupper}
\en
where the first equality is \rlb{A}.
Since we have $\bktmc{\BV}=0$, \rlb{LDupper} is possible only when $I_{\beta(u)}(0)=0$ which means $\phi_\beta(\lambda)\ge0$.

Finally, we note that the condition \rlb{ucond} for the energy density $u$ is automatically satisfied if $u(\beta)$, i.e., the equilibrium energy density (in the infinite volume limit) as a function of the inverse temperature $\beta$, is continuous.  See Fig.~\ref{fig:ubeta}.
The continuity of $u(\beta)$ is guaranteed for quantum spin chains by the general result found, e.g., in \cite{Araki}, and for higher dimensional systems with sufficiently small $\beta$ by the standard results.  See, e.g., \cite{FroelichUeltschi}.

\subsection{Ising model under transverse magnetic field}
\label{s:PIsing}
We finally prove Proposition~\ref{p:TDBIsing}.
The main ingredient of the proof is the correlation inequality
\eq
\bbkt{e^{\lambda\MV}}_{V,\beta}^\mathrm{can}
\le
\exp\Bigl[
V\lambda^2\sum_{y\in\Lambda}\sbkt{\mathsf{S}^{(3)}_x\mathsf{S}^{(3)}_y}^\mathrm{can}_{V,\beta}
\Bigr],
\lb{IsingMain}
\en
which is valid for any $\lambda\in\bbR$, $\beta>0$, and $V$.
When $\tilde{\chi}(\beta)<\infty$, the bound \rlb{IsingMain} is nothing but the desired bound \rlb{elaB} with $\phi_\beta(\lambda)=\lambda^2\,\tilde{\chi}(\beta)$.
This proves Proposition~\ref{p:TDBIsing}.

It remains to show the inequality \rlb{IsingMain}.
Let us be brief since the proof is a combination of standard techniques in rigorous statistical mechanics.

We first rewrite the quantum spin system in $d$ dimension as a classical spin system in $d+1$ dimension, as has been done in many works starting from \cite{Robinson,Ginibre}.
Note that, by using the Lie product formula, the partition function of the quantum model can be written as 
\eq
\mathrm{Tr}[e^{-\beta\Ham}]=\lim_{N\up\infty}Z^{(N)}_V(\beta),
\lb{Lie}
\en
with
\eq
Z^{(N)}_V(\beta)=
\mathrm{Tr}\biggl[
\underbrace{\exp\bigl[-\frac{\beta}{N}\Ham^\mathrm{cl}\bigr]\Bigl\{\prod_x\bigl(1-\frac{\beta h}{N}\mathsf{S}^{(1)}_x\bigr)\Bigr\}
\cdots
\exp\bigl[-\frac{\beta}{N}\Ham^\mathrm{cl}\bigr]\Bigl\{\prod_x\bigl(1-\frac{\beta h}{N}\mathsf{S}^{(1)}_x\bigr)\Bigr\}}_{N}\biggr],
\lb{ZN}
\en
where $\Ham^\mathrm{cl}=-\sum_{x,y\,\,(x>y)}J_{x,y}\,\mathsf{S}^{(3)}_x\mathsf{S}^{(3)}_y$ is the Hamiltonian of the classical Ising model corresponding to \rlb{HamIsingTF}.

Let $\Phi_{\bssigma}$ be the basis state defined in \rlb{Phisigma}.
By inserting $N$ copies of $1=\sum_{\bssigma}\ket{\Phi_{\bssigma}}\bra{\Phi_{\bssigma}}$ into \rlb{ZN}, we see that
\eq
Z^{(N)}_V(\beta)=
\sum_{\bssigma^{(1)},\ldots,\bssigma^{(N)}}
\prod_{n=1}^N\biggl\langle\Phi_{\bssigma^{(n)}}\biggr|\exp\bigl[-\frac{\beta}{N}\Ham^\mathrm{cl}\bigr]\Bigl\{\prod_x\bigl(1-\frac{\beta h}{N}\mathsf{S}^{(1)}_x\bigr)\Bigr\}\biggl|\Phi_{\bssigma^{(n+1)}}\biggr\rangle,
\en
where each $\bssigma^{(n)}$ is summed over all the spin configurations on $\Lambda$.
We set $\bssigma^{(N+1)}=\bssigma^{(1)}$.
Note that $(\bssigma^{(1)},\bssigma^{(2)},\ldots,\bssigma^{(N)})$ may be identified with a spin configuration of the Ising model on the $d+1$ dimensional lattice $\tilde{\Lambda}=\Lambda\times\{1,2,\ldots,N\}$.
We also note that, for $\sigma,\sigma'\in\{+,-\}$
\eq
\bra{\varphi_x^\sigma}\bigl(1-\frac{\beta h}{N}\mathsf{S}^{(1)}_x\bigr)
\ket{\varphi_x^{\sigma'}}
=\begin{cases}
1&\sigma=\sigma'\\
1-\beta h/(2N)&\sigma\ne\sigma',\\
\end{cases}
\en
where the right-hand side can be compactly written as $\exp[\beta J'(\sigma\sigma'-1)]$ with
\eq
J'=-\frac{1}{2\beta}\log\Bigl(1-\frac{\beta h}{2N}\Bigr)=\frac{h}{4N}+O\Bigl(\frac{1}{N^2}\Bigr)\ge0.
\en
This implies an exact equality
\eq
Z^{(N)}_V(\beta)=e^{-\beta J'NV}\,\tilde{Z}^\mathrm{cl}_{\tilde{\Lambda}}(\beta),
\en
where $\tilde{Z}^\mathrm{cl}_{\tilde{\Lambda}}(\beta)$ is the partition function of the classical Ising model on $\tilde{\Lambda}=\Lambda\times\{1,2,\ldots,N\}$ with the ferromagnetic Hamiltonian
\eq
\tilde{H}^\mathrm{cl}_{\tilde{\Lambda}}=-\sumtwo{x,y\in\Lambda}{(x>y)}\sum_{n=1}^N\frac{J_{x,y}}{N}\sigma_{(x,n)}\sigma_{(y,n)}
-J'\sum_{x\in\Lambda}\sum_{n=1}^N\sigma_{(x,n)}\sigma_{(x,n+1)},
\lb{HNcl}
\en
where we denoted sites in $\tilde{\Lambda}$ as $(x,n)$.

By repeating the same procedure for $\mathrm{Tr}[\mathsf{S}^{(3)}_{x_1}\cdots\mathsf{S}^{(3)}_{x_n}\,e^{-\beta\Ham}]$ for arbitrary $x_1,\ldots,x_n\in\Lambda$, one can show that
\eq
\bbkt{\mathsf{S}^{(3)}_{x_1}\mathsf{S}^{(3)}_{x_2}\cdots\mathsf{S}^{(3)}_{x_n}}^\mathrm{can}_{V,\beta}
=2^{-n}\lim_{N\up\infty}\sbkt{\sigma_{(x_1,1)}\,\sigma_{(x_2,1)}\cdots\sigma_{(x_n,1)}}^\mathrm{can, cl}_{\tilde{\Lambda},\beta}
\lb{QC}
\en
where $\sbkt{\cdots}^\mathrm{can, cl}_{\tilde{\Lambda},\beta}$ denotes the canonical correlation in the (classical) Ising model on $\tilde{\Lambda}$ with the Hamiltonian \rlb{HNcl}.

Since \rlb{HNcl} is a ferromagnetic Hamiltonian without magnetic field, Newman's Gaussian inequality \cite{Newman75} (see also \cite{Lebowitz}) states that
\eq
\bbkt{\sigma_{\tilde{x}_1}\sigma_{\tilde{x}_2}\cdots\sigma_{\tilde{x}_n}}^\mathrm{can, cl}_{\tilde{\Lambda},\beta}
\le
\bbkt{Z_{\tilde{x}_1}Z_{\tilde{x}_2}\cdots Z_{\tilde{x}_n}}^\mathrm{Gauss}_{\tilde{\Lambda}},
\lb{Chuck}
\en
for any even $n$ and any $\tilde{x}_1,\tilde{x}_2,\ldots,\tilde{x}_n\in\tilde{\Lambda}$, where $Z_{\tilde{x}}$ (with $\tilde{x}\in\tilde{\Lambda}$) are jointly Gaussian mean zero random variables with $\sbkt{Z_{\tilde{x}}Z_{\tilde{y}}}^\mathrm{Gauss}_{\tilde{\Lambda}}=\bbkt{\sigma_{\tilde{x}}\sigma_{\tilde{y}}}^\mathrm{can, cl}_{\tilde{\Lambda},\beta}$ for any $\tilde{x}, \tilde{y}\in\tilde{\Lambda}$.

Then for any $\lambda\ge0$, we get
\eqa
\Bbkt{\exp\Bigl[\lambda\sum_{x\in\Lambda}\frac{\sigma_{(x,1)}}{2}\Bigr]}^\mathrm{can, cl}_{\tilde{\Lambda},\beta}
&\le
\Bbkt{\exp\Bigl[\lambda\sum_{x\in\Lambda}\frac{Z_{(x,1)}}{2}\Bigr]}^\mathrm{Gauss}_{\tilde{\Lambda}}
\nl&
=\exp\Biggl[\lambda^2\sum_{x,y\in\Lambda}\frac{\bkt{Z_{(x,1)}Z_{(y,1)}}^\mathrm{Gauss}_{\tilde{\Lambda}}}{4}\Biggr]
\nl&
=\exp\Biggl[\lambda^2\sum_{x,y\in\Lambda}\frac{\bkt{\sigma_{(x,1)}\sigma_{(y,1)}}^\mathrm{can, cl}_{\tilde{\Lambda},\beta}}{4}\Biggr],
\lb{expsigZ}
\ena
where the inequality follows by expanding the exponential and using \rlb{Chuck}, the first equality follows from the standard property of Gaussian random variables, and the second equality from the definition of $Z_{\tilde{x}}$.
By using \rlb{QC}, the $N\up\infty$ limit of \rlb{expsigZ} yields
\eq
\Bbkt{\exp\Bigl[\lambda\sum_{x\in\Lambda}\mathsf{S}^{(3)}_x\Bigr]}^\mathrm{can}_{V,\beta}
\le
\exp\Bigl[\lambda^2\sum_{x,y\in\Lambda}\bbkt{\mathsf{S}^{(3)}_x\mathsf{S}^{(3)}_y}^\mathrm{can}_{V,\beta}\Bigr].
\en
Although the bound has been derived assuming $\lambda\ge0$, it is also valid for $\lambda\le0$ since the model is invariant under the global rotation $(\mathsf{S}^{(1)}_x,\mathsf{S}^{(2)}_x,\mathsf{S}^{(3)}_x)\to(\mathsf{S}^{(1)}_x,-\mathsf{S}^{(2)}_x,-\mathsf{S}^{(3)}_x)$ for all $x\in\Lambda$.
Since the translation invariance implies $\sum_{x,y\in\Lambda}\bbkt{\mathsf{S}^{(3)}_x\mathsf{S}^{(3)}_y}^\mathrm{can}_{V,\beta}=V\sum_{x\in\Lambda}\bbkt{\mathsf{S}^{(3)}_o\mathsf{S}^{(3)}_x}^\mathrm{can}_{V,\beta}$, we get the desired \rlb{IsingMain}.

\section{Discussion}
\label{s:discussion}

\subsection{Summary of the paper}
In the present paper, we discussed the foundation of equilibrium statistical mechanics based on quantum mechanics.
We focused on macroscopic isolated quantum systems, and presented a justification of the description of thermal equilibrium in terms of the microcanonical ensemble.

The starting point of our theory is Definition~\ref{d:eq}, which makes clear what we mean by a quantum mechanical pure state representing thermal equilibrium.
Our definition is based on the operational point of view, and is suitable for reproducing thermodynamics.
We then formulated the thermodynamic bound (Definition~\ref{d:tdb}), which ensure that the system, along with the choice of thermodynamic quantities, behaves as a normal thermodynamic system.
In section~\ref{s:examples}, we discussed some important examples where the thermodynamic bound can be proved.

Then, following the standard logic, we have stated in Theorem~\ref{t:thtyp} the typicality of thermal equilibrium, i.e., we have shown that an overwhelming majority of states in the energy shell represent thermal equilibrium.
This is the most important observation in the present paper.
We believe that it provides a strong support to the use of the microcanonical ensemble.

Our results on thermalization (or the approach to thermal equilibrium) is less satisfactory.
By assuming one of the two assumptions, i.e., mild energy distribution in the initial state or the energy eigenstate thermalization, we proved that the time-dependent state $\kpt$ represents thermal equilibrium for sufficiently long and most $t$ (Lemma~\ref{l:th} and Theorems~\ref{t:th1} and \ref{t:th2}).
We stress that the reversible unitary time evolution in an isolated quantum system can describe thermalization, which may appear to be irreversible.
The assumptions in Theorems~\ref{t:th1} and \ref{t:th2} can be, for the moment, stated only for trivial models (see Appendix~\ref{s:toy}), but expected to be valid in a large class of macroscopic quantum systems.
To verify these assumptions in nontrivial quantum many-body systems is a very important open problem.

\subsection{Comparison with classical systems}
\label{s:typclassical}
It may be useful to compare classical and quantum systems in connection with the results in the  present paper.
Let us take the setting described in the beginning of section~\ref{s:what}.

The energy shell $\calS_U$ in a classical system is almost uniquely defined as \rlb{SU}, which is a subspace of the classical phase space.
There are two natural and essentially different quantum counterparts of the energy shell; one is our energy shell $\HVu$ (or $\tHVu$), and the other is the set of energy eigenstates $\{\,\kpj\,|\,j\in\JVu\}$.
It is essential for us to use the former definition in the discussion of typicality.

A physical quantity of a classical system is merely a function $M_V(\cdot)$ on the phase space.
Thus if we take a state $\Gamma$, the only quantity to look at is its value $M_V(\Gamma)$.
In a quantum system, on the other hand, a physical quantity $\MV$ and a pure state $\kph$ determines a probability distribution for the measurement result.
This is why we had several different notions of a pure state representing thermal equilibrium.
In classical case, there is no such variety, and we can say that a state $\Gamma\in\calS_U$ represents thermal equilibrium when
\eq
\Bigl|(M^{(i)}_V(\Gamma)/V)-m^{(i)}(u)\Bigr|\le\delta^{(i)}
\en
for all $i=1,\ldots,n$.  We here followed the notation in section~\ref{s:pure}.

The thermodynamic bound for classical systems is a large deviation upper bound stated for the microcanonical distribution.
The bound for an extensive quantity $M_V(\Gamma)$ can be proved exactly as in section~\ref{s:Pgeneral} if the free energy\footnote{
The function $\phi_\beta(\lambda)$ of \rlb{glim} is obtained by
$\phi_\beta(\lambda)=\beta\{f(\beta,0)-f(\beta,\lambda/\beta)\}-\lambda m$, where $m$ is the equilibrium value of $M_V/V$.
}
\eq
f(\beta,h):=-\lim_{V\uparrow\infty}\frac{1}{\beta V}\log\int d\Gamma\,e^{-\beta\{H_V(\Gamma)-h M_V(\Gamma)\}}
\en
is differentiable in $h$ in an open interval containing $h=0$.

Given the above definition of a state $\Gamma$ representing thermal equilibrium, and the thermodynamic bound for the relevant quantities, the typicality of thermal equilibrium in $\calS_U$ can be proved.
As far as the typicality is concerned, classical systems are easier to treat, and are conceptually simpler.

As for the approach to thermal equilibrium, it seems that the situation in classical systems is essentially different from the quantum case.
Although one can prove certain results when the initial state is described by a probability distribution on $\calS_U$, there are no results when the system starts with a definite initial state $\Gamma(0)\in\calS_U$.
We believe that this reflects an essential difference between the classical and quantum descriptions.

\subsection{Open problems}
\label{s:issues}

\paragraph{Time scale for thermalization:}
In sections~\ref{s:thermalization}, \ref{s:moderate}, and \ref{s:ETH}, we have followed previous works \cite{vonNeumann,GLTZ,Tasaki1998,Reimann,LindenPopescuShortWinter,
GLMTZ09b,Hal2010,ReimannKastner,Reimann2} and proved, under suitable assumptions, that the system thermalizes after a sufficiently long time.
Unfortunately we were not able to make any estimate of the time scale required for thermalization.
This is quite unsatisfactory from a physical point of view, because any results for thermalization is physically meaningless if the required time scale is too large compared with the time scale of thermalization in  nature or in experiments\footnote{
It should be noted however that the observed time scales for thermalization differs considerably depending on the system.
}.

In \cite{GHT13,GHT14short,GHT14long}, Goldstein, Hara and the present author studied the problem of time scale by using the formulation of nonequilibrium subspace introduced in \cite{GLMTZ09b} (see the end of section~\ref{s:comparison}).
In particular it was shown in \cite{GHT14short,GHT14long} that, if one chooses the nonequilibrium subspace in a random manner, then the time required for ``thermalization'' is quite short, of order the Boltzmann time $\hbar/(k_{\rm B}T)$.
In other words, in the (fictitious) space of all the possible systems, the quick decay is a typical property\footnote{
The idea to look for a typical property in the space of systems is due to von Neumann \cite{vonNeumann,GLTZ}. 
}. 
Of course such a quick decay is highly unphysical.
The lesson is that we should not rely on the typicality argument when choosing the system.
This is reasonable since the typicality argument does not take into account properties of realistic Hamiltonian or physical observables\footnote{
But see \cite{Reimann2016} for the discussion about the quick decay in physically realistic situations.
}.

Let us briefly see two simple arguments which show that such a quick decay is unlikely or impossible in some realistic systems.
Suppose that both the Hamiltonian $\Ham$ and an extensive quantity  $\MV$ are the sums (or the integrals) of local quantities.
Then for their commutator, one has $\norm{[\Ham,\MV]}\le aV$, where $a>0$ is a constant.
Then from \rlb{kpt}, one sees that the time-derivative of the time-dependent expectation value $\sbkt{\varphi(t)|\,({\MV}/{V})\,|\varphi(t)}$ is bounded as
\eq
\abs{\frac{d}{dt}\sbkt{\varphi(t)|\,\frac{\MV}{V}\,|\varphi(t)}}
=\frac{1}{V}\,\Bigl|\bbkt{\varphi(t)\bigl|\,[\Ham,\MV]\,\bigr|\varphi(t)}\Bigr|
\le a.
\en
This means that, when the initial expectation value $m_{\rm init}:=\sbkt{\varphi(0)|\,({\MV}/{V})\,|\varphi(0)}$ is different from the equilibrium value $m(u)$, the time required for thermalization is at least $\tau_{\rm min}=|m_{\rm init}-m(u)|/a$ for any $V$.
Note that this is a rigorous lower bound.

When there is a local conserved quantity in the model, one can make use of the Lieb-Robinson bound \cite{LiebRobinson,HastingsKoma,VershyninaLieb} to prove that the required time scale is (as everybody knows) at least of order the linear size of the system.

To conclude we stress that it is quite important to investigate thermalization in concrete and non-trivial many-body quantum systems.
Such a constructive approach is also necessary to understand the conditions required for thermalization, i.e., the moderate energy distribution in nonequilibrium initial states (see section~\ref{s:moderate}) and the energy eigenstate thermalization hypothesis (see section~\ref{s:ETH}).

\paragraph{Characterization of thermal equilibrium:}
Suppose that we take sufficiently many (but finite) extensive quantities $\MV^{(1)}, \MV^{(2)},\ldots$, and construct mutually commuting approximants $\tMV^{(1)}, \tMV^{(2)},\ldots$.
For each $n=1,2,\ldots$, one can define the nonequilibrium projection $\Pneq^{(n)}$ as in \rlb{Pneqn} by referring to the quantities $\tMV^{(1)},\ldots,\tMV^{(n)}$.
In this way we get the criterion $\sbkt{\varphi|\Pneq^{(n)}|\varphi}\le e^{-\alpha V}$ for thermal equilibrium for each $n$.
Noting that $\Pneq^{(n)}\le\Pneq^{(n+1)}$, we see that there is an implication
\eq
\sbkt{\varphi|\Pneq^{(n+1)}|\varphi}\le e^{-\alpha V}
\ \Longrightarrow\ 
\sbkt{\varphi|\Pneq^{(n)}|\varphi}\le e^{-\alpha V}.
\en
The criteria gets stricter as $n$ increases.
It is likely that, as we take into account more quantities, the series of criteria ``converges'' to a single criterion for determining whether $\kph\in\tHVu$ represent thermal equilibrium by all means.
In other words, we expect that there is a self-adjoint operator
\eq
\hQ_{\rm neq}:=\hP[\HVu]\,\Pneq^{(n)}\,\hP[\HVu],
\en
defined by a sufficient set of extensive quantities $\MV^{(1)},\ldots,\MV^{(n)}$, and a pure state $\kph\in\tHVu$ can be definitely said to characterize thermal equilibrium if $\sbkt{\varphi|\hQ_{\rm neq}|\varphi}\le e^{-\alpha V}$.

For the moment we still have no ideas about the nature of the operator $\hQ_{\rm neq}$.
The critical reader might also point out that it is possible that there is nothing like a complete set of quantities, and the criterion gets stronger and stronger as we consider more and more quantities.
We are far from answering such questions.

There is a related question about the applicability of statistical mechanics.
Although statistical mechanics is primarily a machinery for computing the equilibrium values of macroscopic quantities, one can compute the expectation values of not necessarily macroscopic quantities, such as the $n$-point correlation functions of certain local observables.
It is a nontrivial question which of these predictions should be reproduced in realistic systems in thermal equilibrium (or by a pure state $\kph$ representing thermal equilibrium).
It is simply absurd to imagine that the theoretical prediction for the $n$-point function reflects the reality when $n$ is of order the Avogadro constant.
On the other hand we expect that two-point functions from statistical mechanics should be comparable to the result of properly designed experiments (see Appendix~\ref{s:small}).
We still do not know of any criteria that distinguishes the two cases.

See the recent interesting work by Goldstein, Huse, Lebowitz and Tumulka \cite{GHLT} for a related discussion, especially about the distinction between the notions of {\em macroscopic thermal equilibrium}\/ (MATE) and {\em microscopic thermal equilibrium}\/ (MITE).

\appendix

\section{Three toy models}
\label{s:toy}
We shall discuss three simple solvable examples in which the assumptions made in sections~\ref{s:moderate} and \ref{s:ETH} can be easily verified.
Although the results are in a sense trivial, we hope that these elementary examples shed light on the general scenario, and also provide insight to truly nontrivial many body systems.
The material in this section is partly based on our unpublished work \cite{Hal2010}.

\subsection{Independent spins under random magnetic field}
\label{s:spinhx}
Let us start with a trivial example of independent spins under random magnetic field.
In this model,  independent precession of each spin causes the ``approach to equilibrium'' for certain observables.
Although everything is trivial, it may be a good idea to look at the simple (but genuinely quantum) mechanism that realizes the relaxation-like behavior.
Interestingly, the same model also offers a counterexample to the assumptions.

We use the same notation as in section~\ref{s:EIsing}, but here regard $\Lambda$ simply as a set of $V$ sites.
We consider a system of $S=1/2$ spins on $\Lambda$, and take the Hamiltonian
\eq
\Ham=\sum_{x\in\Lambda}^Nh_x\,\mathsf{S}^{(1)}_x,
\lb{HspinInd}
\en
where independent spins are under nonuniform magnetic field.
The local magnetic field $h_x$ is independently drawn from the interval $[-h_0,h_0]$ according to the uniform probability measure, where $h_0>0$ is a fixed constant.

As in \rlb{Sphipm}, we denote by $\ket{\varphi^\pm_x}$ the basis states of the local Hilbert space $\calH_x$.
Then the eigenstates of $\Ham$ is written as
\eq
\ket{\Psi_{\bstau}}=\bigotimes_{x\in\Lambda}\frac{1}{\sqrt{2}}
\Bigl\{\ket{\varphi^+_x}+\tau_x\ket{\varphi^-_x}\Bigr\},
\lb{PsisigmaToy}
\en
where we used the multi-index (or the spin configuration) $\bstau:=(\tau_x)_{x\in\Lambda}$ with $\tau_x=\pm1$.
The corresponding energy eigenvalue is
\eq
E_{\bstau}=\frac{1}{2}\sum_xh_x\,\tau_x.
\lb{Es}
\en
Since $h_x$ are drawn randomly, the energy eigenvalues are nondegenerate with probability one.

As in section~\ref{s:EIsing}, we take the total magnetization (in the direction orthogonal to the magnetic filed) $\MV=\sum_{x\in\Lambda}\hS^{(3)}_x$ as the thermodynamic quantity of interest.
From the symmetry, one has $\bktmc{\MV}=0$ for any $V$ and $u$.

Since each spin independently points upward or downward (in the 3-direction) with probability $1/2$ in the energy eigenstate \rlb{PsisigmaToy}, the probability distribution of $\MV$ is given by the binomial distribution
\eq
\bra{\Psi_{\bstau}}\,\hP[\MV=M]\,\ket{\Psi_{\bstau}}=\frac{1}{2^V}\frac{V!}{N_+!\,N_-!}.
\lb{binomToy1}
\en
Here $N_+$ and $N_-$ are the numbers of up and down spins, respectively, which are determined by $N_++N_-=V$ and $(N_+-N_-)/2=M$.
By recalling the standard large deviation property of the coin toss\footnote{
\label{fn:coin}
It is known for $p\in(1/2,1)$ that $\sum_{N_+\ge pV}2^{-V}{V\choose N_+}\le\exp[-V\{\log2-S_2(p)\}]$.
} \cite{Elis,DemboZeitouni}, \rlb{binomToy1} implies
\eq
\bra{\Psi_{\bstau}}\,\hP\bigl[|\MV|\ge M\bigr]\,\ket{\Psi_{\bstau}}\le2\,e^{-\kappa(\delta)\,V},
\lb{ETHtoyspin}
\en
with $\kappa(\delta)=\log2-S_2[(1/2)+\delta]=2\delta^2+O(\delta^4)$, where $S_2(p)=-p\log p-(1-p)\log(1-p)$ is the binary entropy.
We thus find that the energy eigenstate thermalization hypothesis \rlb{EET} is valid in any range of energy\footnote{
The thermodynamic bound \rlb{tdb} of course follows by summing up \rlb{ETHtoyspin}.
}.

We thus conclude that, as far as one looks at the total magnetization $\MV=\sum_{x\in\Lambda}\hS^{(3)}_x$, the model approaches (thermal) equilibrium from any initial state. 
As we have noted in the beginning, this ``approach to equilibrium'' is nothing but a trivial consequence of independent precession of each spin.

It is interesting to see what happens if we take  $\MV'=\sum_{x\in\Lambda}\hS^{(1)}_x$, which is the total spin in the 1-direction, as the thermodynamic quantity of interest.
In this case we have $\MV'\ket{\Psi_{\bstau}}=(\sum_{x\in\Lambda}\tau_x/2)\ket{\Psi_{\bstau}}$, i.e., the energy eigenstate \rlb{PsisigmaToy} is also the eigenstate of $\MV'$.
Since the energy eigenvalue \rlb{Es} is the sum of the continuously distributed random quantities, it happens in general that two energy eigenvalues $E_{\bstau}$ and $E_{\bstau'}$ which are extremely close to each other have radically different configurations $\bstau$ and $\bstau'$.
As a consequence, the eigenvalue $\sum_{x\in\Lambda}\tau_x/2$ shows an erratic behavior when viewed as a function of the energy eigenvalue $E_{\bstau}$.
See the discussion in  \cite{PH}.
This observation implies that, when one is interested in $\MV'$, the model does not satisfy the energy eigenstate thermalization hypothesis.

This trivial example illustrates how the tendency of the approach to equilibrium can be lost  in a system with a quenched disorder.
Pal and Huse \cite{PH} numerically studied the $S=1/2$ model with the Hamiltonian
\eq
\hH=J\sum_{j=1}^N{\boldsymbol{\mathsf{S}}}_j\cdot{\boldsymbol{\mathsf{S}}}_{j+1}+\sum_{j=1}^Nh_j\,\hS^{(3)}_j,
\en
where $J$ is a constant and $h_j$ is uniformly distributed in $[-h,h]$.
A systematic analysis suggests that the ``localization'' observed above for $J=0$ persists in a model with sufficiently small $|J|$.
In this case, the model lacks the ability to relax to equilibrium by itself.
For large enough $|J|$, the system enters the delocalized phase where it can relax to equilibrium.
See \cite{PH} and references therein for further discussions about the localization in many body quantum systems and its relation to the problem of the approach to equilibrium.
See also \cite{Imbrie} for a recent rigorous result.

\subsection{Free fermions on a double chain}
\label{s:freefermions}
We next discuss a slightly less trivial (but still easily solvable) model of free fermions on a double chain.
We shall confirm that all the assumptions made in Theorems~\ref{t:th1} and \ref{t:th2} are valid in this model.

\begin{figure}
\begin{center}
\includegraphics[width=8cm]{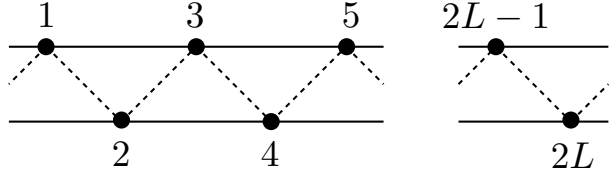}
\end{center}
\caption[dummy]{
The free fermion model defined on a pair of chains is a simple solvable model where we can prove all the assumptions in Theorems~\ref{t:th1} and \ref{t:th2}.
Solid lines represent intra-chain hopping, and dotted lines represent (weak) inter-chain hopping (or coupling).
The model is equivalent to a free fermion on a single chain with nearest and next-nearest neighbor hopping.
}
\label{f:DC}
\end{figure}

\paragraph{The model:}
We consider a free fermion model defined on a double chain as depicted in Fig.~\ref{f:DC}.
The two chains are identified with sets of odd and even integers, respectively, as
\eq
\Lambda_1=\{1,3,\ldots,2L-1\},\quad
\Lambda_2=\{2,4,\ldots,2L\},
\lb{doublechain}
\en
where $L$ is a fixed integer.
We also denote the whole lattice as
\eq
\Lambda=\Lambda_1\cup\Lambda_2=\{1,2,3,\ldots,2L\},
\lb{doublechain2}
\en
and identify the volume with $V=2L$.

For each $x \in\Lambda$, let $\hc_{x}$ and $\hcd_{x}$ be the annihilation and the creation operators, respectively, of a fermion at $x$.
They satisfy the standard canonical anticommutation relations\footnote{
We write $\{\sfA,\sfB\}=\sfA\sfB+\sfB\sfA$.
}
\eq
\{\hcd_{x},\hc_{y}\}=\delta_{x,y},\quad
\{\hc_{x},\hc_{y}\}=\{\hcd_{x},\hcd_{y}\}=0,
\en
for any $x,y\in\Lambda$.
We consider states with $N$ fermions on the lattice.
(We fix $\rho=N/V$ when we make $V$ large.)
The whole Hilbert space is spanned by the states of the form $\hcd_{x_1}\ldots\hcd_{x_N}\vac$, where $x_j\in\Lambda$ with $x_j<x_{j+1}$, and $\vac$ is the normalized state with no fermions in the system.
It satisfies $\hc_x\vac=0$ for any $x$.

We consider the Hamiltonian
\eq
\Ham=\frac{1}{2}\sumx
(e^{i\theta}\hcd_{x}\hc_{x+2}+e^{-i\theta}\hcd_{x+2}\hc_{x})
+\frac{\lambda}{2}\sumx
(e^{i\theta}\hcd_{x}\hc_{x+1}+e^{-i\theta}\hcd_{x+1}\hc_{x}),
\lb{FFHam}
\en
where $\lambda\in(0,1]$ and $\theta\in[0,2\pi)$ are parameters.
The phases $\theta$ is introduced (rather artificially) to avoid degeneracy\footnote{
One can introduce a different phase for the second term.
}.
See Proposition~\ref{p:nondeg}.
We impose periodic boundary conditions, and make identifications $\hc_{2L+1}=\hc_1$ and $\hc_{2L+2}=\hc_1$.

The first term in \rlb{FFHam} represents hopping within each chain, while the second term represents hopping between the two chains.
The model is also interpreted as that on a single chain $\Lambda$ with nearest neighbor and next-nearest neighbor hopping.

\paragraph{Energy eigenstates and eigenvalues:}
Define the set of wave numbers as
\eq
\calK:=\set{\frac{2\pi}{2L}j}{j=1,2,\ldots,2L}.
\lb{setK}
\en
We introduce fermion operators $\had_k$ for $k\in\calK$, which are related with $\hcd_x$ by
\eq
\had_k=\frac{1}{\sqrt{2L}}\sum_{x\in\Lambda}e^{ikx}\hcd_x,\quad
\hcd_x=\frac{1}{\sqrt{2L}}\sum_{k\in\calK}e^{-ikx}\had_k,
\en
and satisfy the anticommutation relations
\eq
\{\had_k,\ha_{k'}\}=\delta_{k,k'},\quad
\{\had_k,\had_{k'}\}=\{\ha_k,\ha_{k'}\}=0,
\en
for any $k,k'\in\calK$.

A standard calculation shows that the Hamiltonian \rlb{FFHam} is diagonalized by using the $\ha$ operators as
\eq
\Ham=\sumk\bigl\{\ez(k)+\lambda\ei(k)\bigr\}\had_k\ha_k,
\lb{FFHam2}
\en
where 
\eq
\ez(k)=\cos(2k+\theta),\quad\ei(k)=\cos(k+\theta).
\lb{FFep}
\en
Take an arbitrary subset $K\subset\calK$ such that $|K|=N$, and define
\eq
\ket{\Psi_K}:=\Bigl(\prod_{k\in K}\had_{k}\Bigr)\vac.
\lb{PsiK}
\en
From \rlb{FFHam2} one finds that $\ket{\Psi_K}$ is an eigenstate of $\Ham$, i.e.,
\eq
\Ham\ket{\Psi_K}=E_K\,\ket{\Psi_K},
\lb{FFHk=Ek}
\en
with the energy eigenvalue
\eq
E_K=\sum_{k\in K}\bigl\{\ez(k)+\lambda\ei(k)\bigr\}.
\lb{EK}
\en
It can be also shown that the corresponding number of states exhibits the standard behavior \rlb{logO} when $U/V$ is sufficiently small.
The energy shell $\HVu$ is spanned by $\ket{\Psi_K}$ with
\eq
u-\Du<\frac{E_K}{V}\le u.
\lb{uDuEJ+J+u}
\en

In sections~\ref{s:moderate} and \ref{s:ETH}, we have assumed that the energy eigenvalues are nondegenerate.
Although nondegeneracy is always achieved by adding a small (random) perturbation to any given Hamiltonian, it is nice to know that nondegeneracy is guaranteed under certain conditions.
By making use of standard results in number theory \cite{Tignol,IR}, we can prove that the present model generically has no degeneracy.
See the end of the section for the proof.

\begin{pro}
\label{p:nondeg}
Let $L>2$ be a prime number with $N<L/2$.
Fix an arbitrary constant $\lambda_0>0$.
For any $\lambda\in(0,\lambda_0]$ except for (at most) a finite number of points, and for any $\theta\in[0,2\pi)$ except for a finite number of points\footnote{
Here the exceptional values of $\theta$ depends on $\lambda$.
}, the energy eigenvalues are nondegenerate, i.e., $E_K=E_{K'}$ implies $K=K'$.
\end{pro}
From a physical point of view, it is absurd to assume that the chain length is equal to a prime number.
We of course do not believe that this is really crucial.
When some of the conditions of the proposition are not satisfied, we still expect the system to exhibit essentially the same behavior although there may be some accidental (and irrelevant) degeneracies in the energy eigenvalues.

\paragraph{Other fermion operators:}
Let $\nu=1,2$ specify one of the two chains.
We denote by  $\mathsf{N}_\nu:=\sum_{x\in\Lambda_\nu}\hcd_{x}\hc_{x}$ the number of fermions on the chain $\nu$.
We also define
\eq
\had_{k,\nu}:=\frac{1}{\sqrt{L}}\sum_{x\in\Lambda_\nu}e^{ikx}\hcd_x,
\lb{adkn}
\en
which creates the state with wave number $k$ only on the chain $\nu$.
We obviously have
\eq
\had_k=\frac{1}{\sqrt{2}}(\had_{k,1}+\had_{k,2}),
\lb{a=a+a}
\en
and
\eq
[\sfN_\nu,\had_{k,\nu'}]=\delta_{\nu,\nu'}\had_{k,\nu}.
\lb{Naa}
\en

For any $k\in\calK$, we denote by $\bar{k}$ the unique element in $\calK$ such that $|k-\bar{k}|=\pi$.
One readily finds $\had_{\bar{k},1}=-\had_{k,1}$ and  $\had_{\bar{k},2}=\had_{k,2}$.
Recalling \rlb{a=a+a}, this implies
\eq
\had_k\had_{\bar{k}}=\frac{1}{2}(\had_{k,1}+\had_{k,2})(\had_{\bar{k},1}+\had_{\bar{k},2})
=\frac{1}{2}(\had_{k,1}\had_{\bar{k},2}+\had_{k,2}\had_{\bar{k},1}).
\lb{akak}
\en
From \rlb{adkn}, we also get the anticommutation relations
\eq
\{\had_{k,1},\ha_{k',2}\}=0,
\quad
\{\had_{k,1},\ha_{k',1}\}=
\begin{cases}
1&k=k',\\-1&\bar{k}=k',\\0&\text{otherwise},\\
\end{cases}
\quad
\{\had_{k,2},\ha_{k',2}\}=
\begin{cases}
1&k=k',\\1&\bar{k}=k',\\0&\text{otherwise},\\
\end{cases}
\en
for any $k,k'\in\calK$.

\paragraph{Energy eigenstate thermalization:}
As for the thermodynamic quantity of interest let us take
\eq
\MV:=\mathsf{N}_1-\mathsf{N}_2,
\lb{FFMV}
\en
which is the difference of the particle numbers in the two chains.
The equilibrium value of $\MV$ is obviously zero by the symmetry.

Let us examine the validity of the energy eigenstate thermalization with respect to $\MV$.
For $K\subset\calK$ with $|K|=N$, define
\eq
K_0:=\set{k\in K}{k\le\pi,\bar{k}=k+\pi\in K}.
\en
We can then write the energy eigenstate \rlb{PsiK} as
\eqa
\ket{\Psi_K}&=\pm\Bigl(\prod_{k\in K_0}\had_{k}\had_{\bar{k}}\Bigr)
\Bigl(\prod_{k\in K\backslash K_0}\had_{k}\Bigr)\vac
\nl
&=
\pm\Biggl(\prod_{k\in K_0}\frac{\had_{k,1}\had_{\bar{k},2}+\had_{k,2}\had_{\bar{k},1}}{2}\Biggr)
\Biggl(\prod_{k\in K\backslash K_0}\frac{\had_{k,1}+\had_{k,2}}{\sqrt{2}}\Biggr)\vac,
\lb{PsiK2}
\ena
where we used  \rlb{akak} and  \rlb{a=a+a}.

Let $N_0=2|K_0|$.
Noting the commutation relation \rlb{Naa}, we find that, in the state \rlb{PsiK2}, $N_0$ fermions are evenly distributed to the two chains, and each of the remaining $N-N_0$ fermions belongs to one of the two chains with independent probability $1/2$.
The probability distribution for the numbers of fermions in the two chains is then
\eqa
&p(N_1,N_2):=\bra{\Psi_K}\,\hP[\mathsf{N}_1=N_1,\mathsf{N}_2=N_2]\,\ket{\Psi_K}
\nl&=\begin{cases}
\dfrac{1}{2^{N-N_0}}\,\dfrac{(N-N_0)!}{\{N_1-\frac{N_0}{2}\}!\,\{N_2-\frac{N_0}{2}\}!}&
\text{when $N_1+N_2=N$, $N_1\ge\dfrac{N_0}{2}$, and $N_2\ge\dfrac{N_0}{2}$}\\
0&\text{otherwise}.
\end{cases}
\lb{pN1N2}
\ena
Given the binomial distribution \rlb{pN1N2} we again see from the standard result in large deviation (see footnote~\ref{fn:coin}) that
\eqa
&\bra{\Psi_K}\,\hP\bigl[\MV\ge V\delta\bigr]\,\ket{\Psi_K}
=\sumtwo{N_1,N_2}{(N_1-N_2\ge V\delta)}p(N_1,N_2)
\nl
&\begin{cases}
\le\exp\biggl[
-V\,(\rho-\rho_0)\biggl\{\log2-S_2\Bigl(\dfrac{1}{2}+\dfrac{\delta}{2(\rho-\rho_0)}\Bigr)\biggr\}\biggr]
&\text{for $\delta\in(0,\rho-\rho_0]$}\\
=0&\text{for $\delta\in(\rho-\rho_0,\rho]$}
\end{cases}
\ena
where $\rho_0=N_0/V$.
Noting the symmetry $\MV\to-\MV$ and the bound
\eq
(\rho-\rho_0)\biggl\{\log2-S_2\Bigl(\dfrac{1}{2}+\dfrac{\delta}{2(\rho-\rho_0)}\Bigr)\biggr\}
\ge
\rho\,\biggl\{\log2-S_2\Bigl(\dfrac{1}{2}+\dfrac{\delta}{2\rho}\Bigr)\biggr\},
\lb{gammaforFF}
\en
we find for any $\delta\in(0,\rho]$ that
\eq
\bra{\Psi_K}\,\hP\bigl[|\MV|\ge V\delta\bigr]\,\ket{\Psi_K}
\le 2 e^{-\kappa(\delta) V},
\lb{ETHFF}
\en
where $\kappa(\delta)$ is defined as the right-hand side of \rlb{gammaforFF}.
So the energy eigenstate thermalization hypothesis \rlb{EET} is valid for any energy eigenstate $\ket{\Psi_K}$.

Note that \rlb{ETHFF} readily implies the thermodynamic bound
\eq
\Bktmc{\hP\bigl[|\MV|\ge V\delta\bigr]}
\le 2e^{-\kappa(\delta)V},
\lb{TDBFF}
\en
for any $u$.

\paragraph{Thermalization:}
Suppose that the conditions for the nondegeneracy in Proposition~\ref{p:nondeg} are satisfied.
Since the energy eigenstate thermalization \rlb{ETHFF} is valid, we see that the conditions for Theorem~\ref{t:th2} are satisfied.
Therefore we have a concrete example in which the approach to thermal equilibrium from an arbitrary initial state $\kpz\in\HVu$ can be proved without any unjustified assumptions.

\paragraph{Nonequilibrium initial states with a moderate energy distribution:}
Note that Theorem~\ref{t:th1}, which states thermalization for certain initial states, does not provide any additional information  when the stronger Theorem~\ref{t:th2} is known to be valid.
It is nevertheless useful to see explicitly that nonequilibrium initial states with a moderate energy distribution (as required in Theorem~\ref{t:th1}) are possible in this model\footnote{
Here we do not make use of Theorem~\ref{t:moderatecondition}, but directly construct examples.
}.

Fix an energy interval $(u-\Du,u]$.
To avoid technical complexity, we assume that the coupling $\lambda$ satisfies $0<\lambda\ll\Du$, and can be neglected when we estimate the energy.
We further simplify the discussion by focusing only on the extreme nonequilibrium where $\MV=\mathsf{N}_1-\mathsf{N}_2$ takes its maximum value $N$.

Take a subset $\tiK\subset\calK$ such that $|\tiK|=N$ and $k\le\pi$ for any $k\in\tiK$.
We also assume that the total energy satisfies
\eq
u-\Du<\frac{1}{V}\sum_{k\in\tiK}\ez(k)\le u,
\en
where we ignored the terms including $\lambda$ in \rlb{EK}.
We denote by $\DVu^\mathrm{neq}$ the total number of $\tiK$ satisfying all these conditions.

We define
\eq
\ket{\Gamma_{\tiK}}:=\Bigl(\prod_{k\in\tiK}\had_{k,1}\Bigr)\vac,
\lb{GammaK1}
\en
which is in the energy shell $\HVu$ (under the assumption $\lambda\ll\Du$), and is an extreme nonequilibrium state with $\MV\ket{\Gamma_{\tiK}}=N\ket{\Gamma_{\tiK}}$.
We thus find that $\DVu^\mathrm{neq}$ is the dimension of the nonequilibrium subspace in this case.
Note that, corresponding to the thermodynamic bound \rlb{tdb}, we have $\DVu^\mathrm{neq}\sim e^{-\gamma V}\DVu$ (which essentially is the definition of $\gamma>0$).

Since $\had_{k,1}=(\had_k-\had_{\bar{k}})/\sqrt{2}$, we can rewrite \rlb{GammaK1} as
\eq
\ket{\Gamma_{\tiK}}=\Biggl(\prod_{k\in\tiK}\frac{\had_k-\had_{\bar{k}}}{\sqrt{2}}\Biggr)\vac.
\lb{GammaK2}
\en
Expanding the product, we see that this state is a linear combination of $2^N$ energy eigenstates \rlb{PsiK}.
This means that $\ket{\Gamma_{\tiK}}$ has the effective dimension $\Deff=2^N$.

The rest is easy.
Consider an initial state
\eq
\ket{\Phi(0)}=\sum_{\tiK}\alpha_{\tiK}\ket{\Gamma_{\tiK}},
\en
where $\tiK$ is summed over the $\DVu^\mathrm{neq}$ subsets $\tiK$ satisfying the above conditions, and all $|\alpha_{\tiK}|$ are nearly equal.
Clearly $\ket{\Phi(0)}$ is in $\HVu$, is an extreme nonequilibrium state, and has the effective dimension
\eq
\Deff\sim 2^{N}\DVu^\mathrm{neq}\sim 2^{N}e^{-\gamma V}\DVu= e^{-(\gamma-\rho\log2)V}\DVu.
\en
By comparing this with \rlb{caV}, we can choose $\eta=\gamma-\rho\log2$.
The condition $\gamma>\eta$ required in Theorem~\ref{t:th1} is thus satisfied.

\paragraph{Proof of Proposition~\protect\ref{p:nondeg}:}
To prove the absence of degeneracy, it is convenient to introduce the standard occupation number description.
With each $K\subset\calK$, we associate a $2L$-tuple $\bsn=(n_j)_{j=1,2,\ldots,2L}$ by
\eq
n_j=\begin{cases}
1&\text{if $\frac{2\pi}{2L}j\in K$},\\
0&\text{if $\frac{2\pi}{2L}j\not\in K$}.\\
\end{cases}
\en
Then the energy eigenvalue \rlb{EK} is written as
\eq
E_K=\sum_{k\in K}\bigl\{\cos(2k+\theta)+\lambda\cos(k+\theta)\bigr\}
=\Re\Bigl[e^{i\theta}\bigl\{\zz(\bsn)+\lambda\,\zi(\bsn)\bigr\}\Bigr],
\en
with
\eq
\zz(\bsn):=\sum_{j=1}^{2L}n_j\,\exp\Bigl[i\frac{2\pi}{L}j\Bigr],\quad
\zi(\bsn):=\sum_{j=1}^{2L}n_j\,\exp\Bigl[i\frac{2\pi}{2L}j\Bigr].
\en

In the following lemma, $\bsn=(n_j)_{j=1,2,\ldots,2L}$ and $\bsn'=(n'_j)_{j=1,2,\ldots,2L}$ denote general $2L$-tuples whose elements are 0 and 1.  We also write $|\bsn|=\sum_{j=1}^{2L}n_j$.
\begin{lemma}
\label{l:zzzz}
Let $L>2$ be prime.
Take any $\bsn$ and $\bsn'$ such that $|\bsn|=|\bsn'|<L/2$.
Then one has $\zz(\bsn)=\zz(\bsn')$ and $\zi(\bsn)=\zi(\bsn')$ simultaneously if and only if $\bsn=\bsn'$.
\end{lemma}

\noindent
{\em Proof of Proposition~\ref{p:nondeg} given Lemma~\ref{l:zzzz}:}\/
The lemma implies that, when $\bsn\ne\bsn'$, the equality $\zz(\bsn)+\lambda\,\zi(\bsn)=\zz(\bsn')+\lambda\,\zi(\bsn')$ may hold only accidentally\footnote{
Note that one inevitably has $\zz(\bsn)\ne\zz(\bsn')$ and $\zi(\bsn)\ne\zi(\bsn')$ when the equality holds.
}, and becomes invalid by an infinitesimal change of $\lambda$.
We thus find that, for any $\lambda\in(0,1]$ except for a finite number of points, the complex quantities $\zz(\bsn)+\lambda\,\zi(\bsn)$ (with all possible $\bsn$ such that $|\bsn|=N$) are all distinct.

Suppose that $\lambda$ is fixed to a non-exceptional value.
The energy eigenvalue $E_K$ may still degenerate if $e^{i\theta}\{\zz(\bsn)+\lambda\,\zi(\bsn)\}$ and  $e^{i\theta}\{\zz(\bsn')+\lambda\,\zi(\bsn')\}$ (with $\bsn\ne\bsn'$) happen to have the same real part.
Such a degeneracy is lifted by infinitesimally changing $\theta$.
Thus degeneracy in $E_K$ can take place only for a finite number of values of $\theta$.~\qedm

\bigskip
We shall state a mathematical lemma which is the essence of Lemma~\ref{l:zzzz}.
Let $\zeta:=e^{i(2\pi/L)}$.
Take an $L$-tuple $\bsm=(m_j)_{j=1,\ldots,L}$ with $m_j\in\bbZ$.
Let $|\bsm|=\sum_{j=1}^L|m_j|$ and $\tiz(\bsm):=\sum_{j=1}^Lm_j\zeta^j$.
It is crucial to note that here $j$ runs from 1 to $L$.

\begin{lemma}
\label{l:Gauss}
One has $\tiz(\bsm)\ne0$ for any $\bsm$ such that $0<|\bsm|<L$.
One also has $\tiz(\bsm)\ne\tiz(\bsm')$ for any $\bsm,\bsm'$ such that $\bsm\ne\bsm'$ and $|\bsm|+|\bsm'|<L$.
\end{lemma}

\noindent
{\em Proof:}\/
This lemma is a straightforward consequence of the classical result by Gauss known as ``the irreducibility of the cyclotomic polynomials of prime index" (see, for example, Chapter 12, Section 3 of \cite {Tignol} or Chapter 13, Section 2 of \cite{IR}).
It implies that the $L-1$ complex numbers $\zeta$, $\zeta^2$, $\ldots$, $\zeta^{L-1}$ are rationally independent, i.e., if $\sum_{n=1}^{L-1}m_n\,\zeta^n=0$ with integers $m_1,\ldots,m_{L-1}$, one inevitably has $m_1=m_2=\cdots=m_{L-1}=0$.

To show the first claim, we note that $|\bsm|<L$ implies that there is $j_1\in\oL$ such that $m_{j_1}=0$.
Then one finds that $\zeta^{-j_1}\,\tiz(\bsm)=\sum_{n=1}^{L-1}\tim_n\,\zeta^n$ with $\tim_n\in\bbZ$.
Since not all of $\tim_n$ are vanishing, we have $\sum_{n=1}^{L-1}\tim_n\,\zeta^n\ne0$, and hence $\tiz(\bsm)\ne0$.

To show the second claim, we observe that $\tiz(\bsm)-\tiz(\bsm')=\tiz(\bsm'')$ with $m''_j=m_j-m'_j$.
Since $0<|\bsm''|\le|\bsm|+|\bsm'|<L$, the first claim shows that $\tiz(\bsm)-\tiz(\bsm')\ne0$.~\qedm

\bigskip

\noindent
{\em Proof of Lemma~\ref{l:zzzz} given Lemma~\ref{l:Gauss}:}\/
We shall rewrite $\zz(\bsn)$ and $\zi(\bsn)$ in the form of $\tiz(\bsm)$.
As for $\zz(\bsn)$, we readily find
\eq
\zz(\bsn)=\sum_{j=1}^{2L}n_j\,\exp\Bigl[i\frac{2\pi}{L}j\Bigr]
=\sum_{j=1}^L(n_j+n_{j+L})\zeta^j.
\lb{zzm}
\en
To deal with $\zi(\bsn)$ we note that $e^{i(\pi/L)j}=\zeta^{j/2}$ for even $j$, and $e^{i(\pi/L)j}=-\zeta^{(j\pm L)/2}$ for odd $j$.
Then we get
\eq
\zi(\bsn)=\sum_{j=1}^{2L}n_j\,\exp\Bigl[i\frac{\pi}{L}j\Bigr]
=\mathop{\sum_{j=1}^L}_{(j\,\text{even})}(n_j-n_{j+L})\zeta^{j/2}
-\mathop{\sum_{j=1}^L}_{(j\,\text{odd})}(n_j-n_{j+L})\zeta^{(j+L)/2}.
\lb{zim}
\en

Take $\bsn$ and $\bsn'$ such that $|\bsn|=|\bsn'|<L/2$.
Then, from the expression \rlb{zzm} and Lemma~\ref{l:Gauss}, we see that $\zz(\bsn)=\zz(\bsn')$ if and only if $n_j+n_{j+L}=n'_j+n'_{j+L}$ for all $j=1,2,\ldots,L$.
Similarly from \rlb{zim} and Lemma~\ref{l:Gauss}, we see that $\zi(\bsn)=\zi(\bsn')$ if and only if $n_j-n_{j+L}=n'_j-n'_{j+L}$ for all $j=1,2,\ldots,L$.
Therefore $\zz(\bsn)=\zz(\bsn')$ and $\zi(\bsn)=\zi(\bsn')$ are simultaneously valid if and only if $\bsn=\bsn'$.~\qedm

\subsection{Toy model for two identical bodies in contact}
\label{s:toytwobody}
Finally we shall see an artificial model of two identical bodies exchanging energy, the situation treated in section~\ref{s:EHC}.
We should warn the reader that, unlike the two simple models that we have discussed in sections~\ref{s:spinhx} and \ref{s:freefermions}, the present example is ``made up'' so that our scenario of typicality and thermalization works perfectly.
Nevertheless we hope that this concrete example will be of help in developing intuitions about nontrivial models.

Let us specify the first subsystem.
We assume that the eigenvalues of the Hamiltonian $\hH^{(1)}$ is $\ez V n$ with $n=1,2,\ldots$, where $\ez>0$ is a fixed constant.
To mimic the behavior of a macroscopic system, we require that each level with $n$ is $\Omega^{(1)}_n$ fold degenerate, where 
\eq
\Omega^{(1)}_n=\exp[Vs_n].
\lb{Omega1Tot}
\en
We assume that the ``entropy density'' $s_n$ is strictly increasing in $n$, and strictly concave, i.e.,
\eq
2s_n>s_{n-1}+s_{n+1},
\lb{sntoy}
\en
for any $n=2,3,\ldots$.
We denote the eigenstate of $\hH^{(1)}$ as $\ket{\psi^{(1)}_{n,j}}$ where $n=1,2,\ldots$, and $j=1,2,\ldots,\Omega^{(1)}_n$.
It satisfies $\hH^{(1)}\ket{\psi^{(1)}_{n,j}}=\ez V n\ket{\psi^{(1)}_{n,j}}$.

The second subsystem is an exact copy of the first, and we denote by $\ket{\psi^{(2)}_{n',j'}}$ the energy eigenstate of the Hamiltonian $\hH^{(2)}$.

For $m=2,3,\ldots$, let $\calN_m:=\{(n,n')\,|\,n,n'\in\{1,2,\ldots\},\  n+n'=m\}$.
Then for any $(n,n')\in\calN_m$, the tensor product $\ket{\psi^{(1)}_{n,j}}\otimes\ket{\psi^{(2)}_{n',j'}}$ is an eigenstate of the noninteracting Hamiltonian $\hH^{(1)}\otimes\mathsf{1}+\mathsf{1}\otimes\hH^{(2)}$ with the eigenvalue $\ez V m$.
The degeneracy of this eigenvalue is given by
\eq
\Omega_m=\sum_{(n,n')\in\calN_m}\Omega^{(1)}_n\,\Omega^{(2)}_{n'}
=\sum_{(n,n')\in\calN_m}\exp[V(s_n+s_{n'})].
\en
We also denote by $\calH_m$ the corresponding $\Omega_m$ dimensional eigenspace.

Suppose that $m$ is even.
Take $(n,n')\in\calN_m$, and write $n=(m/2)+r$ and $n'=(m/2)-r$.
From concavity \rlb{sntoy}, it follows that the quantity $s_n+s_{n'}=s_{(m/2)+r}+s_{(m/2)-r}$ attains its maximum at $r=0$ and decreases strictly\footnote{
{\em Proof:}\/ \rlb{sntoy} implies $s_n-s_{n-1}>s_{n+1}-s_n$.
By repeatedly using this, one finds $s_p-s_{p-1}>s_{q+1}-s_q$ for any $p\le q$.
This means $s_p+s_q>s_{p-1}+s_{q+1}$, which justifies the claim.
} as $r$ deviates from 0.
We thus find that
\eq
\frac{\Omega^{(1)}_{(m/2)+r}\,\Omega^{(2)}_{(m/2)-r}}{\Omega_{m}}
\le
\frac{\Omega^{(1)}_{(m/2)+r}\,\Omega^{(2)}_{(m/2)-r}}{\Omega^{(1)}_{m/2}\,\Omega^{(2)}_{m/2}}
=e^{-\tilde{\kappa}(m,r)\,V},
\lb{OOOO}
\en
for any $|r|\ge1$, where 
\eq
\tilde{\kappa}(m,r):=2s_{m/2}-\{s_{(m/2)+r}+s_{(m/2)-r}\}>0,
\en 
is strictly increasing in $|r|$ (for a fixed $m$).
This bound will be useful below.

We shall design the interaction Hamiltonian $\hH_\mathrm{int}$ so that to leave each subspace $\calH_m$ invariant, and mix up all the basis states in it.
To make the model trivially solvable, we shall go through the following highly artificial construction.
For each $m=2,3,\ldots$, list up all the basis states $\ket{\psi^{(1)}_{n,j}}\otimes\ket{\psi^{(2)}_{n',j'}}$ with $(n,n')\in\calN_m$, and renumber them\footnote{
The numbering of $\ell=1,2,\ldots,\Omega_m$ can be done in an arbitrary manner.
} as $\ket{\Phi_{m,\ell}}$, where $\ell=1,2,\ldots,\Omega_m$.
We then define $\hH_\mathrm{int}$ by
\eq
\bra{\Phi_{m,\ell}}\hH_\mathrm{int}\ket{\Phi_{m',\ell'}}=
\begin{cases}
(V\eo/2)\,e^{i\theta}&\text{if $m=m'$ and $\ell'=\ell+1$}\\
(V\eo/2)\,e^{-i\theta}&\text{if $m=m'$ and $\ell'=\ell-1$}\\
0&\text{otherwise},
\end{cases}
\lb{HintTOY}
\en
where we take the ``periodic boundary condition'' and identify $\ell=\Omega_m+1$ with $\ell=1$.
With this artificial choice, the interaction Hamiltonian $\hH_\mathrm{int}$, restricted on $\calH_m$, is exactly the Hamiltonian of the tight-binding model on a chain of length $\Omega_m$.
The phase $\theta$ is introduced to avoid degeneracy.

Then the energy eigenstate of the full Hamiltonian $\mathsf{H}^{(1)}\otimes\mathsf{1}+\mathsf{1}\otimes\mathsf{H}^{(2)}+\mathsf{H}_\mathrm{int}$ is readily obtained as
\eq
\ket{\Psi_{m,q}}=\frac{1}{\sqrt{\Omega_m}}
\sum_{\ell=1}^{\Omega_m}\exp\Bigl[i\frac{2\pi q\ell}{\Omega_m}\Bigr]\ket{\Phi_{m,\ell}},
\lb{Psimq}
\en
where $q=1,\ldots,\Omega_m$ for each $m=2,3,\ldots$.
The corresponding energy eigenvalue is
\eq
E_{m,q}=V\biggl\{\ez m+\eo\cos\Bigl[\frac{2\pi q}{\Omega_m}+\theta\Bigr]\biggr\}.
\en
By taking $\ez>\eo>0$ and assuming that $\theta/\pi$ is irrational, we find that the energy eigenvalues are nondegenerate.

Take, for simplicity, an even $m$, and choose $u$ and $\Du$ so that $u-\Du=\ez m-\eo$ and $u=\ez m+\eo$.
Then the energy shell $\HVu$ coincides with the subspace $\calH_m$.
We shall again look at the energy difference $\MV=\mathsf{H}^{(1)}\otimes\mathsf{1}-\mathsf{1}\otimes\mathsf{H}^{(2)}$.
By construction each $\ket{\Phi_{m,\ell}}$, which is indeed $\ket{\psi^{(1)}_{n,j}}\otimes\ket{\psi^{(2)}_{n',j'}}$, is an eigenstate of $\MV$, i.e., $\MV\ket{\psi^{(1)}_{n,j}}\otimes\ket{\psi^{(2)}_{n',j'}}=V\ez(n-n')\ket{\psi^{(1)}_{n,j}}\otimes\ket{\psi^{(2)}_{n',j'}}$.
Since the energy eigenstate $\ket{\Psi_{m,q}}$ is a linear combination of all $\ket{\psi^{(1)}_{n,j}}\otimes\ket{\psi^{(2)}_{n',j'}}$ with $(n,n')\in\calN_m$ as in \rlb{Psimq}, we readily see that 
\eq
\bra{\Psi_{m,q}}\,\hP[\MV=V\ez s]\,\ket{\Psi_{m,q}}=
\frac{\Omega^{(1)}_{(m+s)/2}\,\Omega^{(2)}_{(m-s)/2}}{\Omega_{m}}
\le e^{-\tilde{\kappa}(m,s/2)\,V},
\en
where the inequality follows from \rlb{OOOO}.
Then we immediately see that the energy eigenstate thermalization hypothesis \rlb{EET} holds as\footnote{
This, again, implies the thermodynamic bound \rlb{tdb}.
}
\eq
\bra{\Psi_{m,q}}\,\hP[|\MV|\ge V\delta]\,\ket{\Psi_{m,q}}
\le e^{-\kappa(m,\delta)\,V},
\en
where $\kappa(m,\delta)\simeq\tilde{\kappa}(m,\delta/(2\ez))$.

We can thus conclude from Theorem~\ref{t:th2} that the present model exhibits thermalization (in the sense that the energy is almost evenly distributed into the two subsystems) from any initial state.

It is also instructive to see that one can take (initial) states with moderate energy distribution in this model.
From \rlb{Psimq}, one finds that any $\ket{\Phi_{m,\ell}}$ can be written in terms of the energy eigenstates as
\eq
\ket{\Phi_{m,\ell}}=\frac{1}{\sqrt{\Omega_m}}
\sum_{q=1}^{\Omega_m}\exp\Bigl[-i\frac{2\pi q\ell}{\Omega_m}\Bigr]\ket{\Psi_{m,q}}.
\lb{Psimqinv}
\en
Since all the expansion coefficients have the same amplitude, the effective dimension of  any $\ket{\Phi_{m,\ell}}$, which is indeed $\ket{\psi^{(1)}_{n,j}}\otimes\ket{\psi^{(2)}_{n',j'}}$, takes the maximum possible value $\Deff=\DVu$.
This means that there are many nonequilibrium states which satisfies the condition \rlb{caV} with $\eta=0$.
We also see that the condition for Theorem~\ref{t:moderatecondition} is satisfied in this case, and hence an overwhelming majority of nonequilibrium states satisfy \rlb{caV}.

\section{Treatment of non-extensive quantities}
\label{s:small}
In the main body of the present paper we have treated only macroscopic (extensive) quantities $\MV^{(1)},\ldots,\MV^{(n)}$ of a macroscopic system.
Although this is sufficient for our motivation to reproduce equilibrium thermodynamics, one can, if necessary, treat quantities which are not extensive by a slight modification.

\paragraph{Correlation functions:}
We concentrate on a model defined on the $d$-dimensional $L\times\cdots\times L$ hypercubic lattice $\Lambda$ whose sites are denoted as $x,y,\ldots\in\Lambda$, and write $V=L^d$.
We assume that the Hamiltonian $\Ham$ is translationally invariant.

Let $\hf_o$ and $\hg_o$ be operators which act only on a finite number of sites, and denote by $\hf_x$ and $\hg_x$ their translations.
Suppose that one is interested in the correlation function 
\eq
c_r(u):=\lim_{V\up\infty}\Bktmc{\hf_x\,\hg_{x+r}+(\hf_x\,\hg_{x+r})^\dagger},
\en
for $r\in\bbZ^d$.
We shall argue that the value $c_r(u)$ (for limited $r$) can be treated as the equilibrium value of a macroscopic quantity.

For a fixed $r$, we define
\eq
\sfC_{V,r}:=\sum_{x\in\Lambda}\bigl\{\hf_x\,\hg_{x+r}+(\hf_x\,\hg_{x+r})^\dagger\bigr\},
\en
which is regarded as an extensive quantity.
Because of the translation invariance, we have
\eq
c_r(u)=\lim_{V\up\infty}\frac{1}{V}\bktmc{\sfC_{V,r}}.
\en

We then take a sufficiently large subset $\Lambda_0\subset\bbZ^d$ (which is independent of $V$), and include all $\sfC_{V,r}$ with $r\in\Lambda_0$ into the list of macroscopic quantities to consider\footnote{
It is possible to let $\Lambda_0$ depend on $V$, even to let $\Lambda_0=\Lambda$.
We do not go into such extensions, since a finite $\Lambda_0$ is usually sufficient.
}.
We expect that the thermodynamic bound \rlb{tdb} is still valid in general after including $\sfC_{V,r}$.
In fact Proposition~\ref{p:TDBQS} for quantum spin systems extends to this case as it is (with a worse constant).

In this manner we can treat, within our scheme, the correlation function $c_r(u)$ as the equilibrium value of the macroscopic quantity $\sfC_{V,r}$.
All the results about typicality (section~\ref{s:typicality}) and thermalization (sections~\ref{s:thermalization}, \ref{s:moderate}, and \ref{s:ETH}) can be applied as they are when the assumptions are verified.
In most cases it is enough to take $\Lambda_0$ sufficiently large (to exceed the correlation length) in order to recover essential physics described by the correlation function.

\paragraph{Probability distribution in a small system:}
Consider a quantum mechanical system defined on a region (which can be a lattice) with a small volume $V_0$.
Physical quantities in this system should exhibit relatively large fluctuation in thermal equilibrium.
With a little trick (see, e.g., \cite{TasakiSM}), one can also treat the probability distribution of such a fluctuating quantity within our scheme.

Let a self-adjoint operator $\hf$ be the quantity of interest of the small system.
We prepare $N$ identical copies of the small system, and consider a combined system of all the copies.
There are no interactions between small subsystems.
Denoting by $\hf^{(j)}$ the quantity $\hf$ in the $j$-th copy, we define, for $a<b$, the operator
\eq
\sfN_{[a,b]}:=\sum_{j=1}^N\hP\bigl[\hf^{(j)}\in[a,b]\bigr],
\en
which counts the number of copies in which the value of $\hf$ falls in to the interval $[a,b]$.

Let $\sbkt{\cdots}^{\rm mc}_{N,u}$ be the microcanonical expectation corresponding to the energy range $[(u-\Du)N,uN]$ for the whole system.
Then it is easily found that\footnote{
This is a consequence of the equivalence of the microcanonical and the canonical ensembles.
The proof in this case is elementary since the system is a union of noninteracting small parts.
}
\eq
p_{\beta(u)}(a,b):=\lim_{N\up\infty}\frac{1}{N}\sbkt{\sfN_{[a,b]}}^{\rm mc}_{N,u}
\en 
is the probability that the value of $\hf$ falls into $[a,b]$ in a single small system described by the canonical distribution with $\beta(u)$.

Let us divide the spectrum of $\hf$ in to $n$ intervals as $[f_{\rm min},f_{\rm max}]=\bigcup_{j=1}^n[a_{j-1},a_j]$.
We then regard $\sfN_{[a_{j-1},a_j]}$ with $j=1,\ldots,n$ as our macroscopic quantities.
It is again not hard to prove that the thermodynamic bound \rlb{tdb} is valid for any $u$.

From the results in section~\ref{s:typicality}, we can thus essentially recover the probability distribution $p_\beta(a,b)$ from a single pure state of a large system (constructed by combining copies of the original small system).
Unfortunately the large system, as it is, never exhibit thermalization since small parts do not interact with each other.
It is likely that the system thermalizes if we add weak interactions between the small systems, but this is not easy to prove.

\section{Mixed initial state}
\label{s:mixed}
Our results about thermalization in sections~\ref{s:thermalization}, \ref{s:moderate}, and \ref{s:ETH} readily extends to the case where the initial state is a mixed state.
In what follows we only consider density matrices whose supports are the energy shell $\HVu$.
Corresponding to Definitions~\ref{d:eq} or \ref{d:eq2}, we say that a state $\rho$ represents thermal equilibrium if
\eq
{\rm Tr}[\rho\,\Pneq]\le e^{-\alpha V},
\lb{rhoeq}
\en
where $\Pneq$ is defined by \rlb{Pneq1}, \rlb{Pneqnew} or \rlb{Pneqn}  depending on the situation and the treatment.

Take an initial state $\rho(0)$, and let $\rho(t)=e^{-i\Ham t}\rho(0)\,e^{i\Ham t}$ be its time evolution.
Then the statement corresponding to Lemma~\ref{l:th} reads
\begin{lemma}
\label{l:thmix}
Suppose that there is $\tau>0$ and it holds that
\eq
\frac{1}{\tau}\int_0^\tau dt\,{\rm Tr}[\rho(t)\,\Pneq]
\le e^{-(\alpha+\nu) V}.
\lb{taVmix}
\en
Then there exists a collection of intervals $\calG\subset[0,\tau]$ such that $|\calG|/\tau\ge1-e^{-\nu V}$, and we have for any $t\in\calG$ that
\eq
{\rm Tr}[\rho(t)\,\Pneq]\le e^{-\alpha V},
\lb{thermalizationboundmix}
\en
which means that $\rho(t)$ represents thermal equilibrium in the sense of \rlb{rhoeq}.
\end{lemma}

To discuss the extension of Theorem~\ref{t:th1}, it is useful to define two effective dimensions.
The first is
\eq
\Deff^{\rm full}:=\frac{1}{{\rm Tr}[\{\rho(t)\}^2]},
\en
which is the inverse of the purity ${\rm Tr}[\{\rho(t)\}^2]$, and is clearly independent of $t$.
Note that one has $\Deff^{\rm full}=1$ when $\rho(t)$ represents a pure state.

The effective dimension which corresponds to $\Deff$ of \rlb{caV} is defined using the energy eigenstate basis $\{\kpj\}_{j=1,\ldots,\DVu}$ as
\eq
\Deff^{\rm diag}:=\Bigl\{\sum_j\bigl(\bra{\psi_j}\rho(t)\ket{\psi_j}\bigr)^2\Bigr\}^{-1}=\frac{1}{{\rm Tr}[(\rho_{\rm diag})^2]},
\en
where
\eq
\rho_{\rm diag}:=\sum_k\ket{\psi_j}\bra{\psi_j}\rho(t)\ket{\psi_j}\bra{\psi_j}
\en
is the density matrix for the ``diagonal ensemble" obtained by deleting the off-diagonal elements of $\rho(t)$.
Note that $\rho_{\rm diag}$ and $\Deff^{\rm diag}$ are independent of $t$.

These effective dimensions satisfy the bound $\Deff^{\rm full}\le\Deff^{\rm diag}$ because of the monotonicity of the purity\footnote{
{\em Proof}\/:
For an  an arbitrary density matrix $\rho$, one has
$\sum_j\bra{\psi_j}\rho\ket{\psi_j}^2\le\sum_{j,j'}\bra{\psi_j}\rho\ket{\psi_{j'}}\bra{\psi_{j'}}\rho\ket{\psi_j}=\sum_j\bra{\psi_j}\rho^2\ket{\psi_j}$.
}.

We first note that the state is in thermal equilibrium to begin with if $\Deff^{\rm full}$ is sufficiently large.
\begin{theorem}
\label{t:mix1}
If the thermodynamic bound \rlb{tdb} is valid, and one has 
\eq
\Deff^{\rm full}\ge e^{-\eta V}\DVu,
\lb{DDmix}
\en
with $\eta$ satisfying $\gamma-\eta>2(\alpha+\nu)$,
then $\rho(t)$ represents thermal equilibrium for any $t$ (including $t=0$).
\end{theorem}

\noindent
{\em Proof}\/:
From the Schwarz inequality, we get
\eqa
{\rm Tr}[\rho(t)\Pneq]&={\rm Tr}_{\HVu}[\rho(t)\Pneq]
\le\sqrt{
{\rm Tr}_{\HVu}[\{\rho(t)\}^2]\,{\rm Tr}_{\HVu}[\Pneq]
}
\nl
&=\sqrt{\frac{\DVu\bktmc{\Pneq}}{\Deff^{\rm full}}}.
\ena
Then the statement follows exactly as in the proof of Theorem~\ref{t:th1}.
\qedm

Of course it is much more interesting if \rlb{DDmix} is not valid.
The following is a straightforward extension of Theorem~\ref{t:th1}.

\begin{theorem}
If the thermodynamic bound \rlb{tdb} is valid, and one has 
$\Deff^{\rm diag}\ge e^{-\eta V}\DVu$ with $\eta$ satisfying $\gamma-\eta>2(\alpha+\nu)$,
then $\rho(t)$ approaches thermal equilibrium (in the sense of Lemma~\ref{l:th}). 
\end{theorem}

\noindent
{\em Proof}\/:
Note that the nondegeneracy of $E_j$ implies
\eqa
\lim_{\tau\up\infty}\frac{1}{\tau}\int_0^\tau dt\,{\rm Tr}[\rho(t)\,\Pneq]
&=
\lim_{\tau\up\infty}\frac{1}{\tau}\int_0^\tau dt\sum_{j,j'\in\JVu}e^{i(E_j-E_{j'})t}\bra{\psi_{j'}}\rho(0)\ket{\psi_j}\bra{\psi_j}\Pneq\ket{\psi_{j'}}
\nl
&=\sum_{j\in\JVu}\bra{\psi_{j}}\rho(0)\ket{\psi_j}\bra{\psi_j}\Pneq\ket{\psi_{j}}
={\rm Tr}[\rho_{\rm diag}\Pneq].
\ena
Then the rest of the proof is the same as that of Theorem~\ref{t:mix1}.~\qedm

Theorem~\ref{t:th2}, which makes use of the energy eigenstate thermalization hypothesis, also extends to the case with mixed initial state.
We omit the details since the extension is trivial.

\section{Typicality of the canonical expectation values}
\label{s:canonical}
We have exclusively discussed the microcanonical setting in the present paper.
Here we apply some of the techniques in the present paper to the setting where the system of interest is coupled to a heat bath, and show the typicality of  the canonical expectation values.

\paragraph{Setting:}
We consider a  macroscopic quantum system with volume $V$, and denote by $\calH_\mathrm{S}$ and $\hH_\mathrm{S}$ the Hilbert space and the Hamiltonian, respectively.
The system is coupled to a heat bath (reservoir) which itself is a macroscopic quantum system with volume $V_\mathrm{B}=\lambda V$, Hilbert space $\calH_\mathrm{B}$, and Hamiltonian $\hH_\mathrm{B}$.
Here $\lambda$ is a fixed constant which is usually taken to be\footnote{
We do not use the assumption $\lambda\gg1$ in what follows.
The main result is valid even when $\lambda=0$.
See the remark after the theorem.
} $\lambda\gg1$.

The whole system, i.e., the system plus the heat bath, has the volume $\Vtot=(1+\lambda)V$, the Hilbert space $\calH_\mathrm{tot}=\calH_\mathrm{S}\otimes\calH_\mathrm{B}$, and the Hamiltonian
\eq
\hH_\mathrm{tot}:=\hH_\mathrm{S}\otimes\mathsf{1}+\mathsf{1}\otimes\hH_\mathrm{B}+\hH_\mathrm{int},
\lb{HtotB}
\en
where $\hH_\mathrm{int}$ describes the interaction between the system and the bath.
We assume $\norm{\hH_\mathrm{int}}\le h_0V^\zeta$ with constants $h_0>0$ and $0<\zeta<1$.
The energy shell $\HVtu$ for the whole system is defined as in section~\ref{s:MQS} by replacing $V$ with $\Vtot$.

\paragraph{Canonical typicality:}
Popescu, Short, and Winter \cite{PopescuShortWinter} and Goldstein, Lebowitz, Tumulka, and Zangh\`\i\ \cite{GLTZ06} independently stated the important result known as the canonical typicality (see also Sugita \cite{Sugita06}).

Let ${\rho}^\mathrm{mc}_{\Vtot,u}$ be the microcanonical density matrix for the whole system, defined as in \rlb{rhomc} by replacing $V$ with $\Vtot$.
Let $\rho_\mathrm{S}:={\rm Tr}_\mathrm{B}[\,{\rho}^\mathrm{mc}_{\Vtot,u}\,]$ be the reduced density matrix for the system, where $\mathrm{Tr}_\mathrm{B}$ denotes the trace in $\calH_\mathrm{B}$.
It is usually expected (and can be proved with suitable assumptions) that $\rho_\mathrm{S}$ is close to the canonical density matrix for the system with the inverse temperature $\beta(u)$ defined in \rlb{betau}.

Take an arbitrary normalized state $\kph\in\tHVtu$ from the energy shell, and write the corresponding density matrix for the system as $\rho_\varphi:={\rm Tr}_{\rm B}[\,\kph\bra{\varphi}\,]$.
It has been shown that, when the state $\kph$ is sampled from $\tHVtu$ randomly according to the uniform measure as in section~\ref{s:typMain}, the density matrices $\rho_\mathrm{S}$ and $\rho_\varphi$ are very close to each other with probability close to one.
This is the canonical typicality.

Therefore, when one is only interested in physical quantities of the system, it is typical that the state of the system is described by the canonical distribution.
We believe that this provides a satisfactory characterization and justification of the canonical distribution.

\paragraph{Main result:}
To complete the justification of the canonical distribution from the operational point of view, we still need to show that (i)~the result of a single measurement of a macroscopic quantity of the system almost coincides with the expectation value with respect to $\rho_\mathrm{S}$, and (ii)~$\rho_\mathrm{S}$ is indeed close to the canonical density matrix.

Both (i) and (ii) can be done starting from the canonical typicality.
But let us here present a result (which does not make an explicit use of the canonical typicality) which attains the goal directly.

Define the canonical expectation for the system as
\eq
\sbkt{\cdots}^\mathrm{can}_{V,\beta}:=
\frac{{\rm Tr}_{\rm S}[\,(\cdots)\,e^{-\beta\hH_{\rm S}}\,]}
{{\rm Tr}_{\rm S}[\,e^{-\beta\hH_{\rm S}}\,]},
\lb{canVb}
\en
where ${\rm Tr}_{\rm S}[\cdots]$ denotes the trace over the space $\calH_{\rm S}$.
Note that neither $\hH_{\rm int}$ nor $\hH_{\rm B}$ appears in the definition \rlb{canVb}.

We assume that the whole system satisfies the condition for the number of states \rlb{logO}, with $V$ replaced by $\Vtot$.

As in section~\ref{s:pure}, we let $\MV^{(1)},\ldots,\MV^{(n)}$ be extensive quantities and define the nonequilibrium projection $\Pneq$ as \rlb{Pneq1}, \rlb{Pneqnew} or \rlb{Pneqn} , but by replacing $m^{(i)}(u)\simeq\sbkt{\MV^{(i)}}^{\rm mc}_{V,u}/V$ by the canonical expectation value $\sbkt{\MV^{(i)}}^{\rm can}_{V,\beta}/V$.
We then make a crucial assumption that the canonical version of the thermodynamic bound
\eq
\sbkt{\Pneq}^{\rm can}_{V,\beta}\le e^{-\gamma'(\beta)\,V}
\lb{canTDB}
\en
is valid for sufficiently large $V$ with $\gamma'(\beta)>0$.
This is almost the standard large deviation upper bound, which is expected to be valid in general, and can be proved for models treated in sections~\ref{s:EQL} and \ref{s:EIsing}.

\begin{theorem}
\label{t:thcan}
Take the energy density $u$ such that \rlb{ucond} is valid, and let $\beta=\beta(u)$.
Assume that the bound \rlb{canTDB} holds.
We choose a normalized state $\kph\in\tHVtu$ randomly according to the uniform measure as in Theorem~\ref{t:thtyp}.
Then with probability larger than $1-e^{-\nu'V}$, we have
\eq
\sbkt{\varphi|\Pneq\otimes\mathsf{1}|\varphi}\le e^{-\alpha'V},
\lb{canmain}
\en
for sufficiently large $V$, where the constants satisfy $\nu'\simeq\gamma'(\beta)-\alpha'$.
\end{theorem}

The theorem says that, for an overwhelming majority of the states in the energy shell, the result of a single measurement of $\MV^{(i)}$ almost coincides with the canonical expectation value $\sbkt{\MV^{(i)}}^{\rm can}_{V,\beta}$ with probability close to one.
This provides a rather satisfactory justification of the canonical distribution.

As we noted before the ration $\lambda$ need not be large.
This is because the proof makes use of the equivalence of the microcanonical and the canonical ensembles.
When $\lambda$ is large, however, we see that the inverse temperature $\beta$ is essentially determined by the properties and the energy of the heat bath.

\bigskip\noindent
{\em Proof of Theorem~\ref{t:thcan}\/:}\/We shall prove that
\eq
\overline{\sbkt{\varphi|\Pneq\otimes\mathsf{1}|\varphi}}\le e^{-\gamma''V},
\lb{canfin}
\en
where the left-hand side is the average over $\kph\in\tHVtu$ as defined in \rlb{avphi} and $\gamma''\simeq\gamma'(\beta)$.
Then by using the Markov inequality as in the proof of Theorem~\ref{t:thtyp}, we get the desired \rlb{canmain}.

To bound the average, we note that
\eq
\overline{\sbkt{\varphi|\Pneq\otimes\mathsf{1}|\varphi}}=
\sbkt{\Pneq\otimes\mathsf{1}}^{\rm mc}_{\Vtot,u}
\le\eta(u)\Vtot\sbkt{\Pneq\otimes\mathsf{1}}^{\rm can}_{\Vtot,\beta(u)},
\lb{PPeV}
\en
where $\sbkt{\cdots}^{\rm mc}_{\Vtot,u}$ and $\sbkt{\cdots}^{\rm can}_{\Vtot,\beta}$ are the microcanonical and the canonical expectations, respectively, of the whole system with Hamiltonian \rlb{HtotB}.
The equality and the inequality in \rlb{PPeV} follow from \rlb{MC2} and \rlb{A}, respectively.
Then exactly as in \rlb{elaMHC00}, the final expectation is bounded as
\eq
\sbkt{\Pneq\otimes\mathsf{1}}^{\rm can}_{\Vtot,\beta}\le e^{2\beta h_0V^\zeta}\sbkt{\Pneq}^{\rm can}_{V,\beta},
\lb{PPeV2}
\en
where the expectation in the right-hand side is the one defined in \rlb{canVb}.
By combining \rlb{PPeV}, \rlb{PPeV2}, and the assumption \rlb{canTDB}, we get \rlb{canfin}.~\qedm

\bigskip\bigskip
{\small
It is a pleasure to thank Shelly Goldstein and Takashi Hara, my collaborators in closely related topics, for valuable discussions and inspirations, 
Shin Nakano, Yoshiko Ogata, and Luc Rey-Bellet for their indispensable help in mathematical issues,
and
Marcus Cramer,
Fabian Essler,
Tatsuhiko Ikeda,
Joel Lebowitz,
Elliott Lieb,
Takashi Mori,
Hidetoshi Nishimori,
Peter Reimann,
Marcos Rigol,
Takahiro Sagawa,
Udo Seifert,
Akira Shimizu,
and
Yu Watanabe
for useful discussions and comments.
The present work was  supported in part by JSPS Grants-in-Aid for Scientific Research no.~25400407.
}

\end{document}